\documentclass[fleqn,usenatbib]{mnras}
\usepackage{bookmark}
\usepackage{graphicx}
\usepackage[usenames, dvipsnames]{color}
\usepackage{mathptmx}
\usepackage{txfonts}
\usepackage[T1]{fontenc}
\usepackage{ae,aecompl}

\def\Sec{${}^{\prime\prime}$\llap{.}}

\title[Homogeneous photometry of globular clusters]{Homogeneous 
photometry VII. Globular clusters in the {\em Gaia} era.}

\author[P.~B.~Stetson et al.]{
	P.~B.~Stetson$^{1}$\thanks{E-mail:Peter.Stetson@nrc-cnrc.gc.ca},
    E.~Pancino$^{2,3}$,
    A.~Zocchi$^{4}$,
    N.~Sanna$^{2}$,
    M.~Monelli$^{5,6}$\\
$^{1}$Herzberg Astronomy and Astrophysics, National Research Council,
5071 West Saanich Road, Victoria, British Columbia V9E 2E7, Canada\\
$^{2}$INAF -- Osservatorio Astrofisico di Arcetri, Largo Enrico Fermi 5,
I-50125 Firenze, Italy\\
$^{3}$Space Science Data Center, ASI, via del Politecnico snc, I-00133 Roma, Italy\\
$^{4}$European Space Research and Technology Centre (ESA/ESTEC), Keplerlaan 1,
2201 AZ Noordwijk, The Netherlands\\
$^{5}$Instituto de Astrof\'\i sica de Canarias, Calle Via Lactea, E-38205, La 
Laguna, Tenerife, Spain\\
$^{6}$Universidad de La Laguna, Dpto. Astrof\'\i sica, E-38206, La Laguna,
Tenerife, Spain \\}

\begin{document}

\date{Accepted XX month XX. Received YY month YY; in original form ZZ month ZZ}

\pagerange{\pageref{firstpage}--\pageref{lastpage}} \pubyear{2002}

\maketitle

\label{firstpage}

\begin{abstract}  We present wide-field, ground-based Johnson-Cousins {\it
UBVRI\/} photometry for 48 Galactic globular clusters based on almost
90\,000 public and proprietary images. The photometry is calibrated with the
latest transformations obtained in the framework of our secondary standard
project, with typical internal and external uncertainties of order a few
millimagnitudes. These data provide a bridge between existing small-area,
high-precision HST photometry and all sky-catalogues from large surveys like
{\em Gaia}, SDSS, or LSST. For many clusters, we present the first publicly
available photometry in some of the five bands (typically $U$ and $R$). We
illustrate the scientific potential of the photometry with examples of surface
density and brightness profiles and of colour-magnitude diagrams, with the
following highlights: {\em (i)} we study the morphology of NGC\,5904, finding a
varying ellipticity and position angle as a function of radial distance; {\em
(ii)} we show $U$-based colour-magnitude diagrams and demonstrate that no
cluster in our sample is free from multiple stellar populations, with the
possible exception of a few clusters with high and differential reddening or
field contamination, for which more sophisticated investigations are required.
This is true even for NGC\,5694 and Terzan\,8, that were previously considered as
(mostly) single-population candidates. \end{abstract}



\section{Introduction}
\label{sec:intro}

Star clusters, and especially globular clusters (GCs), have always been key
objects: witnesses to the initial epochs of star formation in the Universe,
exceptional dynamical laboratories, and test benches for stellar structure and
evolution theories. For decades, the traditional simple understanding of GCs
considered them to be spherically symmetric, isotropic, non-rotating stellar
systems, each made of a single homogeneous stellar population.  This situation
evolved following the technological revolution started in the 
1980's---charge-coupled device (CCD) cameras on small and large telescopes,
high-resolution and multi-object spectrographs, astronomical data archives---that
has provided a huge amount of information, often of exquisite quality. 

\begin{table*}
\caption{Summary of facilities used, with the range of years covered by the
observations and the total number of images collected in each photometric band,
(n$_o$ are non-standard filters).
More details, including the full image credits, can be found in
Appendix~\ref{sec:credits}.}
\label{tab:datasets}      
\begin{small}           
\centering          
\begin{tabular}{lllrrrrrrr}     
\hline\hline       
Site            & Telescope       & Instrument                           & Dates     &n$_U$& n$_B$& n$_V$& n$_R$& n$_I$&n$_o$ \\
\hline
Apache Point    & ARCSAT 0.5m     & APOGEE                              & 2017      &    0 &   12 &   22 &    0 &    1 &    0 \\
Apache Point    & APO 3.5m        & Arctic                              & 2017      &    0 &   10 &   12 &    0 &    0 &    0 \\
BNAO Rohzen     & BNAO 2.0m       & Photometrics CE200A                 & 2001      &    0 &  212 &  213 &    2 &  168 &    0 \\
Cerro Pach\'on  & SOAR 4.1m       & SOI                                 & 2008-2012 &  276 &  254 &  288 &  144 &  504 &  534 \\
Cerro Tololo    & CTIO 0.9m       & TI1/2, Tek, Tek2k, RCA              & 1986-2012 &   39 & 2377 & 2630 &  416 &  930 &  876 \\
Cerro Tololo    & CTIO 1.0m       & Y4KCam                              & 2006-2013 &   29 & 2144 & 2654 &  349 & 1110 & 1362 \\
Cerro Tololo    & CTIO 1.3m       & ANDICAM                             & 2005-2008 &    0 &  270 &  294 &    0 &  257 &    0 \\
Cerro Tololo    & CTIO 1.5m       & TI1/2, Tek, Tek2k, Site2K           & 1987-1998 &   16 &  234 &  234 &   20 &  143 &    0 \\
Cerro Tololo    & CTIO 4.0m       & TI1/2, Mosaic1/2, Tek2k, DECam      & 1983-2016 &  457 & 1325 & 1688 &  178 & 1838 & 2958 \\
DAO Victoria    & DAO 1.8m        & SITE1                               & 1995-1996 &    0 &    7 &   12 &    8 &    8 &    0 \\
ESO La Silla    & Dutch 0.9m      & Tektronic 33                        & 1997      &    0 &    0 &  167 &    0 &  166 &    0 \\
ESO La Silla    & Danish 1.5m     & MAT/EEV, DFOSC                      & 1995-2005 &    8 & 2689 & 5079 & 2114 &  890 &  117 \\
ESO La Silla    & ESO/MPI 2.2m    & WFI                                 & 1999-2012 & 2456 & 4000 & 5072 & 1096 & 3647 &  184 \\
ESO La Silla    & NTT 3.6m        & EMMI, SUSI, EFOSC                   & 1992-2014 &  184 & 1991 & 1159 &  321 &  819 &  108 \\
ESO Paranal     & VLT 8.0m        & VIMOS, FORS1, FORS2                 & 1999-2012 &  200 &  304 &  692 &  243 &  306 &    0 \\
Kitt Peak       & KPNO 0.9m       & RCA, T2kA                           & 1984-1999 &    0 &  179 &  188 &    0 &   93 &   92 \\
Kitt Peak       & KPNO 2.1m       & CCD, T1kA                           & 1992-1994 &    0 &    1 &   74 &    1 &   77 &    0 \\
Kitt Peak       & WIYN 3.5m       & WIYNMiniMos                         & 2001-2014 &    0 &    0 &   12 &   18 &   30 &   18 \\
Kitt Peak       & KPNO 4.0m       & RCA, Mosaic1, T2kB                  & 1984-2011 &    0 &  213 &  231 &   48 &  180 & 2075 \\
La Palma        & JKT 1.0m        & GEC, EEV, SITE, TEK                 & 1991-2003 &    5 &  137 &  240 &  118 &  190 &    2 \\
La Palma        & INT 2.5m        & GEC, EEV, Tek, WFC, Patterson       & 1986-2018 & 1157 & 3129 & 2575 &  852 & 2344 &  320 \\
La Palma        & NOT 2.6m        & CCD7                                & 1993-2001 &    0 &   45 &   78 &    0 &   86 &    0 \\
La Palma        & TNG 3.6m        & OIG, GEC, E2V, Loral                & 1990-2009 &   26 &  113 &  176 &    0 &   97 &    0 \\
Las Campanas    & Warsaw 1.3m     & 8k MOSAIC                           & 2003-2008 &    0 &  983 & 1077 &    0 &  112 &    0 \\
Las Campanas    & Magellan 6.5m   & MagIC, IMACS                        & 2006-2009 &    0 &   91 & 7359 &    0 &  128 &    0 \\
MMO Nantucket   & MMO 0.4m        & Roper PVCAM                         & 2014      &    0 &    0 &  146 &  145 &    0 &    0 \\
Mauna Kea       & UH 2.2m         & CCD                                 & 1986      &    0 &    2 &    2 &    1 &    3 &    0 \\
Mauna Kea       & CFHT 3.6m       & Lick, HRCam, CFH12K, MegaPrime      & 1984-2014 &    5 &  736 & 1208 &  383 &  764 &  520 \\
Mauna Kea       & Subaru 8.2m     & SuprimeCam                          & 2008      &    0 &  250 &  270 &    0 &    0 &    0 \\
McDonald        & LCOGT 1.0m      & CCD camera                          & 2013      &    0 &    0 &  120 &    0 &  109 &    0 \\
Siding Spring   & LCOGT 1.0m      & CCD camera                          & 2014      &    0 &    0 &   28 &    0 &   22 &    0 \\
Siding Spring   & AAT 3.9m        & CCD1                                & 1991      &    0 &    6 &    0 &   28 &    0 &    0 \\
Sutherland      & SAAO 1.0m       & CCD 986                             & 2006      &   56 &   56 &   68 &    1 &   70 &    0 \\
Sutherland      & LCOGT 1.0m      & CCD camera                          & 2013-2014 &    0 &    0 &  449 &    0 &  380 &    0 \\
San Pedro M\'artir & SPM 0.84m    & E2V-4290                            & 2014-2018 &   19 &  631 &   28 &   28 &   29 &    0 \\
San Pedro M\'artir & SPM 2.14m    & E2V-4290                            & 2018      &    9 &   13 &   13 &   14 &   14 &    0 \\
Teide           & IAC80 0.8m      & CAMELOT                             & 2017-2018 &   22 &   37 &   31 &   28 &   31 &    0 \\
\hline
\end{tabular}
\end{small}
\end{table*}

Deviations from spherical symmetry have been detected for many Galactic
\citep{geyer83,white87,chen10} and extra-Galactic
\citep{elson87,davoust90,han94,barmby07,vandenbergh08,wang13} GCs, motivating a
treatment of these systems that takes into account their flattening, and demanding
that the cause of the observed morphology be determined. Rotation both in the plane
of the sky \citep{bellini17,gaia_gc,bianchini18,milone18} and along the line of
sight \citep[e.g., see][]{fabricius14,boberg17,cordero17,lanzoni18}, pressure
anisotropy \citep{vanleeuwen00,watkins15}, and effects due to  interaction with the
tidal field of the host galaxy \citep{drukier98,kupper10,dacosta12} have also been
detected in many GCs, pointing out the need for a more realistic description of
their dynamics that would allow us to finally assess the formation mechanism of
these systems and to fully understand their evolution. The ESA space mission {\em
Gaia} \citep{gaia}, in particular, by providing access to the full six-dimensional
phase space of positions and velocities for large numbers of stars in clusters,
will finally allow us to probe the full complexity of their structural, kinematic,
and chemical properties through time.  

Moreover, GCs are now known to host Multiple Populations (MPs)
\citep{kraft94,gratton12,bastian17}. The first convincing evidence for MPs comes
from abundance differences among their stars, in particular anti-correlations in
a set of light elements (C, N, O, and Na; often also Mg, Al, Li, and F) that
have been found in GCs with adequate data. Generally, these variations are {\it
not\/} accompanied by a net chemical evolution: heavy elements and the total
C+N+O abundance are constant in most GCs, except for a minority
\citep[$\simeq20$\%, see][]{milone17} of anomalous GCs (see also
Section~\ref{sec:clean}). The probable origin of these variations is found in
the CNO cycle, converting hydrogen into helium, with its hotter Ne-Na and Mg-Al
sub-cycles \citep{denisenkov89,kraft94}. Unfortunately, even the most successful
scenarios proposed so far \citep{decressin07,dercole08} appear to suffer from
major drawbacks \citep{bastian13,renzini15}, and the problem can be considered
unsolved at the present time. Many spectroscopic investigations have explored
the chemistry of GCs in great detail, and large spectroscopic surveys such as
APOGEE  \citep{meszaros15} and Gaia-ESO \citep{pancino17b} are complementing the
limited spectroscopic capabilities of {\em Gaia} for abundance determinations.

The complex chemistry of GCs is reflected in their photometric properties.
Multiple evolutionary sequences have been observed in some colour-magnitude
diagrams (CMDs), starting with the ground-based discovery of the anomalous red
giant branch (RGB) in $\omega$~Cen \citep{lee99b,pancino00}.  Since then,
space-based imagery has made it possible to routinely obtain exquisitely fine
details in the colour-magnitude and colour-colour diagrams of most globular
clusters \citep{piotto15,milone17}:  multi-band photometry involving
ultraviolet filters is especially effective thanks to the $U$-band's sensitivity
to CN abundance, allowing us to study the relative behaviour, proportion, and
distribution of the MPs \citep{marino08,lardo11,sbordone11,milone17}.

However, space-based photometry, which is the ideal means to study the crowded
central regions of GCs, typically covers a relatively small area, generally
amounting to no more than one or two half-light radii, depending on the GC. On the
other hand {\em Gaia}---which offers extremely precise and accurate photometry over
the whole sky---only contains three broad-band colours ($G$, $G_{BP}$, and
$G_{RP}$), that are not able to separate MPs. Finally, large surveys like SDSS
\citep{abolfathi18}, which offer multiband, all-sky photometry and could nicely
complement space-based photometry, have not employed the point-spread-function
(PSF) fitting techniques that are required to obtain the best results in crowded
stellar fields \citep[][]{an08}.

\begin{table*}
\caption{For the GC sample considered in this paper, we list our new central
coordinates (Section~\ref{sec:centroids}); the metallicity [Fe/H], reddening
E(B--V), total luminosity M$_{\rm{V}}^{tot}$, and apparent visual distance modulus
(M--m)$_{\rm{V}}$ from the \citet{harris10} catalogue; the total number of images
analyzed per GC (n$_{\rm{images}}$); the number of RR~Lyrae variables catalogued
per GC (n$_{\rm{RR}}$); the X$_0$ and Y$_0$ GC centroids
(Section~\ref{sec:centroids}); and the priority ranking based on the criteria
described in Section~\ref{sec:gc}.} 
\label{tab:clusters}
\begin{small}
\centering
\begin{tabular}{llrrrrrrrrrrr}
\hline\hline
Cluster &  & RA         & Dec        & [Fe/H] & E(B--V) & M$_{\rm{V}}^{tot}$& (M--m)$_{\rm{V}}$ & n$_{\rm{images}}$ 
  & n$_{\rm{RR}}$ & X$_0$ & Y$_0$ & Rank \\
        &  & (hh mm ss) & (dd mm ss) & (dex)  & (mag)	& (mag)             & (mag)             & 
  &               & (pix) & (pix) & \\
\hline  
NGC\,104  & 47 Tuc     & 00 24 03.63 & --72 04 46.6 & --0.76 & 0.04 &  --9.42 & 13.37 & 2553 &   2 &  +303 &   +55 &   1 \\
NGC\,288  &            & 00 52 44.98 & --26 35 04.8 & --1.24 & 0.03 &  --6.74 & 14.83 & 1178 &   2 &  +235 & --444 &  21 \\
NGC\,1261 &            & 03 12 15.97 & --55 12 57.6 & --1.35 & 0.01 &  --7.81 & 16.10 & 2313 &  23 &  +492 &  +221 &  19 \\
NGC\,1851 &            & 05 14 06.82 & --40 02 47.0 & --1.22 & 0.02 &  --8.33 & 15.47 & 3068 &  48 &    +6 &    +3 &   7 \\
NGC\,1904 & M\,79      & 05 24 11.31 & --24 31 29.3 & --1.57 & 0.01 &  --7.86 & 15.59 & 2356 &  10 &   +11 &   --2 &  10 \\
NGC\,2298 &            & 06 48 59.37 & --36 00 20.0 & --1.85 & 0.14 &  --6.30 & 15.59 &  999 &   4 &    +2 &   --1 &  68 \\
NGC\,2808 &            & 09 12 05.43 & --64 51 53.9 & --1.15 & 0.22 &  --9.39 & 15.59 & 4023 &  18 & --156 &  +179 &  18 \\
E\,3      &            & 09 20 55.81 & --77 16 57.9 & --0.80 & 0.30 &  --2.77 & 14.12 &  380 &   0 &   +62 & --386 &  82 \\
NGC\,3201 &            & 10 17 36.35 & --46 24 41.1 & --1.58 & 0.23 &  --7.46 & 14.21 & 2224 &  86 &  +188 & --362 &  26 \\
NGC\,4147 &            & 12 10 06.34 &  +18 32 32.4 & --1.83 & 0.02 &  --6.16 & 16.48 & 1209 &  15 & --106 &   +70 &  62 \\
NGC\,4372 &            & 12 25 50.56 & --72 39 16.3 & --2.09 & 0.39 &  --7.77 & 15.01 &  450 &   0 &   +23 &   +17 &  42 \\
NGC\,4590 & M\,68      & 12 39 28.00 & --26 44 35.9 & --2.06 & 0.05 &  --7.35 & 15.19 & 1202 &  42 &     0 &   --1 &  20 \\
NGC\,4833 &            & 12 59 35.26 & --70 52 30.2 & --1.80 & 0.32 &  --8.16 & 15.07 & 1394 &  20 &  +103 & --395 &  24 \\
NGC\,5024 & M\,53      & 13 12 54.91 &  +18 10 06.8 & --1.99 & 0.02 &  --8.70 & 16.31 & 1686 &  63 &   --5 &   --3 &  14 \\
NGC\,5053 &            & 13 16 26.63 &  +17 42 00.9 & --2.29 & 0.04 &  --6.72 & 16.19 & 1114 &  10 &  +196 & --379 &  58 \\
NGC\,5139&$\omega$~Cen & 13 26 47.15 & --47 28 49.5 & --1.62 & 0.12 & --10.29 & 13.97 & 8611 & 198 &  --17 &  --11 &   5 \\
NGC\,5272 & M\,3       & 13 42 11.54 &  +28 22 42.9 & --1.50 & 0.01 &  --8.88 & 15.07 & 3461 & 241 &   +16 &    +9 &   4 \\
NGC\,5286 &            & 13 46 25.76 & --51 22 15.0 & --1.67 & 0.24 &  --8.61 & 15.95 & 2466 &  53 &  --10 &   +10 &  29 \\
NGC\,5634 &            & 14 29 36.95 & --05 58 36.1 & --1.88 & 0.05 &  --7.69 & 17.16 &  819 &  19 &   --5 &   --1 &  52 \\
NGC\,5694 &            & 14 39 36.87 & --26 32 30.6 & --1.86 & 0.09 &  --7.81 & 17.98 &  740 &   3 &  +285 & --364 &  67 \\
IC\,4499  &            & 15 00 18.15 & --82 12 55.6 & --1.60 & 0.23 &  --7.33 & 17.09 & 1870 &  99 &  +253 &  +334 &  83 \\
NGC\,5824 &            & 15 03 59.06 & --33 04 06.3 & --1.85 & 0.13 &  --8.84 & 17.93 & 1924 &  26 &    +7 &   --1 &  54 \\
NGC\,5904 & M\,5       & 15 18 33.38 &  +02 04 52.0 & --1.27 & 0.03 &  --8.81 & 14.46 & 3820 & 130 &  +340 & --197 &   3 \\
NGC\,5927 &            & 15 28 00.96 & --50 40 12.3 & --0.37 & 0.45 &  --7.80 & 15.81 &  292 &   0 &    +6 &    +9 &  64 \\
NGC\,5986 &            & 15 46 03.27 & --37 47 10.0 & --1.58 & 0.28 &  --8.44 & 15.96 & 1035 &  10 &   --2 &     0 &  36 \\
Pal\,14   & AvdB       & 16 11 01.11 &  +14 57 31.0 & --1.52 & 0.04 &  --4.73 & 19.47 &  465 &   0 &  --55 &    +2 & 116 \\
NGC\,6121 & M\,4       & 16 23 35.24 & --26 31 33.3 & --1.20 & 0.36 &  --7.20 & 12.83 & 5927 &  50 &  +270 & --378 &  38 \\
NGC\,6101 &            & 16 25 47.01 & --72 12 08.7 & --1.82 & 0.05 &  --6.91 & 16.07 &  794 &  18 &   +98 & --391 &  49 \\
NGC\,6205 & M\,13      & 16 41 41.50 &  +36 27 38.0 & --1.54 & 0.02 &  --8.70 & 14.48 & 1269 &   9 &	 0 &	+1 &   2 \\
NGC\,6218 & M\,12      & 16 47 14.38 & --01 56 54.1 & --1.48 & 0.19 &  --7.32 & 14.02 & 1324 &   2 &   --2 &   --2 &  23 \\
NGC\,6254 & M\,10      & 16 57 09.19 & --04 06 01.5 & --1.52 & 0.28 &  --7.48 & 14.08 & 4494 &   2 &	+3 &   --4 &  28 \\
NGC\,6341 & M\,92      & 17 17 07.73 &  +43 08 03.4 & --2.28 & 0.02 &  --8.20 & 14.64 & 1560 &  17 &	+5 &   --8 &   6 \\
NGC\,6366 &            & 17 27 44.45 & --05 04 46.0 & --0.82 & 0.71 &  --5.77 & 14.97 &  506 &   1 &  +294 & --212 &  94 \\
NGC\,6656 & M\,22      & 18 36 23.28 & --23 54 28.1 & --1.64 & 0.34 &  --8.50 & 13.60 & 3438 &  31 &  --42 &  --15 &  17 \\
NGC\,6712 &            & 18 53 04.66 & --08 42 08.4 & --1.01 & 0.45 &  --7.50 & 15.60 &  494 &  15 &  +333 &   +21 &  70 \\
NGC\,6752 &            & 19 10 53.43 & --59 59 08.4 & --1.54 & 0.04 &  --7.73 & 13.13 & 8703 &   0 &  --69 &   +97 &   8 \\
NGC\,6760 &            & 19 11 12.60 &  +01 01 41.6 & --0.52 & 0.77 &  --7.86 & 16.74 &  207 &   0 &  +328 &  --13 &  90 \\
NGC\,6809 & M\,55      & 19 39 58.78 & --30 57 50.4 & --1.81 & 0.08 &  --7.55 & 13.87 & 1772 &  15 &   --8 &   --7 &  16 \\
Terzan\,8 &            & 19 41 43.94 & --33 59 59.6 & --2.00 & 0.12 &  --5.05 & 17.45 &  404 &   3 &  +465 &  +121 & 117 \\
Pal\,11   &            & 19 45 14.62 & --08 00 27.7 & --0.39 & 0.35 &  --6.86 & 16.66 &  689 &   0 &	+4 &   --1 &  92 \\
NGC\,6838 & M\,71      & 19 53 46.53 &  +18 46 42.7 & --0.73 & 0.25 &  --5.60 & 13.79 & 1266 &   1 &  --94 &	+1 &  65 \\
NGC\,6934 &            & 20 34 11.02 &  +07 24 16.8 & --1.54 & 0.10 &  --7.46 & 16.29 &  743 &  79 &   --7 &	+2 &  53 \\
NGC\,6981 & M\,72      & 20 53 27.64 & --12 32 04.4 & --1.40 & 0.05 &  --7.04 & 16.31 & 1807 &  44 &   --4 &	+9 &  57 \\
NGC\,7006 &            & 21 01 29.37 &  +16 11 15.1 & --1.63 & 0.05 &  --7.68 & 18.24 & 1780 &  64 &   +50 & --348 &  66 \\
NGC\,7078 & M\,15      & 21 29 58.43 &  +12 10 02.9 & --2.26 & 0.10 &  --9.17 & 15.37 & 2292 & 164 &  +356 &  +648 &  11 \\
NGC\,7089 & M\,2       & 21 33 27.37 & --00 49 20.0 & --1.62 & 0.06 &  --9.02 & 15.49 & 1185 &  38 &	+7 &   +84 &   9 \\
NGC\,7099 & M\,30      & 21 40 22.05 & --23 10 43.7 & --2.12 & 0.03 &  --7.43 & 14.62 &  616 &   7 &   +32 &  --13 &  15 \\
NGC\,7492 &            & 23 08 26.47 & --15 36 39.4 & --1.51 & 0.00 &  --5.77 & 17.06 &  350 &   3 &  +337 &	+2 &  75 \\
\hline
\end{tabular}
\end{small}
\end{table*}

The goal of the present study is therefore to provide multiband, wide-field,
ground-based photometry to complement the upcoming data from {\em Gaia} and large
photometric and spectroscopic surveys, based on a large collection of publicly
available and proprietary images accumulated by the first author over many years.
By bridging the gap between the space-based photometry of GC centres and the large
surveys covering their outer parts, this catalogue will allow for a complete
coverage of the extent of these systems, making it possible to study them with
unprecedented detail.

This paper is organized as follows: in Section~\ref{sec:data} we describe the
sample selection, data sources, and image pre-reductions; in Section~\ref{sec:phot}
we describe the photometric measurement and  calibration process; in
Section~\ref{sec:cats} we illustrate the final catalogue content; in
Section~\ref{sec:cmd} we present some preliminary results; in
Section~\ref{sec:concl} we summarize our findings and draw conclusions.



\section{Sample and data reductions}
\label{sec:data}

The photometry presented in this paper is based on a large data
collection\footnote{http://www3.cadc-ccda.hia-iha.nrc-cnrc.gc.ca/en/community/STETSON/index.html},
containing more than half a million astronomical images obtained with CCDs for
popular science targets like dwarf galaxies, open clusters, globular clusters,
supernova hosts, photometric standard fields, and other targets of wide
astronomical interest \citep{photo1}.  The majority of these images are
public-domain data downloaded from astronomical data archives, although there are
also some proprietary images obtained by ourselves or donated by colleagues. Some
of these latter date back to the days before astronomical archives existed:  the
earliest CCD images in the data set are from January 8, 1983 (see
Table~\ref{tab:datasets} and Appendix~\ref{sec:credits}), and we continue to
accumulate data even now.

PSF photometry is extracted from these digital images with DAOPHOT and ALLFRAME,
and is aperture-corrected and merged using the associated suite of ancillary
software packages \citep{daophot1,daophot2,stetson94}.  The resulting photometry
in the {\it UBVRI\/} filters is homogeneously calibrated to the photometric
system defined by \citet{landolt92a} employing colour transformations optimized
independently for each observing run, each night, and each CCD used for
observation.  In the process we define a large selection of secondary
photometric standard stars in the {\it UBVRI\/} filters to help register the
many different observing runs to a common photometric system. Some results from
the collection have already been published for dwarf galaxies
\citep{smecker94,stetson98,stetson14}, GCs
\citep{stetson05a,stetson05b,bergbusch09,viaux13}, and Open Clusters 
\citep[OCs,][]{stetson03,brogaard12}, particularly in the ``Homogeneous
photometry" series of papers.

The database is constantly evolving with the addition of new images, the definition
of new secondary standards, and the refinement of the photometric indices for
existing standards. Occasionally a star that has been used as a standard is
discarded as evidence of intrinsic variability becomes significant.  Thus the
photometric catalogues published by us here and elsewhere are subject to revision
over time, but this takes the form of incrementally improved precision and accuracy
in the published photometric indices. Constant referral back to Landolt's original
photometric results \citep{landolt92a} safeguards against secular drift of our
photometric system over time.  In any given publication, the results we present are
the best we can do at that particular time and we make every attempt to provide an
objective description of the instantaneous quality of the data. So far, none of the
claimed results of the previous studies have been challenged in any way by our
subsequent improvements in the data.

Here we present photometry for 48 GCs (Section~\ref{sec:gc}) obtained from the
current image collection (Section~\ref{sec:datasets}) using the latest
photometric solutions (Section~\ref{sec:photcal}). The present catalogues
supersede previous publications, such as those for 47~Tuc by
\citet{bergbusch09}, for M\,15 by \citet{stetson94}, or for the GC collection in
the SUMO project \citep{monelli13}.  The previously published results are fully
consistent, within the claimed uncertainties, with those presented here.

The major strengths of these catalogues are the wide field of view, the large
number of images used, the multiband optical {\it UBVRI\/} photometry, and the
robust, precise photometric calibrations based, typically, on hundreds of
photometric standard-star observations in each filter on each CCD on each night.
To illustrate: the current data set contains some 7\,508 night-chip-filters of
calibrated observations. Among these, the median number of photometric standard stars observed per
night per chip per filter is 1\,236.  

For the central regions of the target clusters the ground-based photometry
presented here has less precision and less depth than space-based photometry
\citep{sarajedini07,piotto15} due to stellar crowding in the central parts.  On
the other hand, due to the use of the somewhat more venerable {\it UBVRI\/}
bandpasses, our results may be more directly and accurately compared to many
legacy studies in the historical literature.  Furthermore, it would require only
minor effort to find stars in common between our catalogues and the published
space-based catalogues to establish relative transformations and merge the datasets.
We have not done that here because it is not essential for our immediate
purposes and would unnecessarily expand this article.  

   \begin{figure*}
   \vspace{-0.5cm}
   \includegraphics[trim={1cm, 0, 0, 0},width=6cm]{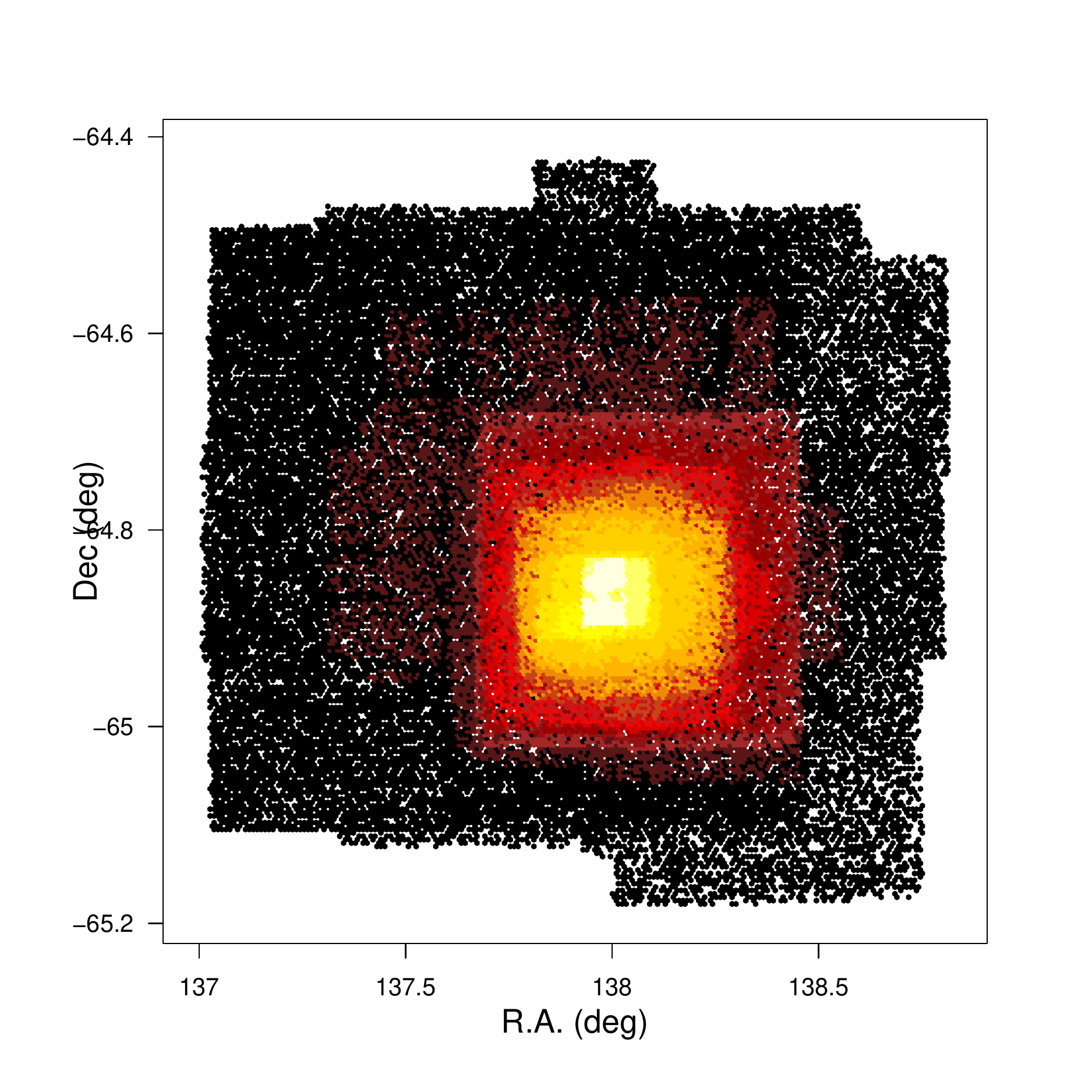}\includegraphics[trim={1cm, 0, 0,  0},width=6cm]{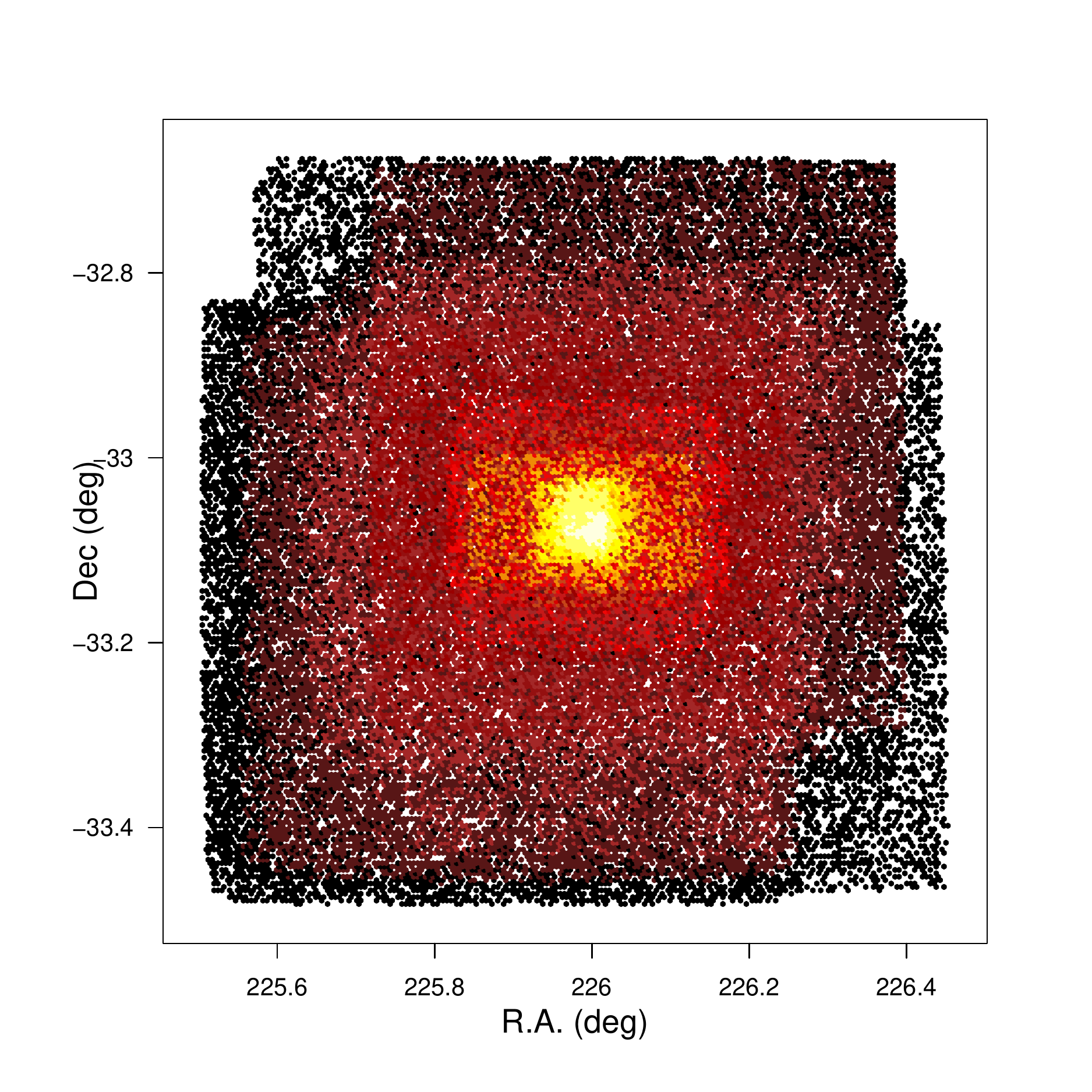}\includegraphics[trim={1cm, 0, 0, 0},width=6cm]{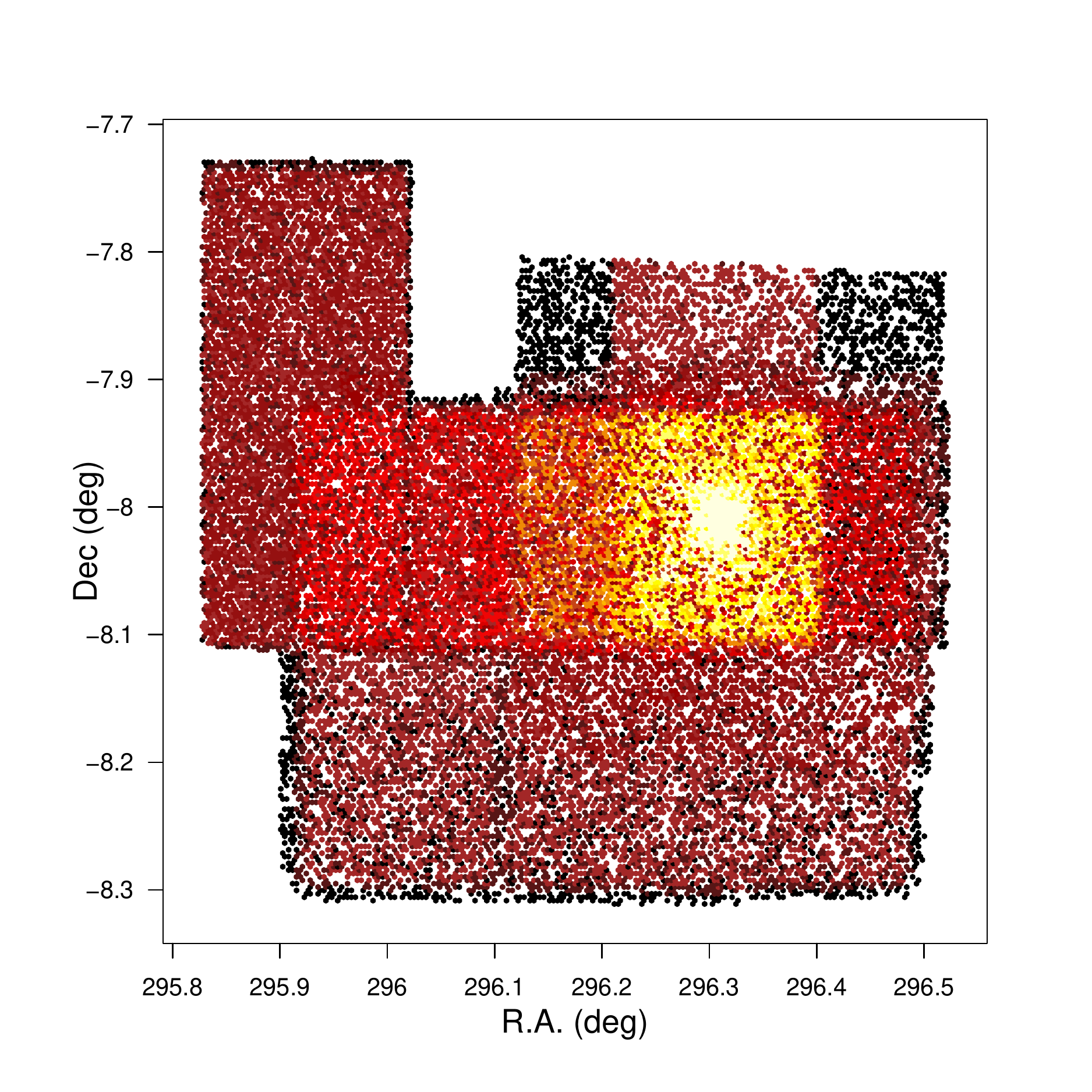}
   \vspace{-0.3cm}
      \caption{Examples of binned maps, with bins colored according to the
      maximum number of exposures (black lowest and yellow highest, varying as
      detailed below). {\em Left  panel:} NGC\,2808, one of the most covered
      GCs, with almost 4\,000 images used in the central regions (see
      Table~\ref{tab:clusters}). {\em Center panel:} NGC\,5824, with
      almost 2\,000 images in the central regions and a more uniform
      coverage. {\em Right panel:} Palomar~11, with its peculiar image
      arrangement and $\simeq$650 images in the center.}
   \label{fig:heat}
   \end{figure*}


\subsection{Cluster selection}
\label{sec:gc}

Initially, we prioritized GCs in the \citet{harris96,harris10} catalogue based on
their global properties, according to general criteria of scientific utility that
should also make these GCs particularly suitable for study with the first {\em
Gaia} data releases: {\em (i)} low reddening, and thus also low differential
reddening, to make the CMD sequences easier to interpret; {\em (ii)} high total
luminosity, to have better number statistics for model comparisons; {\em (iii)} low
apparent distance modulus, to get better photometry at any given absolute stellar
magnitude. By combining these three criteria, interstellar reddening is penalized
twice because it makes the colours harder to interpret and extinction makes the
stars appear fainter and more difficult to measure with precision. The number of RR~Lyrae
stars was also considered as an additional criterion, but we found that it did not
change the cluster ranking significantly. 

Given the above prioritization, the GC choice was then finalized based on the
availability of sufficient images obtained under photometric observing
conditions\footnote{Our working definition of ``photometric'' observing
conditions is somewhat vague, but in general it means during the course of the
night the r.m.s.\ dispersion of the fitting residuals for the standard stars observed is $\lesssim
0.03$\,mag, and the derived extinction coefficients seem reasonable.  In our
experience the extinction coefficients for a 2\,000\,m mountain are $\sim 0.48$
in $U$, 0.25 in $B$, 0.14 in $V$, 0.105 in $R$, and 0.06 in $I$.  La~Palma, in
particular, occasionally has significantly larger extinction coefficients due to
dust blown from the Sahara.  The actual r.m.s.\ residual observed in each filter
on each night is included as a component of the uncertainty associated with each
resulting magnitude measurement, which is passed along to subsequent stages of the
analysis.}, the essential criterion being the possibility of deriving an accurate
absolute calibration (see Section~\ref{sec:photcal} for more details) in {\it
all five} of the {\it UBVRI} photometric bandpasses. The availability of
$U$-band photometry is important because it is very sensitive to the CN
molecular bands around 388\,nm, especially in red giants, which makes it a good
indicator of multiple populations in GCs \citep{sbordone11}. The availability of
the $R$ bandpass is less essential because for normal stars its information is
largely redundant with that from the $V$ and $I$ bandpasses.  Nevertheless, in
order to somewhat limit the scope of this paper, we include in our present
sample only clusters with fundamentally calibrated photometry in all five
bandpasses. 

This leads to the regrettable exclusion of high-priority GCs like NGC\,6723 (ranked
\#12 out of 150) or NGC\,6541 (\#13), because they lack the $R$ filter, but it allows
the inclusion of lower-priority GCs like Terzan~8 and Pal\,14.  We anticipate there
will be a future paper like this one that will provide catalogues for clusters
currently lacking the photometric $R$ band---a few of them, indeed, will have
calibrated $R$-band photometry by the time the next paper is ready, since we are
continuing to accumulate data. The final list of GCs considered here, sorted by
right ascension, is reported in Table~\ref{tab:clusters}, along with relevant GC properties.


\subsection{Data pre-reduction}
\label{sec:datasets}

The photometric catalogues of the 48 GCs presented here are obtained from
84\,106 individual CCD images (not including standard fields) nominally in the
Johnson-Cousins passbands (see Figure~\ref{fig:heat} and
Table~\ref{tab:clusters}), plus an additional 9\,166 images obtained in other
filters; these latter do not contribute to the calibrated photometry of this
paper, but there were included in the reductions for the information they can
provide toward the completeness of the star catalogues and the precision of the
astrometry. The CCD images were collected with a variety of telescopes and
cameras, as summarized in Table~\ref{tab:datasets}; in all, data from 1\,327
nights divided among 390 observing runs are employed here. More details on the data
sources and the full image credits are given in Appendix~\ref{sec:credits}.  

In a small minority of cases, notably data from the Mosaic cameras on the CTIO and
KPNO 4m telescopes and from MegaCam on the CFHT 3.6m telescope, and occasionally data
contributed by colleagues, pre-processing of the images had already been performed. In
the remaining cases pre-processing was done employing the bias frames, flat-field
frames, etc.\ which had also been provided by the archives.  If such calibration
images were not available, we have not attempted to use the science images. Data were
reduced with in-house routines and IRAF\footnote{IRAF is distributed by the National
Optical Astronomy Observatories, which are operated by the Association of Universities
for Research in Astronomy, Inc., under cooperative agreement with the National Science
Foundation.} tasks \citep{iraf1,iraf2}. All images were corrected for direct current
(DC) bias offset, trimmed to the good imaging section, and corrected with a
two-dimensional bias masterframe. When suitable long/short flat-field images were
available (rare), two-dimensional corrections for shutter-timing errors were also
applied. For the most part dark-current correction was not performed under the
assumption that dark current can be treated as a component of the sky brightness,
which does not interest us and is in any case removed in the photometric analysis. On
the very rare occasions when the raw images had no overscan region, the master bias frames
removed the mean bias level as well as any stable two-dimensional electronic pattern;
to the extent that DC bias levels fluctuated during the night, this also can be
considered a component of the instantaneous sky brightness, which has no consequence
for our photometric results.

Whenever both high-signal dome and sky flat-field images were available, we used
the dome pattern to remove small-scale sensitivity variations, and the sky pattern to remove
large-scale variations, by constructing a master flat $ F = D \times \langle S/D
\rangle$, where $D$ is the master dome flat, $S$ is the master sky flat, and
$\langle S/D \rangle$ is a median-smoothed ratio of the sky and dome masterflats
with a smoothing scale of order a seeing disk. Finally, bad-pixel maps were
derived from median stacks of large numbers of science images and were used to
mask the science images before photometric processing, both for images
preprocessed by us and for those that had been preprocessed before we acquired
them.  


\section{Photometry}
\label{sec:phot}

   \begin{figure*}
   \vspace{-0.5cm}
   \includegraphics[trim={1cm, 0, 0, 0},width=6cm]{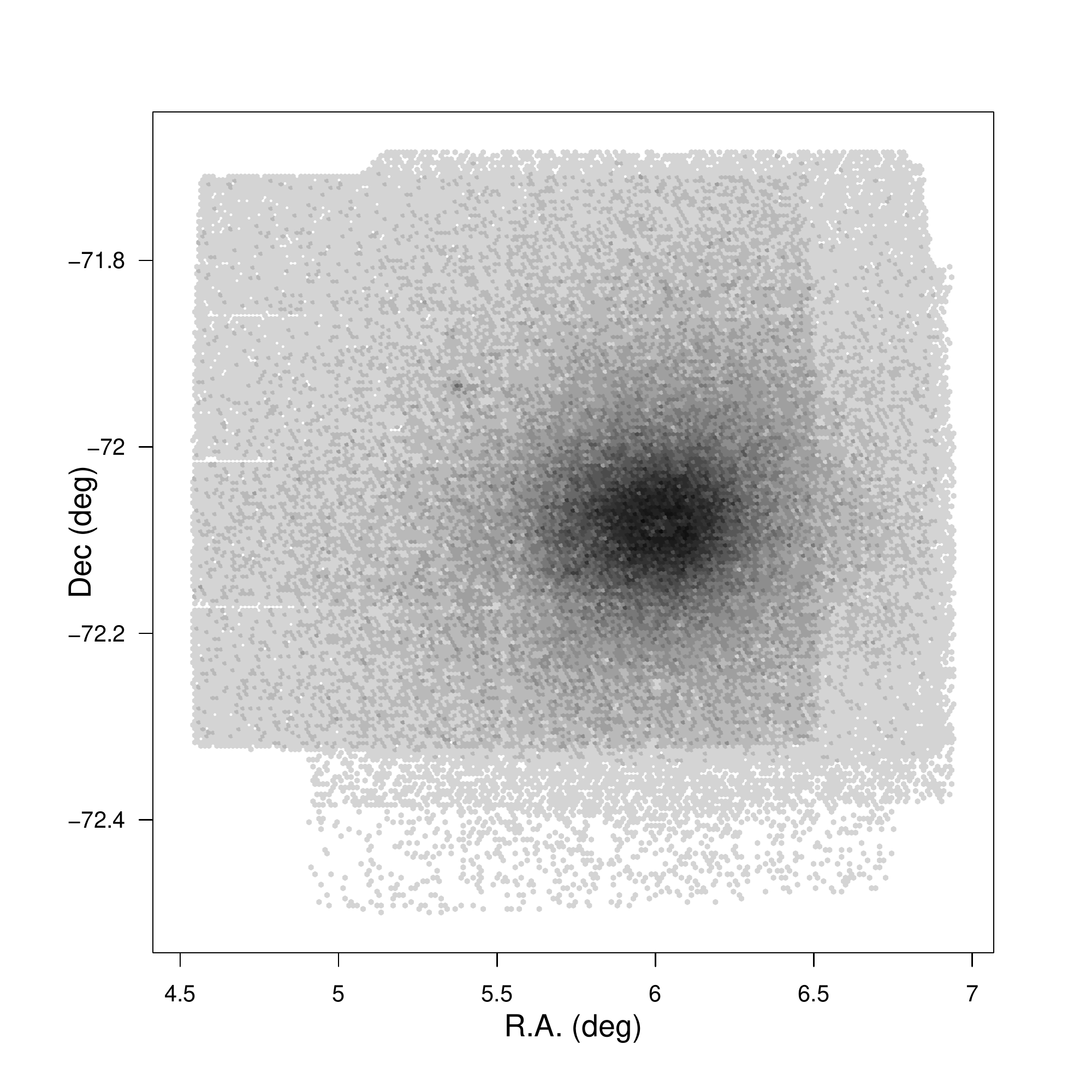}\includegraphics[trim={1cm, 0, 0, 0},width=6cm]{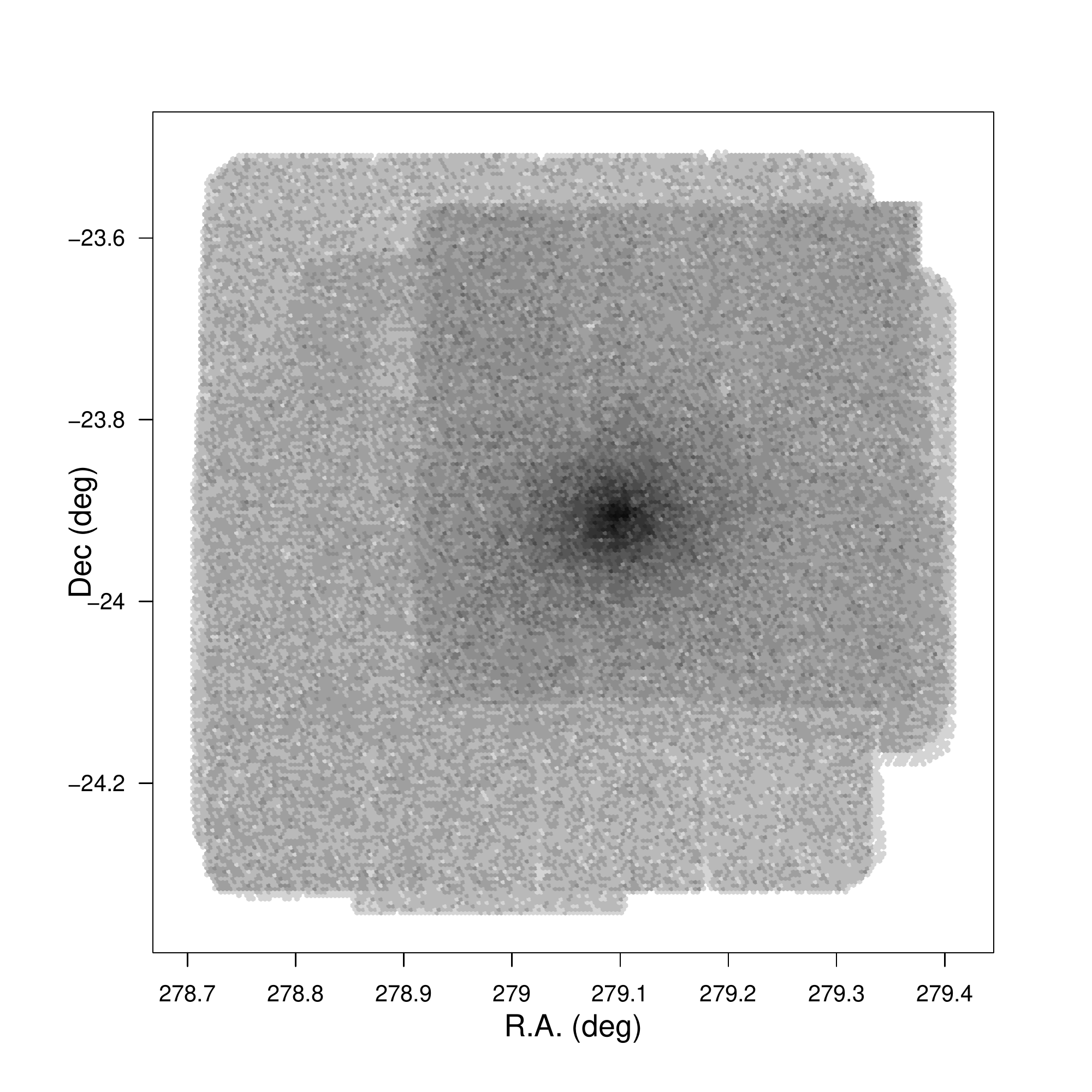}\includegraphics[trim={1cm, 0, 0, 0},width=6cm]{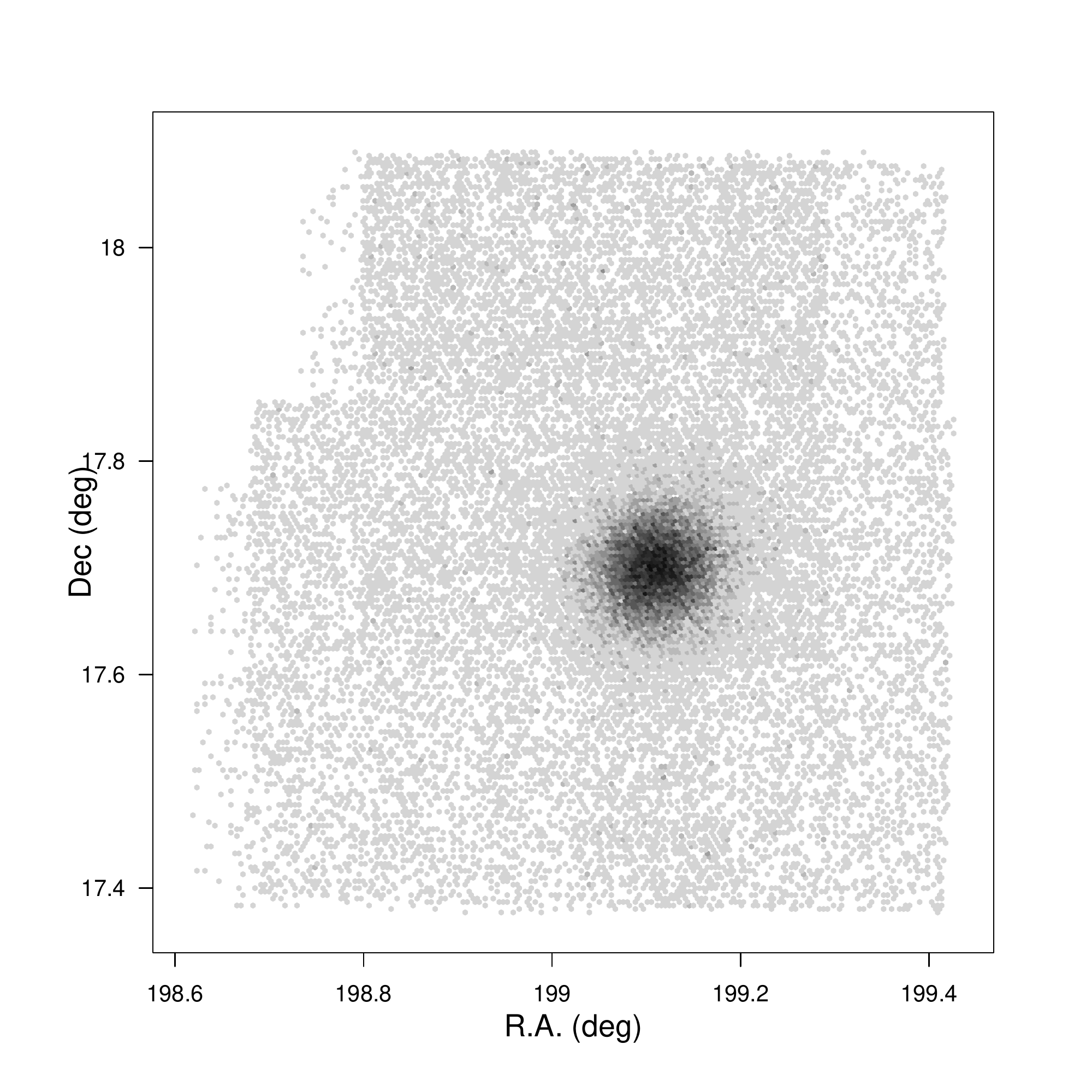}
   \vspace{-0.3cm}
      \caption{Examples of binned star density maps. {\em Left panel:}
      47~Tucanae. The decrease in star counts caused by crowding is visible in
      the center, as well as the image footprints of mosaic cameras. The
      overdensity on the upper left of the cluster center is Bologna~A
      \citep{bellazzini05}. A hole caused by a very bright star is also apparent
      right below the GC central overdensity. {\em Center panel:} M\,22, one of
      the most observed GCs in the sample. {\em Right panel:} the simpler case
      of NGC\,5053, whose photometry is based on fewer stars and CCD exposures.} 
   \label{fig:maps}
   \end{figure*}

Instrumental magnitudes were measured with the PSF-fitting packages DAOPHOT, ALLSTAR,
and ALLFRAME \citep{daophot1,daophot2,stetson94} following procedures initially
described in \citet{daophot1}, and amplified in, e.g., \citet{stetson94}. Candidate
stars were identified in the images, initial stellar brightness and sky brightness
estimates were derived from synthetic-aperture photometry, model PSFs were derived from
the brightest and most isolated of the unsaturated stars in each image, and an initial
pass of PSF photometry was performed.  After this, stars overlooked by the
star-finding algorithm were added by hand, and improved PSFs were obtained from the
original images after all stars {\it except\/} the PSF stars had been  provisionally
subtracted from the images.  Almost always the PSF was considered to vary
quadratically with position within the images except when the number of available bright, isolated
stars was insufficient (sometimes the case for early, small-format CCDs), or the
fitting residuals gave convincing evidence that a quadratically varying model was not
needed.  Corrections relating PSF magnitudes to synthetic-aperture magnitudes were
derived from growth-curve analysis of the PSF stars in images from which all other
stars had been subtracted.

Multiple images of the same science field were registered to a common shared geometric
reference system employing third-order polynomials (10 parameters in each of $x$ and
$y$), again excepting a few early, small-format images where the number of stars was
small and the higher-order distortions negligible.  Then final PSF magnitudes were
measured with the ALLFRAME routine imposing a common star list consistently to all
overlapping frames.  The simultaneous exploitation of information from all available
images ultimately allows for deeper and more precise photometry, with the deepest and
best-seeing images providing better positional constraints on faint detections as well
as self-consistent deblending in crowded regions.  


\subsection{Photometric calibration}
\label{sec:photcal}

The goal of our calibration procedure is to transform the measured instrumental
magnitudes to a photometric system as close to that of \citet{landolt92a} as
possible. This is done using venerable methodology dating back to the beginnings of
photoelectric photometry that takes into account both color and extinction
corrections (see Appendix~\ref{sec:color}).  Our approach also allows for some
variation of photometric zero-points (ZPs) across the face of each two-dimensional
detector (see Section~\ref{sec:illum}), determined independently for each observing
night and CCD. The average magnitudes and standard errors for each star are then
obtained by robust statistical methods, with the procedure described in detail in
Appendix~\ref{sec:color}.

The final photometric calibration for each of our science targets is then
performed in steps. Initially, candidate secondary photometric standards
\citep{standards1} are selected from each cluster catalogue by an automatic
filtering process followed by visual inspection to identify stars with
well-formed images that are also well isolated from neighbors. These are
provisionally calibrated on the basis of Arlo Landolt's standards
\citep{landolt92a}; stars with multiple concordant observations are considered
valid secondary photometric standards and their results are added to our library
of standard-star indices. This process is repeated for every science target
field for which we have multiple observations obtained under photometric
observing conditions, particularly when they came from multiple observing runs. 
This applies to {\it all\/} suitably observed targets: stars in OCs and GCs,
stars in and near dwarf galaxies, stars in the fields of supernova hosts, stars
in Landolt fields that were not themselves included in \citet{landolt92a}, and
all other targets meeting the basic acceptance criteria. At present, a candidate
is accepted as a secondary standard if it satisfies the following criteria: {\em
(i)} having at least five observations obtained under photometric observing
conditions; {\em (ii)} a computed standard error of the mean magnitude in a
given passband of $<$0.02~mag. Only passbands satisfying both these conditions
are accepted for any given star, and a star is retained only if at least two of
the five target magnitudes satisfy the criteria; magnitudes in  other passbands
not meeting these criteria are set to null values.  Additionally, any stars that
show evidence of intrinsic variability with r.m.s.\ dispersion $\geq$0.05~mag
considering all passbands together are rejected.  

All these secondary standards are then used to augment the primary standards in
refitting the calibration equations described in Appendix~\ref{sec:color} for
all CCDs on all nights of all observing runs. This recomputation does not
necessarily improve the absolute reference of our global average photometric
system to that of \citet{landolt92a}. It does, however, help to ensure that each
individual calibration equation for each CCD on each night of each run has been
referenced to the {\it same\/} photometric system, and random photometric
calibration errors resulting from small-number statistics are greatly reduced. 
This is especially beneficial for nights and runs where only a few Landolt
standards were observed over a limited range of airmasses. It also greatly
enhances the consistency even of runs where significant numbers of Landolt
standards were observed. But especially it allows us, for the first time, to
exploit data for our highly desirable science targets from runs where no Landolt
standards were observed. To gauge the scale of the improvement, when
\citet{landolt73} photometry is adjusted for trivial ZP displacements and
combined with the \citet{landolt92a} data, the resulting number of fundamental
\citet{landolt92a} standards that meet our acceptance criteria ($n\geq 5$,
$\sigma\leq0.02$ in at least two of five filters) are 427, 451, 456, 301, and
295 stars in $U$, $B$, $V$, $R$, and $I$, respectively. {\it As of this
writing\/}, our current working set of standards---Landolt's merged with our own
list of secondary standards---contains 64\,356, 164\,435, 180\,661, 88\,139, and
155\,241 stars in $U$, $B$, $V$, $R$, and $I$, respectively. If we count only
stars with $n\geq 100$ in a given filter, Landolt's merged catalogue contains
seven stars in {\it UBV} and three stars in {\it RI}; our combined list of
primary and secondary standards contains 597, 15\,109, 22\,895, 3\,487, and
11\,931 in $U$, $B$, $V$, $R$, and $I$, respectively.  These numbers are
constantly growing as we continue to incorporate additional data.  Accordingly,
the last significant digit in each measured  magnitude can change with time, but
there is no secular drift of our system as we constantly monitor and remove any
discrepancy with the fundamental Landolt standards, and the claimed standard
error that we associate with each measurement is a realistic description of the
instantaneous state of our knowledge.

Once the transformation-equation coefficients for the various
run/night/chip/filter combinations have all been redetermined on the basis of
the augmented standard-star list, somewhat relaxed acceptance criteria are
adopted in choosing the local standards to be used in calibrating each
individual science field. The criteria for {\it local\/} standards are {\em (i)}
having at least three observations obtained under photometric observing
conditions; {\em (ii)} a computed standard error of the mean magnitude of
$<$0.04~mag; {\em (iii)} no evidence of variability with r.m.s. $\geq$0.05~mag
based on all available filters.  At this point we are assuming that all the
transformation constants {\it except\/} the individual ZPs of the various images
are known with a precision vastly better than the uncertainties of our
individual instrumental magnitudes.  This final step is intended to 
remove any unanticipated systematic ZP errors due to, for instance, unmodeled
departures from the temporal and directional constancy of the derived extinction
coefficients, or inadequacies in the derived corrections from a relative PSF
magnitude scale to an absolute aperture-defined magnitude scale for each image. 
The individual photometric ZP of each image is redetermined via a robust
weighted averaging technique that iteratively reduces the weight of discrepant
observations \citep[see lecture 3b by][]{stetson89} producing a ZP that is
optimally consistent with the preponderance of available calibrators.  
Accepting all other parameters as given and redetermining the average photometric
ZP for each individual CCD image at this point is a last step in ensuring that
all observations are on a common photometric system: ``common'' with a tolerance
that is tighter than the tolerance to which it is ``correct.'' Stated
differently, redetermining the zero-point of every image at this stage produces
the tightest possible principal sequences in the CMD, but does not affect the
absolute placement of those sequences within colour-magnitude space.   

   \begin{figure}
   \centering
   \includegraphics[width=\columnwidth]{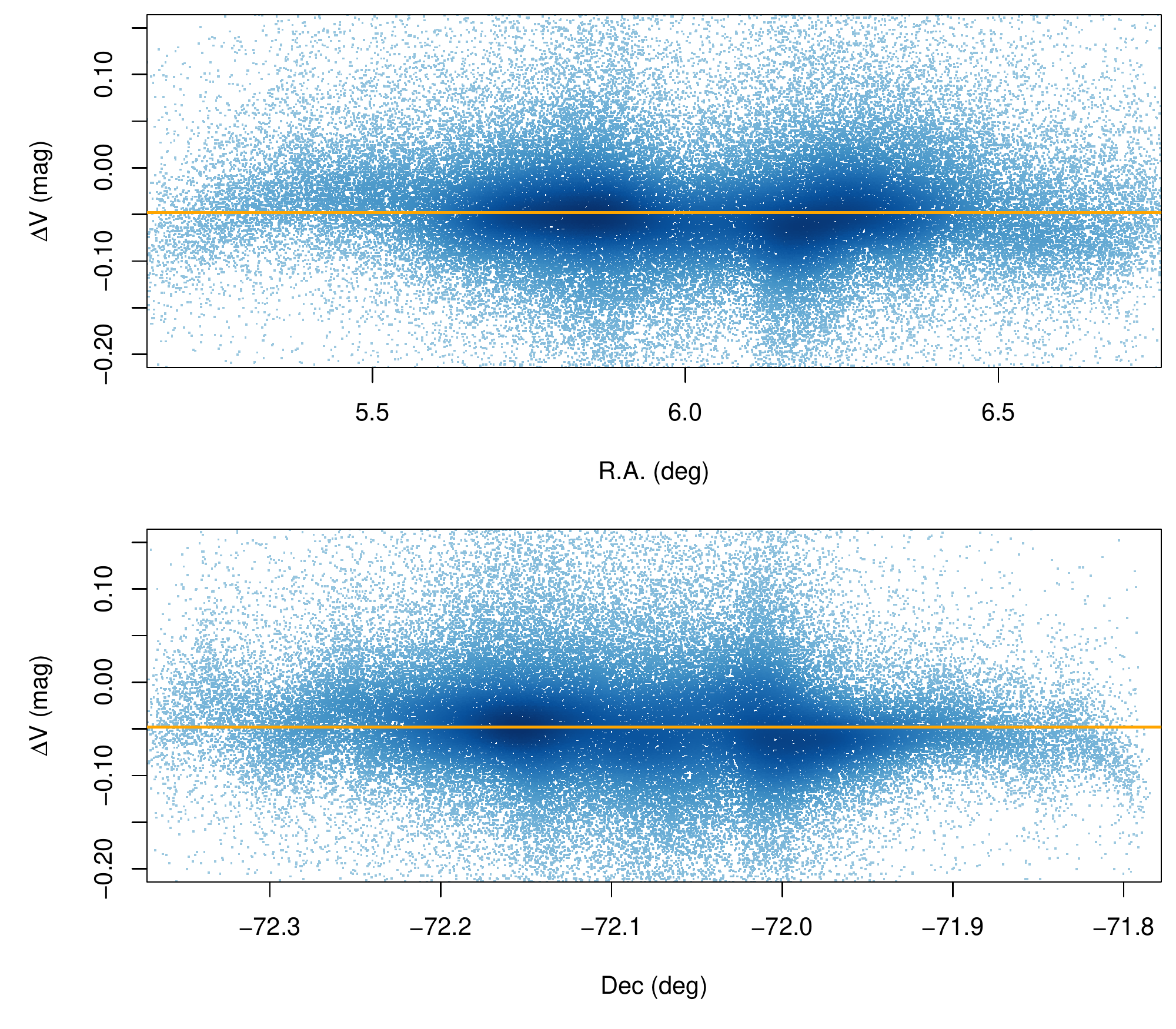}
   \vspace{-0.5cm}
      \caption{Example of ZP variations in wide field imagers. Our photometry of
      47~Tucanae is compared with that of \citet{kaluzny98} and plotted as a
      function of RA (top panel) and Dec (bottom panel). The vertical stripes
      correspond to the overlap of different exposures in the \citet{kaluzny98}
      photometry. Part of the wavy patterns, especially at the borders, are caused
      by light concentration (see Section~\ref{sec:illum}) in one or both
      catalogues.} 
   \label{fig:illum}
   \end{figure}

To give an idea of the numbers involved, {\it as of this writing\/} the 48 GCs
discussed here have a total of 61\,514 local
secondary standards. The ZPs of the 84\,106 individual CCD images\footnote{We
reduce and analyze each CCD separately even in case of mosaic cameras
(Section~\ref{sec:color}). For example, with WFI each exposure
produces 8 individual CCD images.} have been redetermined from a median of 155
local standards each. Among these, the median standard deviation of the
magnitude residual for an individual star is 0.009~mag in U, 0.004~mag in B,
0.003~mag in V, 0.007~mag in R, and 0.005~mag in I. Using this approach, the
robust weighted average results in a relative ZP whose reliability is not quite
as good as the uncertainty of a single magnitude measurement divided by the
square root of 155. In fact the median uncertainty in the derived photometric ZP
of a typical individual image is independently estimated {\it ex post facto} to be
$\simeq0.0015$~mag considering the r.m.s.\ repeatability of measurements in all
filters and for all targets.  The uncertainty contributed to the average
calibrated magnitudes for a given star by the calibration itself is therefore of
the order of this number divided by the square root of the number of images
where that star was measured in that bandpass.  The actual limiting precision
for a given average magnitude for a given star is therefore normally dominated
by readout noise, photon statistics, and crowding errors in the individual
images.  We recall that errors in the instantaneous atmospheric extinction were
removed by the redetermination of the ZPs of individual exposures.  


\subsection{Position-dependent photometric errors}
\label{sec:illum}

In all cases the present photometry includes wide- (mosaic) as well as
narrow-field (single-CCD) images (Table~\ref{tab:datasets}). The final combined
areal coverage of our calibrated photometry ranges from a maximum extent as
small as 35$\times$35\,arcmin for NGC$\,$6101 to as much as
80$\times$80$\,$arcmin for NGC$\,$288 (see other examples in
Figure~\ref{fig:maps}).  In the crowded center of each cluster in our catalogue,
where many images from many different instruments overlap (see
Figure~\ref{fig:maps}), any position-dependent photometric errors that remain after the
minor spatial corrections that we have applied should be small, and should be
further beaten down as the square root of the number of runs, nights, and dither
offsets.  They will certainly be less significant than systematic radial
gradients in photometric bias produced by crowding in the cluster centers;
crowding errors that affect the measured magnitudes should have a smaller effect
on the colors that are derived from them, but users of these data should still
be aware of possible small systematic errors in the colors of stars in the
crowded cluster centers.

The outermost regions of each field can suffer in principle from minor
large-scale ZP variations, aggravated by the fact that these regions typically
are contained in the smallest number of CCD images obtained during the smallest
number of independent observing runs. An example is given in
Figure~\ref{fig:illum}, where our photometry of 47~Tuc is compared with that by
\citet{kaluzny98}, who report CCD linearity effects and calibration problems for
fainter stars. The observed $V$ magnitude differences in the external regions
can reach $\simeq$0.07--0.08~mag. If we humbly assign half of the difference to each
study, we can place a generous upper limit to the maximum ZP variations of our
photometry of $\lesssim$0.04~mag in the external regions ($\gtrsim$0.5~deg from
the center). In fact, we believe that our careful calibration procedures produce
results that are considerably better than this.  Fortunately, any systematic
photometric errors in the outer parts of each field affect primarily stars
unassociated with the cluster.

\begin{table*}
\caption{Summary of literature comparisons (see also Figure~\ref{fig:lit} and
Section~\ref{sec:lit}). The available passbands are indicated, along with the
number of stars and GCs in common with our catalogue.}
\label{tab:lit}
\centering
\begin{footnotesize}
\begin{tabular}{@{}l@{}r@{~~}r@{~~}l l@{}r@{~~}r@{~~}l}
\hline\hline       
Reference & Bands & n$_{\star}$ & GCs & Reference & Bands & n$_{\star}$ & GCs\\
\hline 
\citet{alvarado95}$^a$            & $UBVRI$ &      19 & {\em 2 GCs}  & \citet{melbourne00}               & $BVI$   &  3\,237 & NGC\,4833    \\
\citet{alves01}                   & $BV$    &  5\,104 & NGC\,5986    & \citet{mclaughlin06}              & $UV$    &  3\,757 & NGC\,104 \\
\citet{anderson08}                & $VI$    &358\,120 & {\em 36 GCs} & \citet{mochejska02}               & $UBV$   & 11\,841 & NGC\,6121 \\
\citet{bellazzini02b}             & $VI$    &  1\,634 & NGC\,5634    & \citet{monaco04}                  & $BVI$   &136\,753 & NGC\,6656 \\
\citet{bellini09}                 & $UBVRI$ &339\,783 & NGC\,5139    & \citet{nardiello15}               & $UBVI$  &  3\,939 & NGC\,6121 \\
\citet{bergbusch96}               & $BV$    &     763 & NGC\,7099    & \citet{odewahn92}                 & $BVR$   &     133 & {\em 2 GCs} \\
\citet{buonanno87}                & $BV$    &  1\,331 & NGC\,7492    & \citet{ortolani90}                & $BV$    &  2\,444 & {\em 2 GCs} \\
\citet{cohen11}                   & $BV$    &  9\,683 & NGC\,6101    & \citet{pickles10}$^b$             & $UBVRI$ &  4\,794 & {\em 41 GCs} \\
\citet{cote95}                    & $BV$    &     315 & NGC\,3201    & \citet{pollard05}                 & $BVI$   & 19\,811 & NGC\,6254 \\
\citet{delafuente15}              & $VI$    &  1\,061 & E\,3         & \citet{rees93}                    & $BV$    &     480 & NGC\,5904 \\
\citet{ferraro90}                 & $BV$    &  5\,420 & NGC\,2808    & \citet{rey01}                     & $BV$    &  3\,949 & {\em 2 GCs} \\
\citet{ferraro97}                 & $BVI$   & 17\,877 & NGC\,5272    & \citet{rey04}                     & $BV$    & 76\,113 & NGC\,5139 \\
\citet{feuillet14}                & $BVI$   & 39\,169 & NGC\,7078    & \citet{rosenberg00a,rosenberg00b} & $VI$    &193\,788 & {\em 29 GCs} \\
\citet{geffert00}                 & $BV$    &  4\,030 & NGC\,6838    & \citet{saha05}                    & $BVRI$  &  1\,617 & Pal\,14 \\
\cite{guarnieri93}                & $BV$    &  3\,371 & NGC\,6205    & \citet{samus95}                   & $BVRI$  &  1\,368 & NGC\,5286 \\
\citet{holland92}                 & $BV$    &     320 & Pal\,14      & \citet{sarajedini95}              & $BVI$   &     776 & NGC\,5053 \\
\citet{hilker06}                  & $BV$    &     141 & Pal\,14      & \citet{sariya12}                  & $UBVI$  & 12\,435 & NGC\,6809 \\
\citet{kaluzny98}                 & $VI$    & 86\,523 & NGC\,104     & \citet{sariya15}                  & $BVI$   &  2\,326 & NGC\,6366 \\
\citet{kravtsov97}                & $UBV$   &  2\,410 & NGC\,1904    & \citet{sollima05}                 & $BVI$   & 94\,205 & NGC\,5139 \\
\citet{kravtsov09}                & $UBVI$  & 12\,025 & NGC\,3201    & \citet{thomson12}                 & $UBVI$  &  5\,852 & NGC\,6752 \\
\citet{kravtsov10}                & $UBVI$  &  5\,124 & NGC\,1261    & \citet{walker94}                  & $BVI$   &     525 & NGC\,4590 \\
\citet{kravtsov14}                & $UBVI$  & 16\,339 & NGC\,6752    & \citet{walker98}                  & $BVI$   &  4\,043 & NGC\,1851 \\
\citet{lee99a}                    & $BV$    & 17\,366 & NGC\,7089    & \citet{wang00}                    & $BVRI$  &      98 & NGC\,4147 \\
\citet{lewis06}                   & $VI$    &  5\,559 & Pal\,11      & \citet{vanleeuwen00}              & $BV$    &  8\,237 & NGC\,5139 \\
\citet{libralato14}               & $BVR$   & 55\,156 & {\em 2 GCs}  & \citet{white70}$^a$               & $UBV$   &      43 & {\em 5 GCs} \\
\citet{lynga96}                   & $BVR$   &  2\,886 & NGC\,5139    & \citet{zaritsky02}                & $UBVI$  &  2\,267 & NGC\,104 \\
\citet{marconi01}                 & $BVI$   &  7\,921 & NGC\,6101    \\
\hline
\multicolumn{4}{l}{$^a$Photoelectric sequences. $^b$Synthetic fitted magnitudes.}\\
\end{tabular}
\end{footnotesize}
\end{table*}

   \begin{figure}
   \centering
   \includegraphics[trim={0.7cm, 0, 0, 0},width=8.8cm]{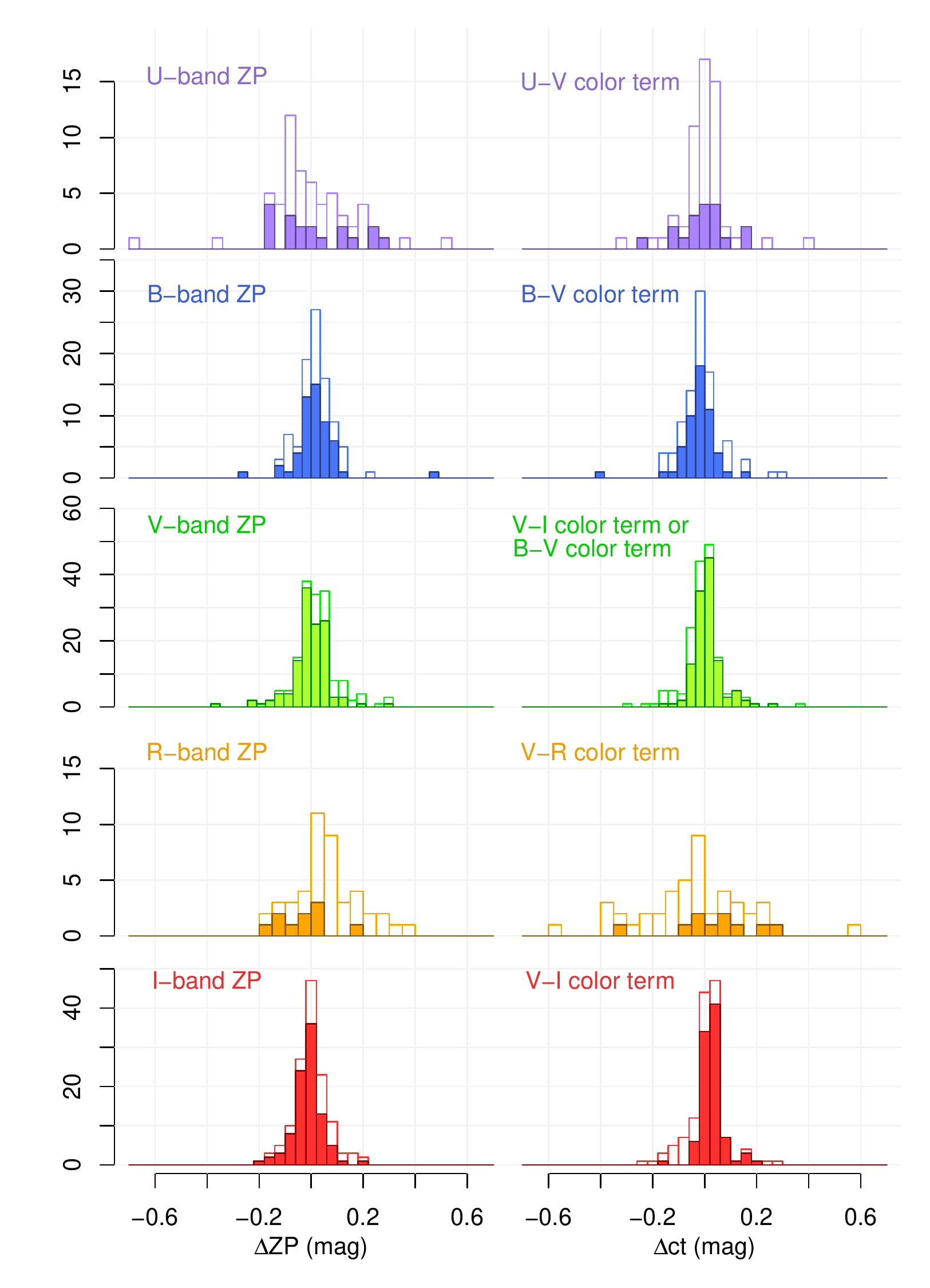}
   \vspace{-0.3cm}
      \caption{Results of the literature comparisons (see Section~\ref{sec:lit}
      and Table~\ref{tab:lit}). Each row refers to a different photometric
      passband, as indicated. The left panels report histograms of the magnitude
      ZPs differences of the literature comparison, the right panels the
      corresponding color-term differences. Empty histograms include the fitted
      synthetic magnitudes by \citet{pickles10}, filled histograms report only
      the observed CCD measurements and photoelectric sequences.} 
   \label{fig:lit}
   \end{figure}


\subsection{Literature comparisons}
\label{sec:lit}

We compared our photometric catalogues with the literature. Whenever possible,
we preferred published photometry performed with techniques suited for crowded fields
(viz., PSF photometry). In spite of the several hundreds of papers devoted to GC
photometry in the literature, we found that only a limited number of studies
actually published electronic catalogues. Moreover, some of the first CCD
studies of the 1970's and 1980's published lists of measured magnitudes but did
not publish any coordinates.

After some initial tests, we also excluded HST photometry based on WFPC2
\citep[e.g.,][]{piotto02} because of the tiny field of view and too different
spatial resolution. Among the HST studies, we chose the ACS survey of GCs
\citep{sarajedini07,anderson08} because of its wider field, but we had to cut
out the GC centers to allow for a meaningful comparison with our ground-based
data. We did not compare with the HST UV Legacy Survey \citep[{\em early data
release},][]{piotto15} because their $U$-band data (F336W) are not yet
calibrated accurately enough to the Johnson system.  There is also concern that
the transformation equations between the HST filter and the ground-based $U$
system may depend on stellar temperature, reddening, metallicity, and gravity in some
complex fashion, due to the inclusion of molecular bands, the Balmer convergence
and jump, and the H and K lines of ionized calcium within the various filters'
bandpasses.  This is a potential topic for future, more detailed studies. Given
the paucity of $U$ and $R$ measurements in the published literature, we also
included in the comparisons the \citet{pickles10} synthetic fitted magnitudes of
Tycho\,2 stars in the observed GC fields and some photoelectric sequences
\citep{white70,alvarado95}.  The list of literature studies finally considered
is presented in Table~\ref{tab:lit}. We did not include in the comparisons
previously published versions of our own photometry.

We cross-correlated our catalogues with the literature with the CataPack
software\footnote{http://davide2.bo.astro.it/$\sim$paolo/Main/CataPack.html}. We
then used a weighted linear model to fit the magnitude differences as a function
of color, recognizing that our comparisons are biased towards giant stars, which
generally have smaller measurement uncertainties than dwarfs, due to better
photon statistics and smaller susceptibility to crowding effects. The results of
all literature comparisons are summarized in Figure~\ref{fig:lit}, where we plot
histograms of the ZPs and the color terms of the magnitude differences. 

The median ZP differences (our minus the literature) and the related
semi-interquartile ranges are: $ZP_U$=--0.008$\pm$0.088, $ZP_B$=0.015$\pm$0.036,
$ZP_V$=--0.003$\pm$0.035, $ZP_R$=--0.026$\pm$0.058, and
$ZP_I$=--0.009$\pm$0.028~mag. These figures compare well with the estimated
accuracies provided in Section~\ref{sec:photcal}. The highly discrepant
comparisons are just a few. Concerning the $U$ band, one should note that the
filters used with various instruments can have significant differences from the
standard Landolt $U$ transmission \citep[with some surprising effects,
see][]{momany03}, and that there are few literature sources to compare with. For
the $R$ band there are even fewer literature studies.

The median values and semi-interquartile ranges of the corresponding color terms
are: $C_{(U,U-V)}$=--0.002$\pm$0.044, $C_{(B,B-V)}$=--0.015$\pm$0.027,
$C_{(V,B-V),(V,V-I)}$=0.001$\pm$0.019, $C_{(R,V-R)}$=0.055$\pm$0.062, and
$C_{(I,V-I)}$=0.023$\pm$0.021~mag. In a minority of cases we observe that the color
terms change visibly for bluer stars at the turnoff or on the upper main
sequence, or for the horizontal branch (HB) stars. We ascribe this to an
incomplete representation of surface gravities in the standards used to
calibrate some of the literature photometry. In very few cases a small but
perceptible second-order term was apparent from the plots, but we did not
attempt to quantify it. The most striking case of a second-order term was in the
comparison with the Pal~11 photometry by \citet{lewis06}; the data for that
study were not obtained under photometric conditions and were calibrated by
comparison with other photometric samples.

For one specific GC, NGC\,6760, we found no dedicated literature study and the GC
appears only in the \citet{piotto02} sample. In general, literature photometry
covers much smaller areas than the results we present here (with few exceptions:
$\omega$~Cen, 47~Tuc, M~22), and very few studies included more than two or three
bands. Additionally, the sample sizes of the literature studies are one order of
magnitude (sometimes two) smaller than ours, although several contain comparable
numbers of stars \citep[for example][]{monaco04,bellini09}. The majority of our
sample GCs has no previous $U$ or $R$ photometry publicly available in the
literature.

\begin{table}
\caption{Column-by-column description of the final catalogue, described in
details in Section~\ref{sec:cats}. The table is published in its entirety
electronically and at CDS. The most updated version of each catalogue will also
be available at the CADC.}
\label{tab:cats}
\centering
\begin{tabular}{lrl}
\hline\hline       
Column     & Units      & Description \\
\hline  
Cluster    &            & GC name \\
Star       &            & Star ID (unique for each GC) \\
X          & (sec)      & X coordinate in arcseconds \\
Y          & (sec)      & Y coordinate in arcseconds \\
$U$        & (mag)      & Johnson U magnitude \\
$\sigma_U$ & (mag)      & U magnitude error \\
n$_U$      &            & Number of U measurements \\
$B$        & (mag)      & Johnson B magnitude \\
$\sigma_B$ & (mag)      & B magnitude error \\
n$_B$      &            & Number of B measurements \\
$V$        & (mag)      & Johnson V magnitude \\
$\sigma_V$ & (mag)      & V magnitude error \\
n$_V$      &            & Number of V measurements \\
$R$        & (mag)      & Cousins R magnitude \\
$\sigma_R$ & (mag)      & R magnitude error \\
n$_R$      &            & Number of R measurements \\
$I$        & (mag)      & Cousins I magnitude \\
$\sigma_I$ & (mag)      & I magnitude error \\
n$_I$      &            & Number of I measurements \\
$\chi$     &            & DAOPHOT's $\chi$ parameter \\
sharp      &            & DAOPHOT's sharp parameter \\
vary       &            & Welch-Stetson variability index \\
weight     &            & Weight of variability index \\
RA         & (hh mm ss) & Right Ascension \\
Dec        & (dd mm ss) & Declination \\
\hline
\end{tabular}
\end{table}


\section{The catalogue}
\label{sec:cats}

The final photometric catalogue for each GC was obtained by robust weighted
averages of the calibrated magnitudes obtained from the individual images
(Section~\ref{sec:photcal}), taking into account the formal uncertainties in the
individual magnitudes based upon readout noise, photon statistics, the quality of
the profile fits, and the image-to-image repeatability \citep[see][and
Appendix~\ref{sec:color} for more details]{daophot1,daophot2,stetson94}.

The combined catalogue format is summarized in Table~\ref{tab:cats} and reports
for each passband the final average magnitude, its standard error of the mean
($\sigma$), and the number of images employed in that passband. Each star in the
catalogue is uniquely identified by specifying both the GC name and the star ID.
The $x$ and $y$ pixel coordinates in arcseconds relative to an arbitrary origin
are also tabulated. The $y$-axis ($x$ = 0) is a great circle of right ascension,
and the $x$-axis ($y$ = 0) is tangent to a parallel of declination at the origin. 
The $x$ axis increases east and $y$ increases north.  These coordinates are
accurately on the system of the {\em Gaia} DR1 (Section~\ref{sec:astrometry}) and
should be precise enough to allow unambiguous cross-identification with other
ground-based studies:  the uncertainty of each position is $\sim$1\% of the seeing
disk. Other relevant quantities listed in Table~\ref{tab:cats} are detailed in the
following sections.


\subsection{Quality indicators}
\label{sec:qc}

The catalogues are based on images obtained with different instruments and under
different observing conditions, and the images cover diverse areas on the sky. Thus,
the quality of the provided measurements can differ significantly for similar
stars at different locations in a cluster.  Different parameters can be used to
judge the quality of measurements for each star and---depending on the specific
science goal---to select the most useful ones. A detailed discussion about how to
select the best star list using the provided parameters is given by
\citet{stetson88}; here we briefly summarize the parameters presented in our
catalogue.  

   \begin{figure}
   \includegraphics[trim={0, 0, 0, 0},width=\columnwidth]{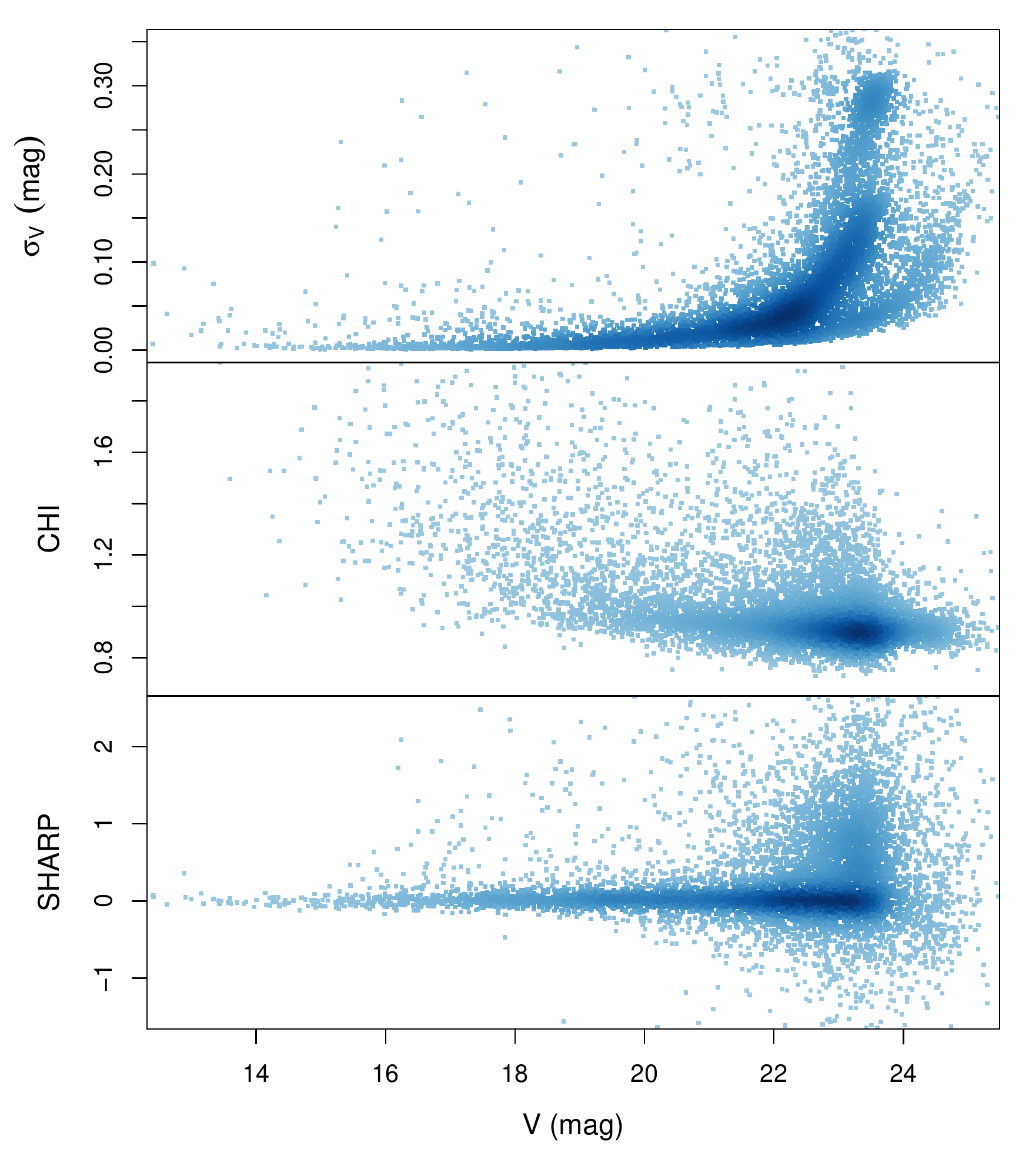}
   \vspace{-0.3cm}
      \caption{Example of image quality indicators for NGC~5694, plotted as a
      function of $V$ magnitude (see Section~\ref{sec:qc} for details). {\em Top
      panel:} the substructure in $\sigma_V$ at the faint magnitude end is caused
      by the varying depth of the used CCD images. {\em Center panel:} the bright
      stars deviate more from the typical PSF, but disturbed PSFs are apparent at
      all magnitudes. {\em Bottom panel:} at the faint magnitude end there are
      many cosmic ray hits and defective pixels (negative values) and diffuse
      objects (positive values).} 
   \label{fig:qpars}
   \end{figure}

The most obvious parameters are $\sigma$, the standard error associated with each magnitude
measurement \citep[see][for more details]{stetson88} and $n$, the number of independent
measurements used to determine it. In addition, unlike most published
photometry, the catalogue presented here contains all five Johnson-Cousins
bands. This allows for colour-colour selections, that can sometimes separate in a very
efficient way giants from dwarfs, or GC stars from field stars, or normal stars from abnormal ones or non-stellar
objects, or badly measured stars from well measured ones.

\begin{figure}
   \includegraphics[trim={0, 0, 0, 0},width=\columnwidth]{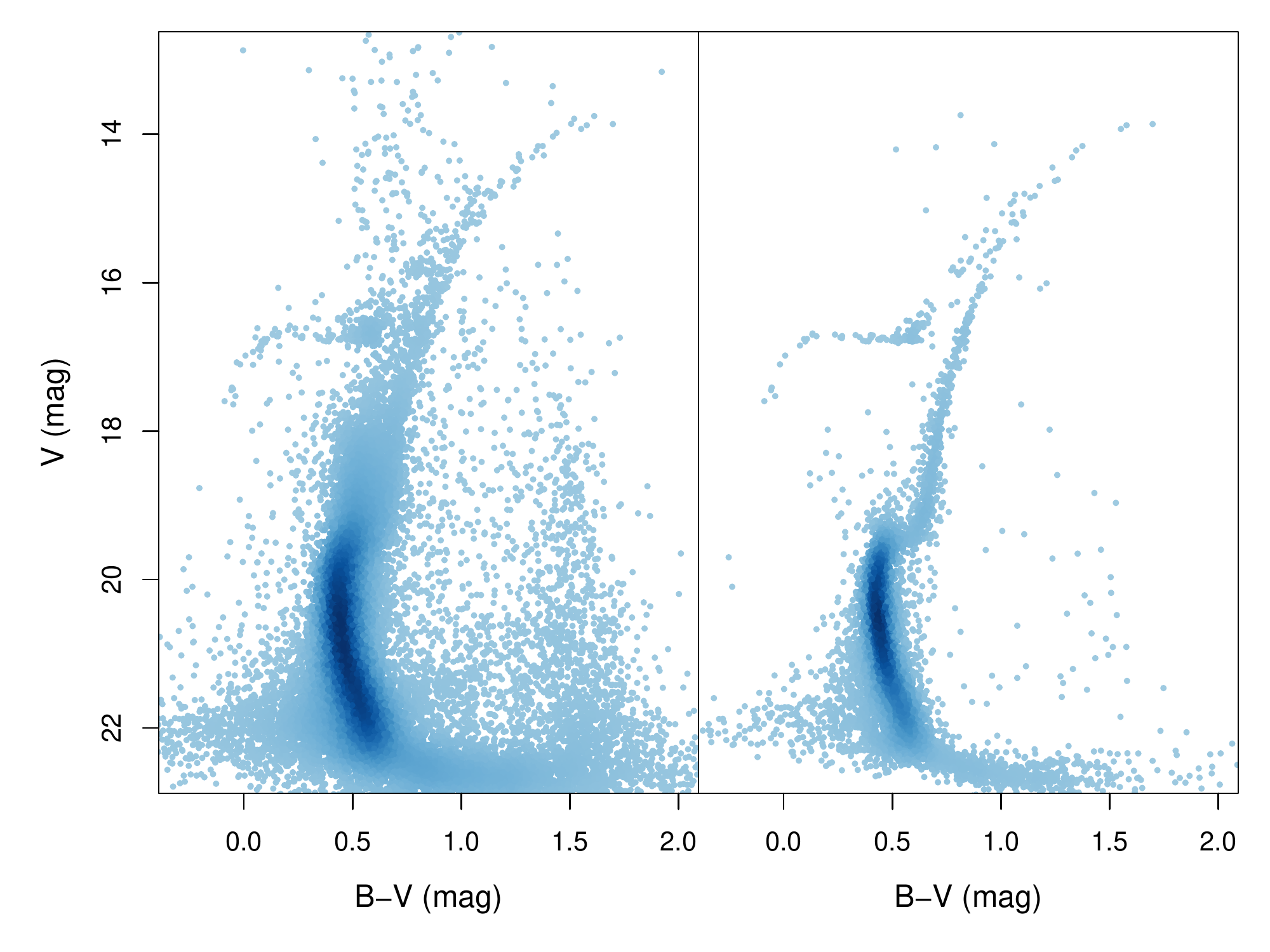}
   \vspace{-0.3cm}
      \caption{Example of CMD cleaning for NGC\,1261. The left panel shows the
      $V$,$B-V$ CMD of the entire catalogue, the right panel an example selection,
      excluding stars within the core radius and outside 80\% of the tidal radius
      from \citet{harris10}, and also excluding stars outside the range
      -0.3$<$sharp$<$0.3, with $\chi>$3, or with $\sigma_V$ and
      $\sigma_B$>0.2~mag.} 
   \label{fig:cleaning}
   \end{figure}

Two additional quality parameters, specific to the DAOPHOT/ALLSTAR/ALLFRAME
codes, are $\chi$ and {\it sharp}, described in detail in \citet{stetson88} and
illustrated in Figure~\ref{fig:qpars}. The $\chi$ parameter measures the
observed pixel-to-pixel scatter in the profile fits compared to the expected
scatter from readout noise and photon statistics. The {\it sharp} parameter is a
measure of how much of the badly modeled light is concentrated in the central pixels
compared to the surrounding ones. It takes positive values for apparently
extended objects (like background galaxies, lumps in nebulosity, or haloes
around bright objects) and negative values for more pointed objects (like
defective pixels or cosmic ray hits), as shown in Figure~\ref{fig:qpars}. An
example of CMD cleaning using these parameters can be found in
Figure~\ref{fig:cleaning}.

\subsection{Cluster centroids}
\label{sec:centroids}

The GC centroids listed in Table~\ref{tab:clusters} were recomputed in a
consistent way with the procedure described in details by \citep{stetson98}. We
suggest using these estimates, rather than literature centroids, in any procedure
that requires a precise, self-consistent centroid determination, such as for
example selection of stars in annuli to build star density profiles and the like
(see Section~\ref{sec:profiles}).

The adopted procedure can be briefly summarized as follows. A virtual circular
aperture is superimposed on the star catalogue down to a certain magnitude
limit. The median $x$ and $y$ coordinates of the objects contained within the
circular aperture are determined, and the aperture is shifted on the catalogue
until the median coordinates of the objects contained in the circle coincide
with the center of the circle. The exercise is repeated for aperture radii
scaled by the fourth root of two, typically from 150 to 600 arcseconds. Two runs
of this procedure are performed, one on the disks defined by the range
(0,$r_n$), and another on the annuli ($r_{n-1},r_n$). The overall range of the
various centroid determinations is typically of order a few arcseconds, and the
medians of the derived $x$ and $y$ centroids are adopted as the cluster
photocenter, presuming that crowding errors, incompleteness, and other relevant
effects will mostly be symmetric. 

The samples enclosed by the various radii are not statistically independent, but
the consistency observed even when smaller radii and various different magnitude
limits are adopted indicates that the method is repeatable at the
arcsecond level.


\subsection{Astrometric calibration on {\em Gaia}}
\label{sec:astrometry}

The relative astrometric calibration of star positions was performed using {\em
Gaia} DR1 \citep[the first data release,][]{gaia1} as a reference.

The RA and Dec in {\em Gaia} DR1 are projected from the sphere onto an ($X$,$Y$)
plane by mapping position angle and angular distance relative to a chosen
reference point on the sky to $\theta$ and $\rho$ in the plane. The angular
coordinates are then transformed to $X$ and $Y$ by simple plane geometry. This
differs from a gnomonic projection in that it preserves a constant scale in the
radial direction to arbitrarily large angles, and the distortion in the
circumferential direction is reduced.  Specifically, if a gnomonic projection
were applied to a field comprising half the sky, the scale of the projection
would be an infinite number of planar units per arcsecond at a radial distance of 90
degrees from the tangent point.  For the projection we use here, at a radial
distance of 90 degrees from the reference point the scale of our planar $(X,Y)$
coordinates is one unit per arcsecond in the radial direction, and $\pi/2$ units
per arcsecond in the circumferential direction.  The observed ($x$,$y$)
coordinates in the individual CCD images are then transformed to this master
($X$,$Y$) catalogue using bivariate cubic polynomials with 20 plate constants.
The transformed observed positions can then be converted to right ascension and
declination in the {\em Gaia} DR1 system by reversing the mapping of the polar
coordinates.

In a few cases, as illustrated e.g. in Figure~12 by \citet{gaia1}, there were
large gaps in the catalogues for some GCs as downloaded from Gaia DR1. We
therefore also downloaded all available Digitized Sky Survey images of each
field from the Canadian Astronomy Data
Centre\footnote{http://www.cadc-ccda.hia-iha.nrc-cnrc.gc.ca/en/}. The
photographic coverage was generally selected to extend several arcminutes beyond
the CCD coverage. Positions and magnitude indices for stars identified in these
images were determined using the software described in \citet{stetson79}. The
photographic results were mapped onto the {\em Gaia} catalogue as just
described, and then the CCD data were mapped onto the merged {\em
Gaia}/photographic catalogue.

   \begin{figure}
   \includegraphics[trim={0 0 0 0},width=\columnwidth]{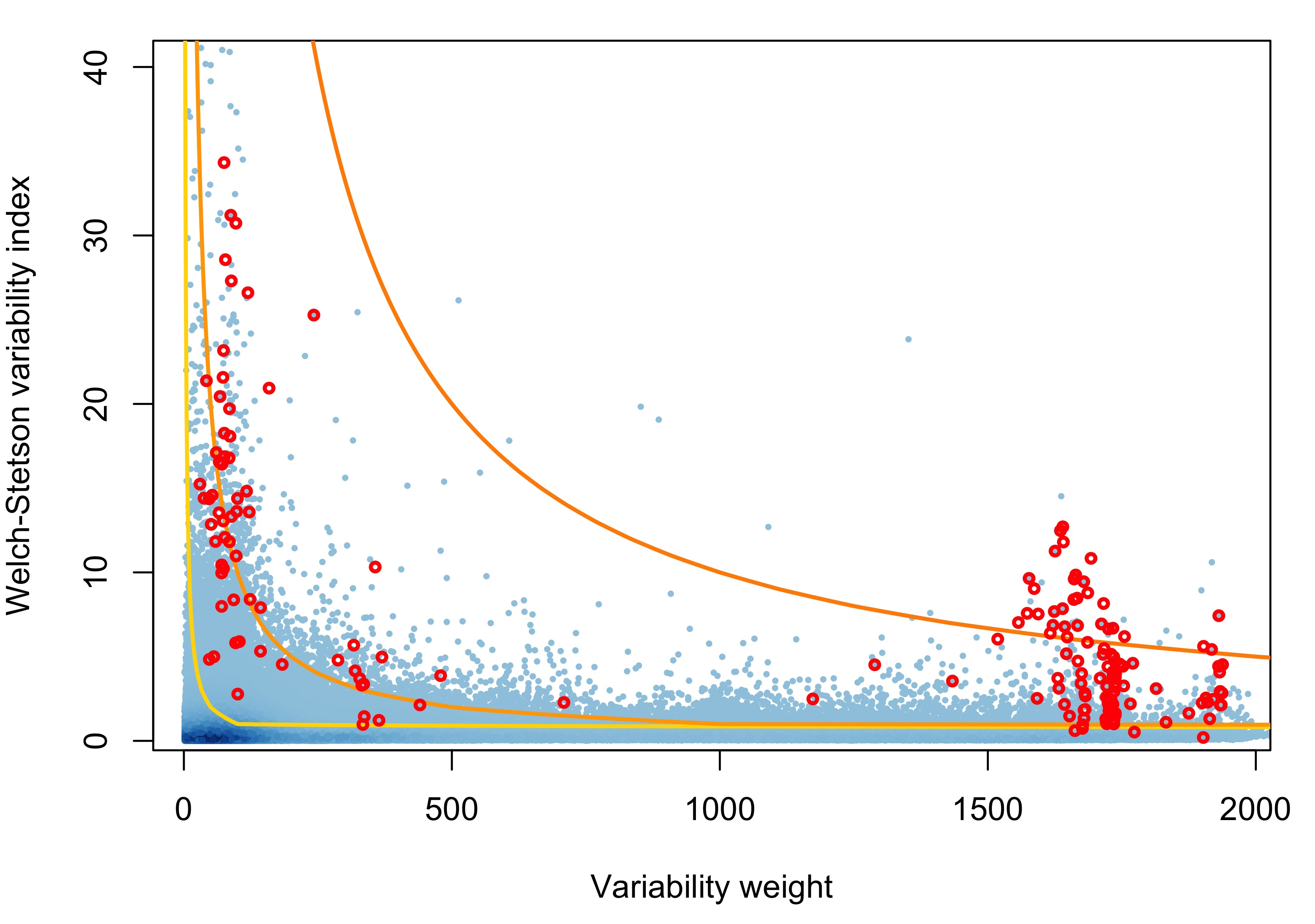}
   \vspace{-0.3cm}
      \caption{Behaviour of the variability indicators for the GC with the
      largest number of RR~Lyrae variables: $\omega$~Centauri. The plot shows a
      zoom into the Welch-Stetson index versus variability weight plane of the
      entire catalogue (blue dots). The lines of evidence equal to 2 (gold), 3
      (orange), and 4 (tomato) are overplotted for reference. Known RR~Lyrae
      variables characterized by \citet{braga16} are overplotted in red:
      they mostly have variablity weight above $\simeq$30 and evidence higher
      than $\simeq$2.} 
   \label{fig:vary}
   \end{figure}

   \begin{figure*}
   \centering
   \includegraphics[width=0.32\textwidth]{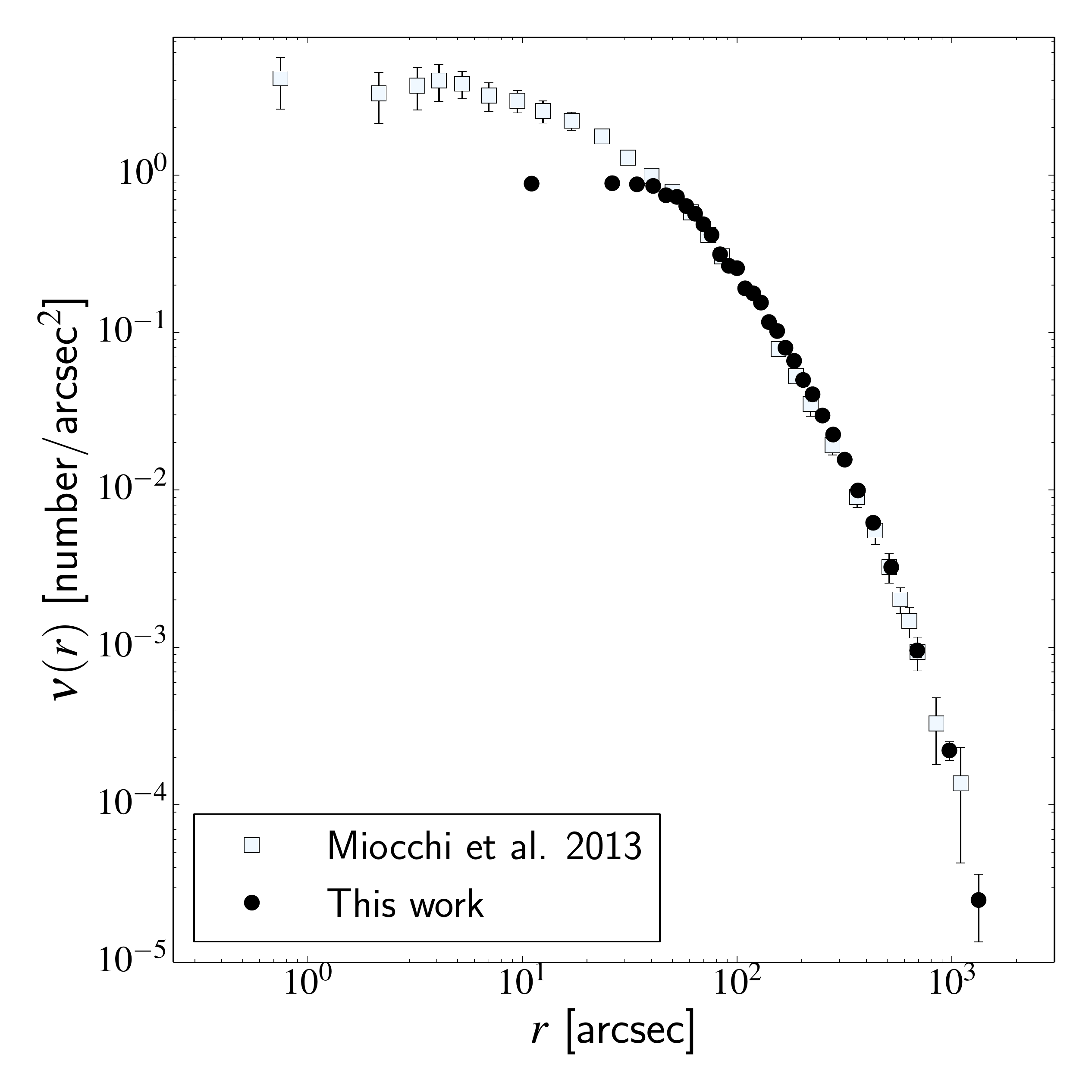}
   \includegraphics[width=0.32\textwidth]{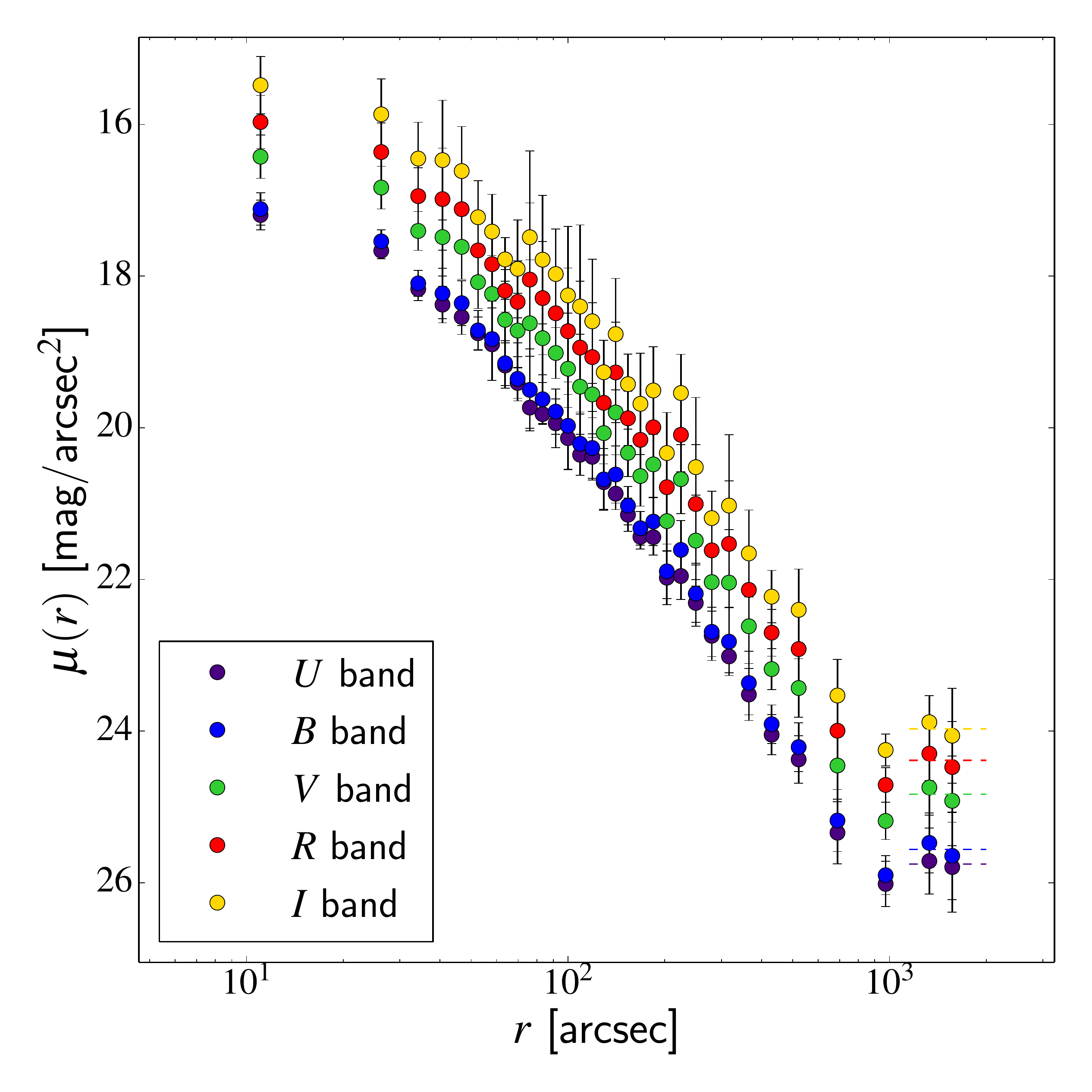}
   \includegraphics[width=0.32\textwidth]{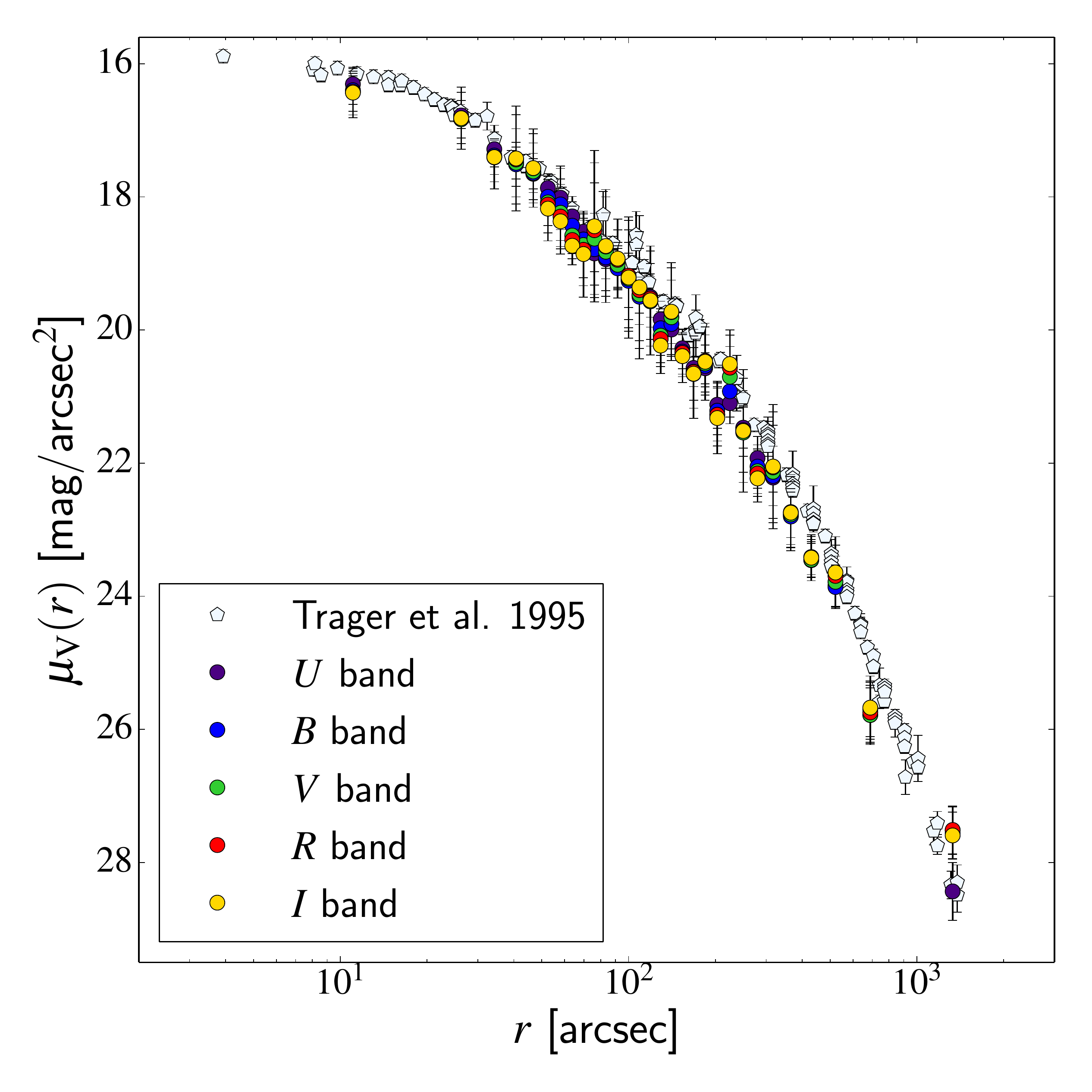}
      \caption{Example of number density and surface brightness profiles for the case
      of M\,5. {\em Left panel:} background-subtracted number density profile $\nu$,
      as a function of the projected distance from the center, $r$ (black circles)
      compared to the profile by \citet[][white squares]{miocchi13}. {\em Center
      panel:} surface brightness profile $\mu$ as a function of $r$. Violet, blue,
      green, red, and yellow circles represent the profiles calculated in the $U$,
      $B$, $V$, $R$, and $I$ bands, respectively. The dashed lines indicate the
      background level estimated for each band. {\em Right panel:} the surface
      brightness profiles, coloured as in the center panel, background-subtracted and
      opportunely shifted using the integrated color indices (see
      Section~\ref{sec:profiles}), and compared with the profile presented by
      \citet[][white pentagons]{trager95}.}
   \label{fig:profiles}
   \end{figure*}

\subsection{Variability indicators}
\label{sec:var}

The photometric catalogue published here (Table~\ref{tab:cats}) does not contain
epoch data or light curves for variable stars; in a few cases these are
published elsewhere, for example for NGC\,4147, $\omega$\,Cen, and M\,4
\citep{stetson05b,braga16}. However, it contains useful information that can
give an idea of which stars have a higher probability of being variable. The
first indicator provided is the Welch-Stetson (W/S) variability index
\citep[][]{welch93}, based on the correlation of the normalized residuals
relative to the {\it UBVRI\/} average magnitudes for observations obtained close
in time. The second is the weight, equal to the number of residual pairs plus
one-half the number of singleton residuals used to build the W/S index. A useful
derived parameter is the variability {\em evidence}, which we define as the
logarithm of the W/S index times its weight. 

An example of the behaviour of the three indicators is shown in
Figure~\ref{fig:vary}. As can be seen, for low values of weight the W/S
parameter provides a large number of false positives and even relatively high
values can be caused by image defects, crowding compounded with seeing
variations, and other complications. Stars of the highest variability index {\it
and\/} weight are the most likely to be real variables. Therefore, the
``evidence'' parameter can be an efficient way to pre-select and to rank stars
for further examination: if stars are ranked in order of decreasing evidence and
examined individually in that order, then by the time the investigator
encounters (say) twenty false positives in a row it can be considered that the
point of diminishing returns has been reached.


\section{Results}
\label{sec:cmd}

As mentioned above, the photometric catalogues provided here represent a
significant improvement over previously published ground-based catalogues
(Section~\ref{sec:lit}) in terms of area, photometric bands, number of GCs
homogeneously analyzed, number of stars per GC, and photometric quality. It will
have a greater scientific impact once combined with proper motions from {\em Gaia}
and abundances from spectroscopic surveys
\citep{carretta09a,carretta09b,meszaros15,pancino17b}, but we believe that it is
already a valuable resource as it is.  In the following sections, we show some
selected new results, to illustrate the quality of the photometry and its
potential for a rich scientific harvest.  We have ideas for studies that we intend
to perform, but we are making these data available to the community in the hopes
that others will find uses that we ourselves have not yet considered.


\subsection{Cluster profiles} 
\label{sec:profiles}

Determining the structural properties of GCs is the first step in understanding
their formation and evolution. The simplest observables for this purpose are the
number-density and surface-brightness profiles, containing information on the
spatial distribution of stars within each GC. 

The only detailed compilation of surface brightness profiles for a large number of
GCs is still the one by \citet[][containing 125 GCs]{trager95}, often computed
combining surface brightness profiles extracted from mid-1980's CCD images and
even older star counts on photographic plates. Care needs to be taken when using
these profiles for mass-segregated clusters, for which the shapes of the number
density and surface brightness profiles could differ significantly. Recently,
surface brightness profiles for the innermost parts of 38 clusters have been
measured with HST photometric data \citep{noyola06} and number density profiles
for 26 Galactic globular clusters were presented by \citet{miocchi13}. For some
GCs, additional number-density and/or surface-brightness profiles have been
calculated individually from high-quality photometry, for example: NGC\,2419
\citep{bellazzini07}; NGC\,6388 \citep{lanzoni07}; $\omega$ Cen \citep{noyola08};
and M\,92 \citep{dicecco13}, among others. However, the only compilations of
homogeneously computed profiles for a large number of globular clusters are the
three studies mentioned above.

Here we consider M\,5, one of the GCs in our sample with the most data and for which
the images in all bands cover a similar extent, as an example of how the current
version of our photometry database---of which we present the first 48 GCs
here---could be used to compute high-quality surface-brightness and number-density
profiles of GCs homogeneously. We consider stars brighter than $V = 19$~mag and
fainter than $V = 12$~mag, to ensure the catalogue is roughly complete at all
radii and to avoid fluctuations due to very bright stars.  

We computed the number-density profile by considering several concentric annuli of
variable width, each containing 700 stars, to minimise the fluctuations in the
profiles. The profiles are centered on the (X$_0$,Y$_0$) centroids computed as
described in Section~\ref{sec:centroids} and listed in Table~\ref{tab:clusters}.
Each annulus was divided into eight subsectors, where the number density was
computed as the number of stars divided by the subsector's area. For each annulus,
the number density and its error were computed as the mean and standard deviation
of its subsectors' densities. We estimate the background density by considering
the outermost regions, and we subtract this from the density profile. The
resulting background-subtracted number density profile, $\nu(r)$, with $r$ equal
to the mean distance from the center for a given annulus, is shown in
Figure~\ref{fig:profiles} (left panel) where it is compared with the profile
provided by \citet{miocchi13}, showing excellent agreement. The discrepancy in the
innermost part is due to crowding effects in our catalogue, as previously
discussed.

\begin{figure*}
  \centering
  \includegraphics[width=0.245\textwidth]{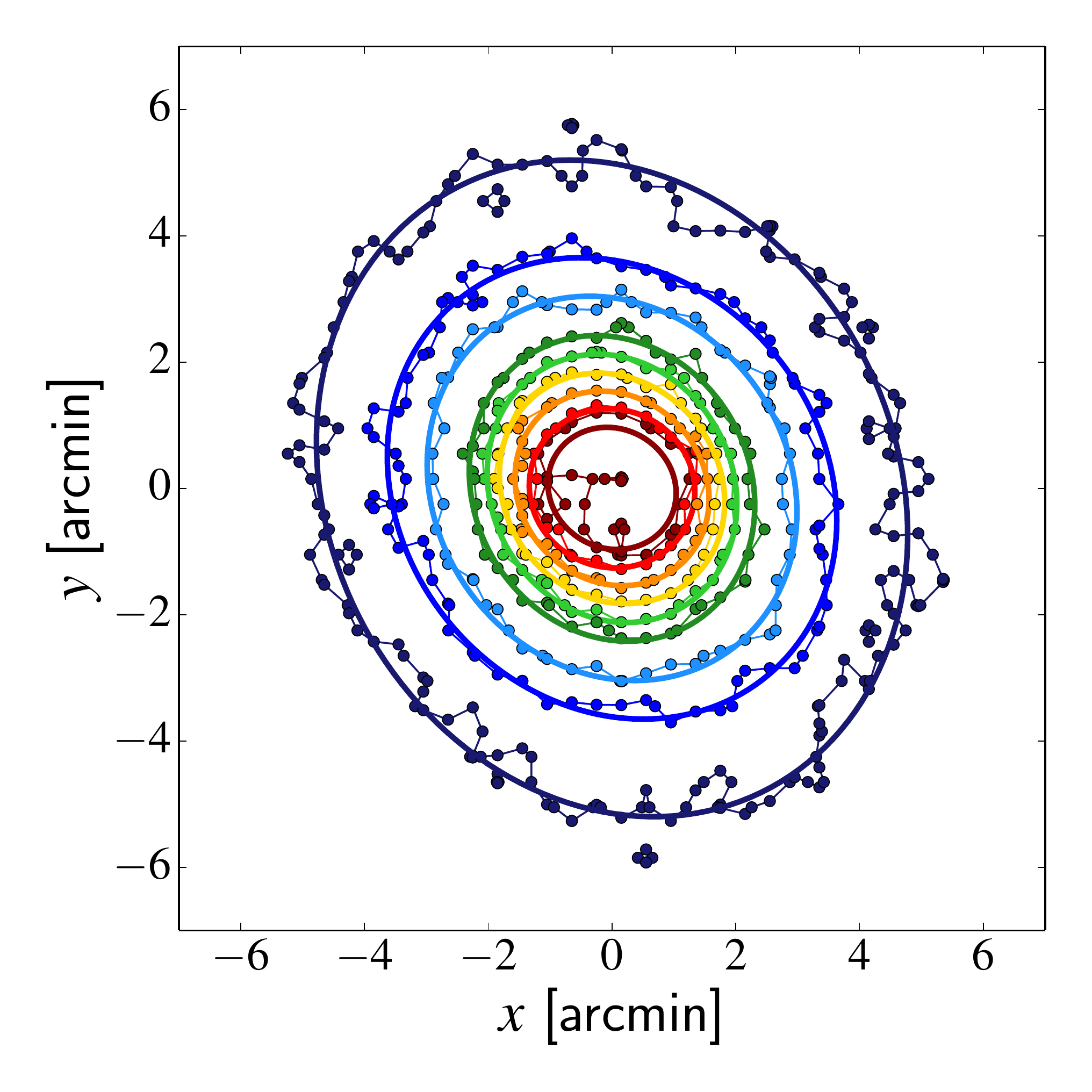}
  \includegraphics[width=0.245\textwidth]{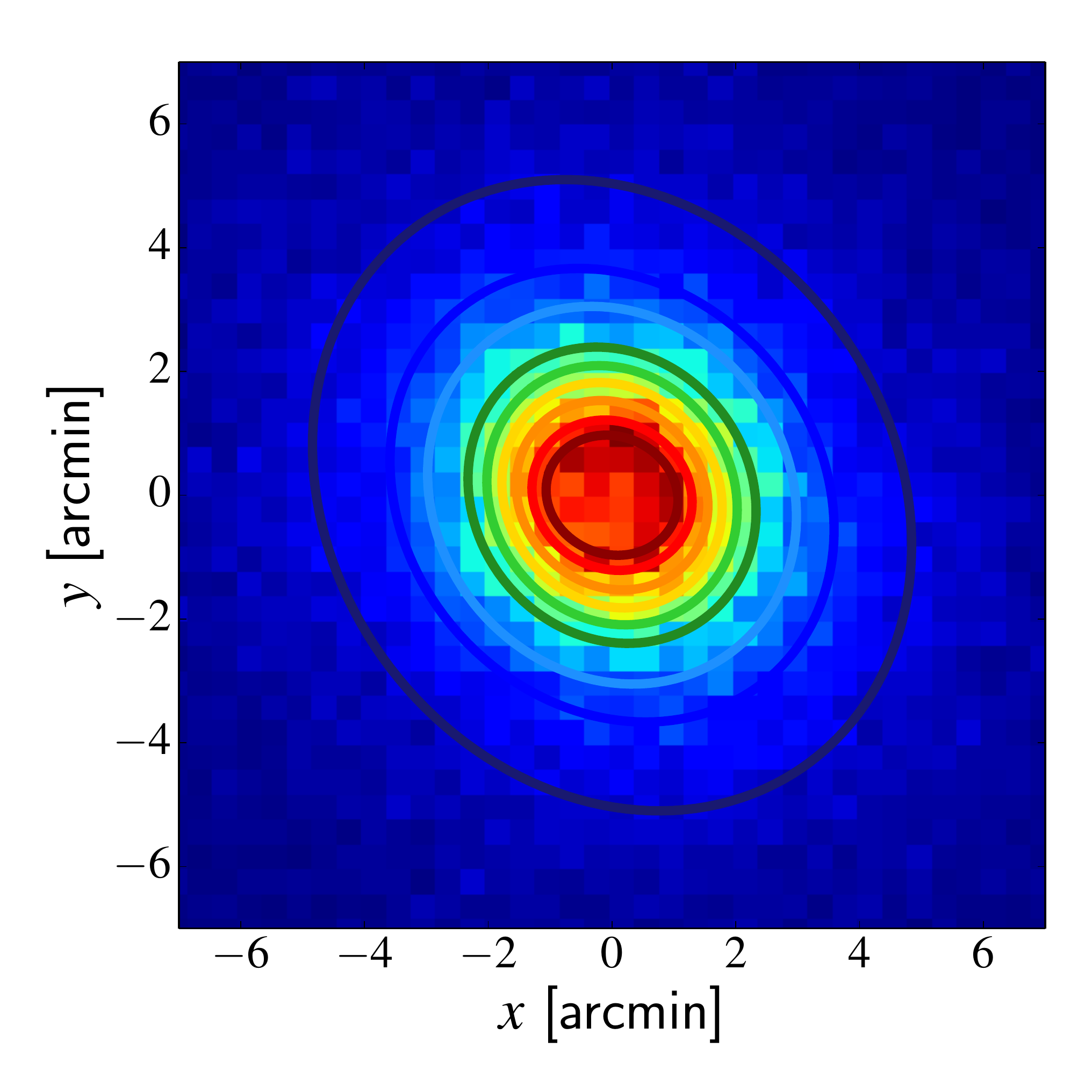}
  \includegraphics[width=0.245\textwidth]{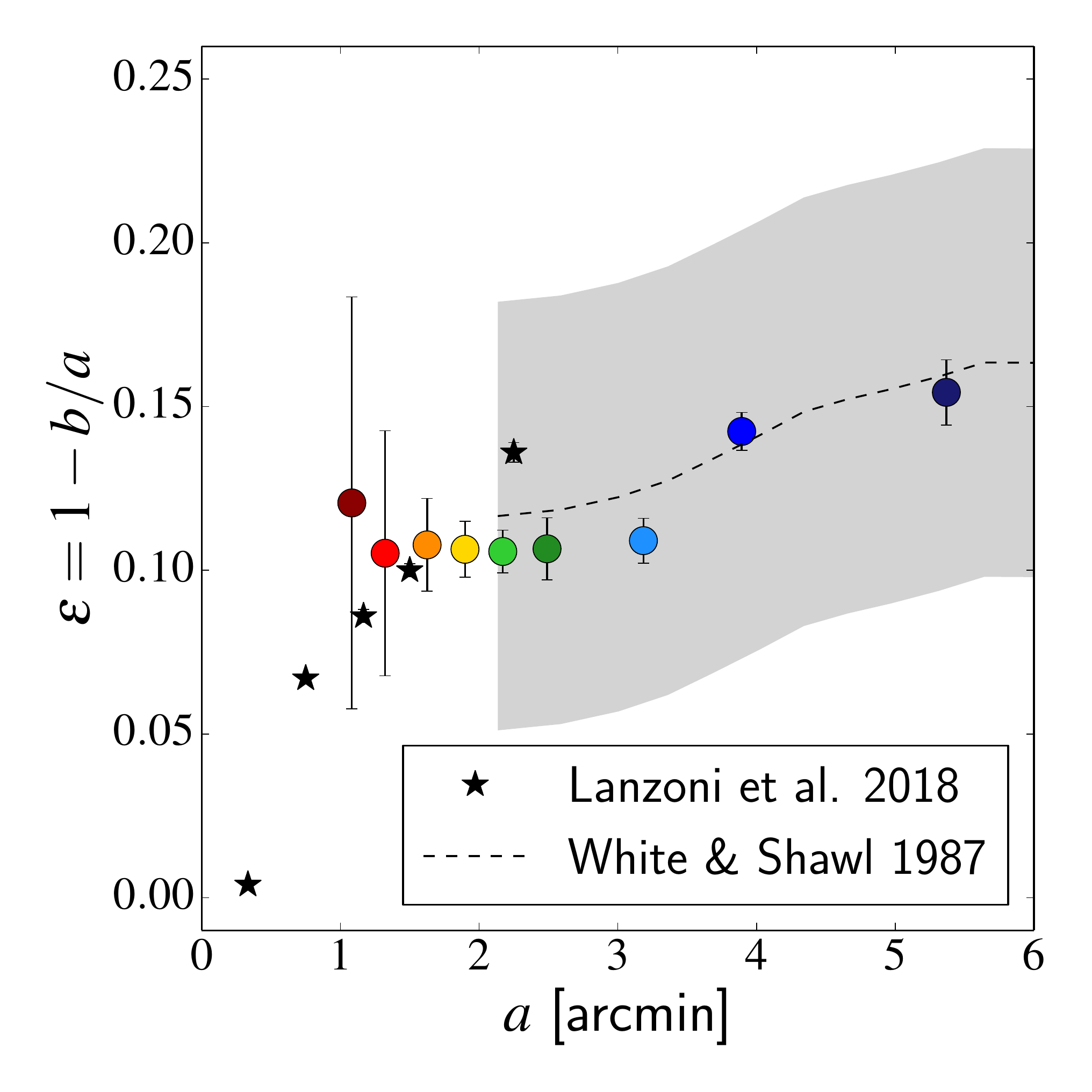}
  \includegraphics[width=0.245\textwidth]{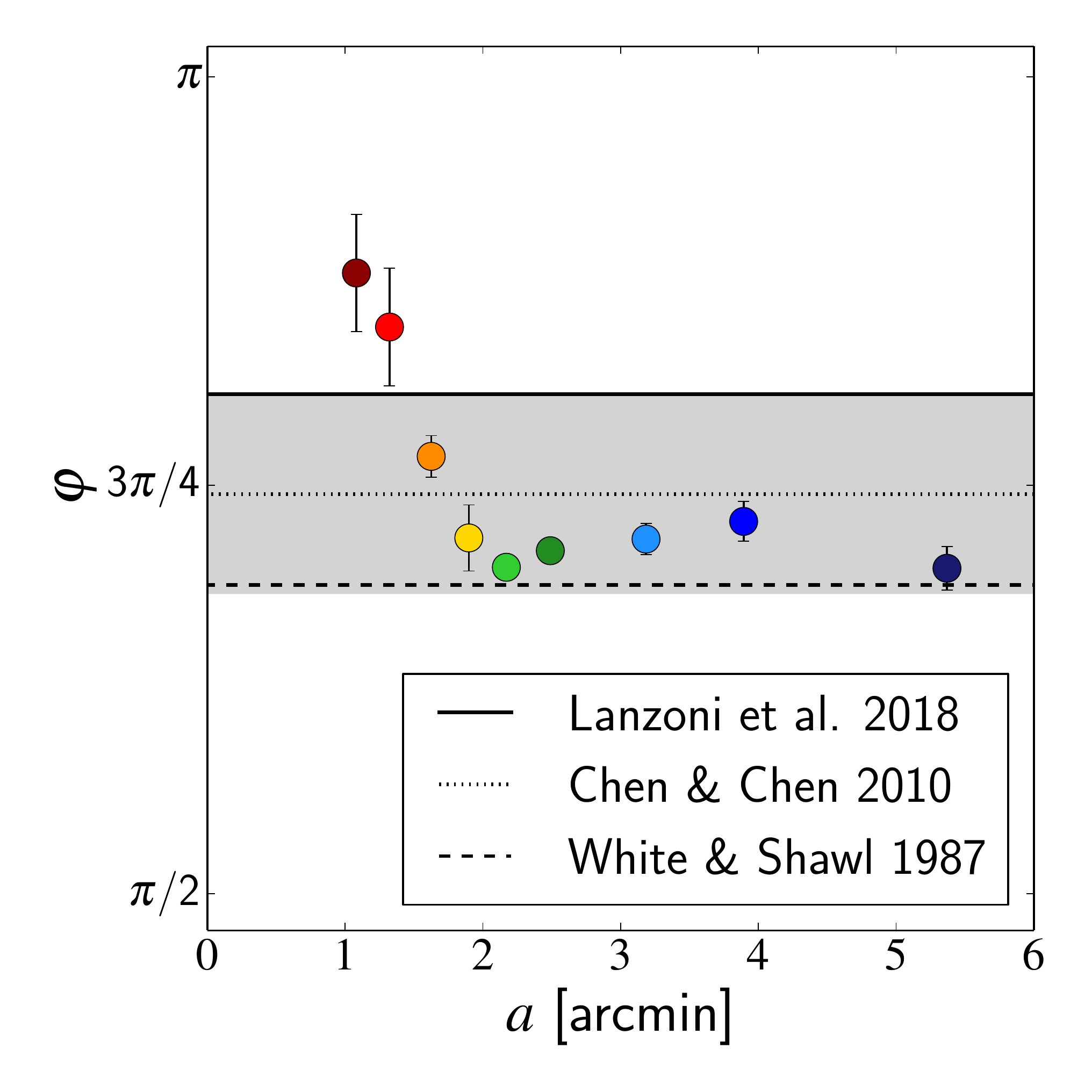} 
    \caption{Morphology of NGC\,5904 (M\,5). In all panels, colours consistently
    refer to the selected levels of the projected number density, listed on
    Table~\ref{tab:ellipses}. {\em Left panel:} Isodensity contours (dots and
    thin solid lines) and their corresponding best-fit ellipses (thick solid
    lines) for a particular choice of the grid (Section~\ref{sec:ellipses}).
    {\em Center-left panel:} the same best-fit ellipses are overplotted on the
    density map of M\,5, where each box corresponds to a cell of the chosen
    grid. {\em Center-right panel:} ellipticity profile $\varepsilon(a)$ as a
    function of the semi-major axis of the ellipses, coloured as in the
    preceding panels. The black stars report the ellipticity estimated by
    \citet{lanzoni18} and the dashed line the ones by \citet{white87}, with
    their errors represented by the grey shaded area. {\em Right panel:}
    position angle $\varphi(a)$ as a function of the semi-major axis of the
    ellipses. The dashed, solid, and dotted lines correspond, respectively, to
    the values determined by \citet{white87}, \citet{lanzoni18}, and
    \citet{chen10}, with the uncertainty on the last one shown as a grey shaded
    area.}
  \label{fig:ellipses}
\end{figure*}

We also computed the surface-brightness profiles, $\mu(r)$, in each of the {\it
UBVRI\/} photometric bands, by summing the corresponding flux in each subsector
and then proceeding in the same way as for the number density, including the
background subtraction. In this case, we also took into account the
uncertainties in the individual luminosities to compute the standard error for
each point in the profiles. Figure~\ref{fig:profiles} (center panel) shows the
resulting surface-brightness profiles. As can be seen, the profiles in the
different bands have very similar shapes, and they overlap within the errors
(Figure~\ref{fig:profiles}, right panel) when shifted to the $V$ band using the
integrated color indices provided in the \citet{harris10} catalogue: $U-V =
0.89$, $B-V = 0.72$, $V-R = 0.45$, and $V-I = 0.95$~mag. Indeed, the same figure
shows a comparison between our shifted surface-brightness profiles and the
surface-brightness profile provided by \citet{trager95}, which is quite
satisfactory. 

The profiles computed in this section are only presented as a function of the
distance from the centre, assuming that the GC is spherically symmetric. Even
though this is a common assumption, often adopted to describe GCs (and in
particular it was adopted in the literature profiles we compare ours to), in
Section~\ref{sec:ellipses} we will show that this is not necessarily true. With
the present catalogues, it will be possible to compute density and surface
brightness profiles without assuming spherical symmetry, and we plan to do this in
the future.


\subsection{Cluster shape} 
\label{sec:ellipses}

For a long time, GCs were considered spherically symmetric systems, and the
observational profiles are commonly reported as a function of the distance from
the centre only. While this is reasonable for GCs that do not exhibit perceptible
flattening, for many other GCs, including M\,5 (Section~\ref{sec:profiles}),
significant deviations from spherical symmetry were detected
\citep{geyer83,white87,chen10}. Unfortunately, because of incomplete sampling of
different regions of GCs, the few available estimates of the global ellipticity
could differ significantly from one work to another, as in the case of
\citet{white87} and \citet{chen10}. A homogeneous study of a large number of
well-covered GCs is still sorely missing. 

The photometry presented here has the potential to provide the basis for such a
study, so we used NGC\,5904 (M\,5) as a test case for a method to quantify the
deviation of the GC morphology from spherical symmetry\footnote{The method is
loosely based on those employed by \citet{danilov94} and \citet{pancino03}.}, that
we will apply to our entire catalogue in the near future. For this GC,
\citet{white87} report an ellipticity equal to $0.14$. We consider all the stars
brighter than $V$=21~mag and we do not correct for the field population, because
it is reasonable to assume it to be uniform across the extent of the cluster.

We start by considering a grid covering the area occupied by NGC\,5904, made of
$0.4^\prime \times 0.4^\prime$ cells, defined in the $x$ and $y$ coordinates
(Section~\ref{sec:cats}) and centered on the (X$_0$,Y$_0$) centroid of NGC\,5904
(see Table~\ref{tab:clusters} and Section~\ref{sec:centroids}). The size of the
cells is small enough to provide a good representation of the GC and large enough
to avoid excessive fluctuations due to the discrete nature of the dataset. For
each cell, identified by the position of its centre on the plane, we compute the
number density and we interpolate it as a function of $x$ and $y$\footnote{We use
the number density and not the surface brightness because the former is more
stable: when evaluated on a grid, the surface brightness fluctuates a lot more and
makes it difficult to obtain reliable contours.}. We consider several contours for
the interpolating function, corresponding to selected fractional values $f_{\rm
peak}$ of the peak value, in particular we use $f_{\rm peak}$= 0.85, 0.8, 0.7,
0.6, 0.5, 0.4, 0.25, 0.17, and 0.08. We then determine the parameters of the
ellipse that provides the best fit for each of the contours thus
obtained\footnote{To determine the ellipse best-fit parameters, we use the code by
http://nicky.vanforeest.com/misc/fitEllipse/fitEllipse.html.}: the semi-major axis
$a$, the semi-minor axis $b$, the counterclockwise angle of rotation from the
$x$-axis $\varphi$, and the position of the centre of the ellipse $(x_{\rm
e},y_{\rm e})$ . 


\begin{table}
\caption{Best-fit parameters identifying the ellipses corresponding to the selected
number density contours for M\,5: semi-major axis $a$; semi-minor axis $b$;
counterclockwise angle of rotation from the $x$-axis to the major axis of the
ellipse $\varphi$ (note that the position angle recorded here also corresponds to
the angle between the North and the semiminor axis of the ellipse, going towards
East; therefore, the angle between the North and the semimajor axis of the ellipse
can be found by simply subtracting $\pi/2$ from the values of $\varphi$);
ellipticity $\varepsilon = 1 - b/a$; position of the centre of the ellipse $(x_{\rm
e},y_{\rm e})$; fraction of total stars contained within the corresponding contour
$f_{\rm int}$; and value of the density on the contour expressed as a fraction of
its peak value $f_{\rm peak}$.} 
\label{tab:ellipses}
\centering
\begin{scriptsize}
\begin{tabular}{ccccccccccc}
\hline 
\hline
$a$ & $b$ & $\varphi$ & $\varepsilon$ & $(x_{\rm e},y_{\rm e})$ & $f_{\rm int}$ & $f_{\rm peak}$ \\
($^{\prime}$) & ($^{\prime}$) & (rad) &  & ($^{\prime\prime}$,$^{\prime\prime}$) &  &  \\
\hline
$1.12 \pm 0.06$ & $0.94 \pm 0.05$ & $2.73 \pm 0.09$ & $0.16 \pm 0.06$ &  (6.0,6.3) & 0.10 & 0.85 \\
$1.33 \pm 0.03$ & $1.19 \pm 0.04$ & $2.64 \pm 0.11$ & $0.11 \pm 0.03$ &  (2.2,3.9) & 0.14 & 0.80 \\
$1.63 \pm 0.02$ & $1.45 \pm 0.01$ & $2.39 \pm 0.05$ & $0.11 \pm 0.01$ & (-0.2,2.5) & 0.20 & 0.70 \\
$1.89 \pm 0.01$ & $1.70 \pm 0.01$ & $2.25 \pm 0.07$ & $0.10 \pm 0.01$ & (-1.3,3.3) & 0.26 & 0.60 \\
$2.17 \pm 0.01$ & $1.94 \pm 0.01$ & $2.20 \pm 0.02$ & $0.10 \pm 0.01$ & (-0.5,3.1) & 0.32 & 0.50 \\
$2.49 \pm 0.02$ & $2.22 \pm 0.02$ & $2.24 \pm 0.02$ & $0.11 \pm 0.01$ &  (0.6,2.5) & 0.37 & 0.40 \\
$3.18 \pm 0.02$ & $2.84 \pm 0.02$ & $2.25 \pm 0.03$ & $0.11 \pm 0.01$ &  (0.8,2.2) & 0.49 & 0.25 \\
$3.89 \pm 0.02$ & $3.34 \pm 0.01$ & $2.29 \pm 0.04$ & $0.14 \pm 0.01$ &  (1.0,1.6) & 0.57 & 0.17 \\
$5.37 \pm 0.06$ & $4.54 \pm 0.02$ & $2.20 \pm 0.04$ & $0.15 \pm 0.01$ & (-1.1,2.3) & 0.70 & 0.08 \\
\hline
\hline
\end{tabular}
\end{scriptsize}
\end{table}


We repeated this procedure several times, shifting the position of the grid:
we considered 60 different positions for the grid, each time shifting it
horizontally and vertically by 3${}^{\prime\prime}$. For each contour, the final values of the
parameters and their errors are listed in Table~\ref{tab:ellipses}, and
correspond to the mean and standard deviation of the best-fit values obtained at
each centring of the grid. This procedure allows us to prevent 
fluctuations in the number density due to the discrete nature of the sample
from influencing the morphology determination.

The two left panels of Figure~\ref{fig:ellipses} show the results in graphical
form, for one particular choice of the underlying grid. It can be seen that the
innermost profile ($f_{\rm peak} = 0.85$) is irregular, and has the shape of a
horseshoe, probably caused by stochastic effects and possibly by incompleteness
due to stellar crowding in the GC center; the best-fit ellipse in this case needs
to be taken with caution. 

   \begin{figure*}
   \centering
   \includegraphics[width=0.325\textwidth]{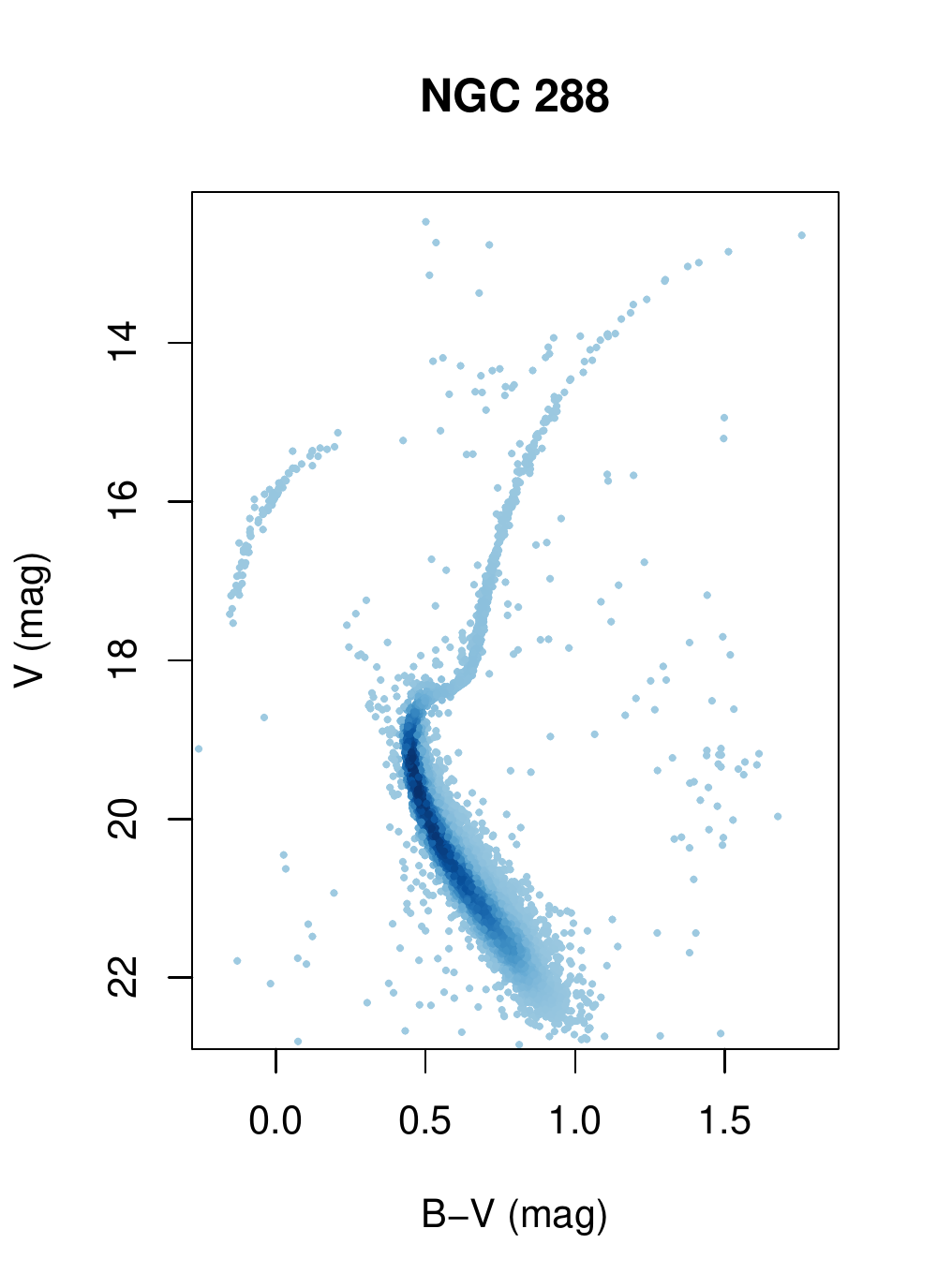}
   \includegraphics[width=0.325\textwidth]{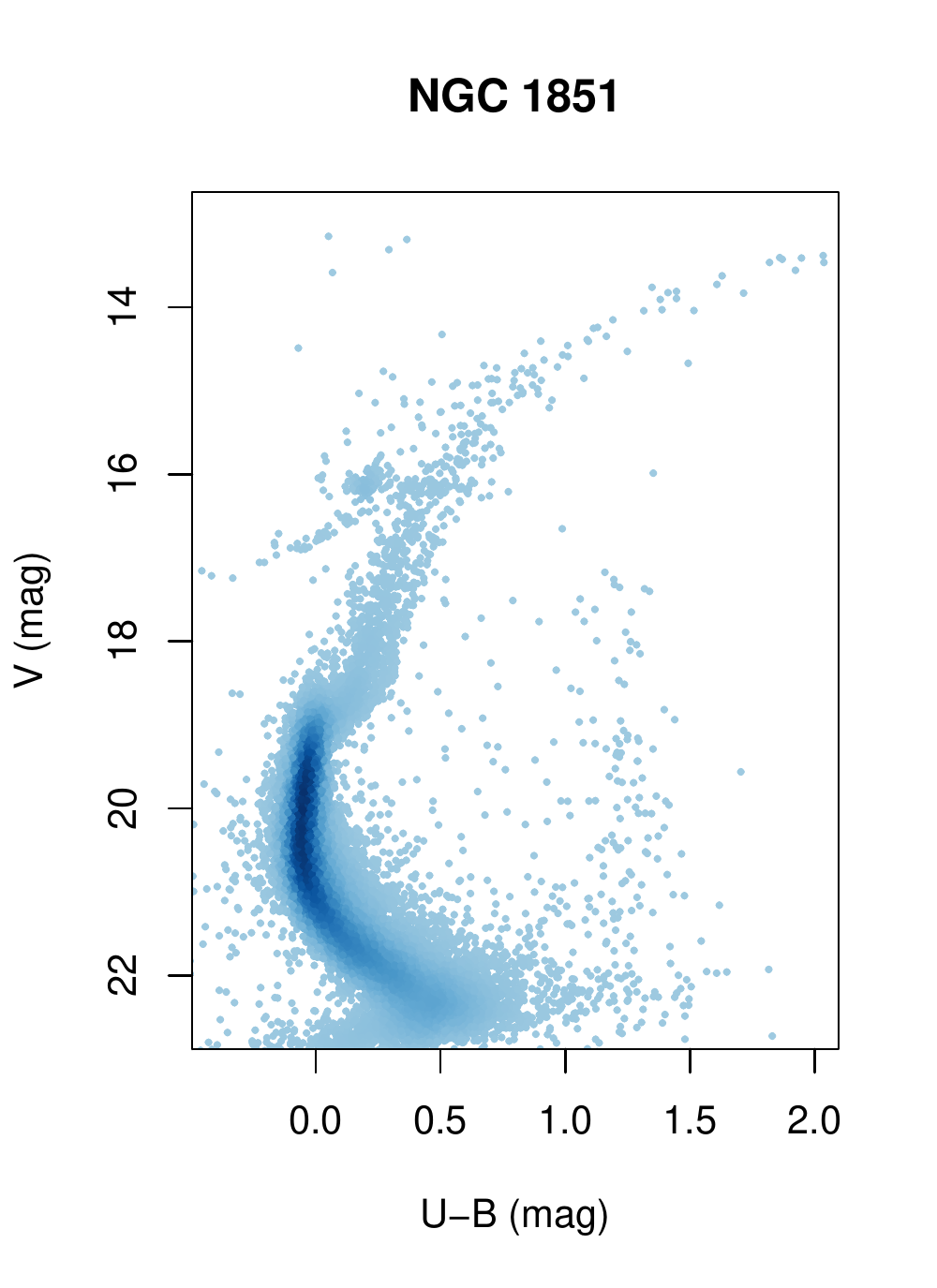}
   \includegraphics[width=0.325\textwidth]{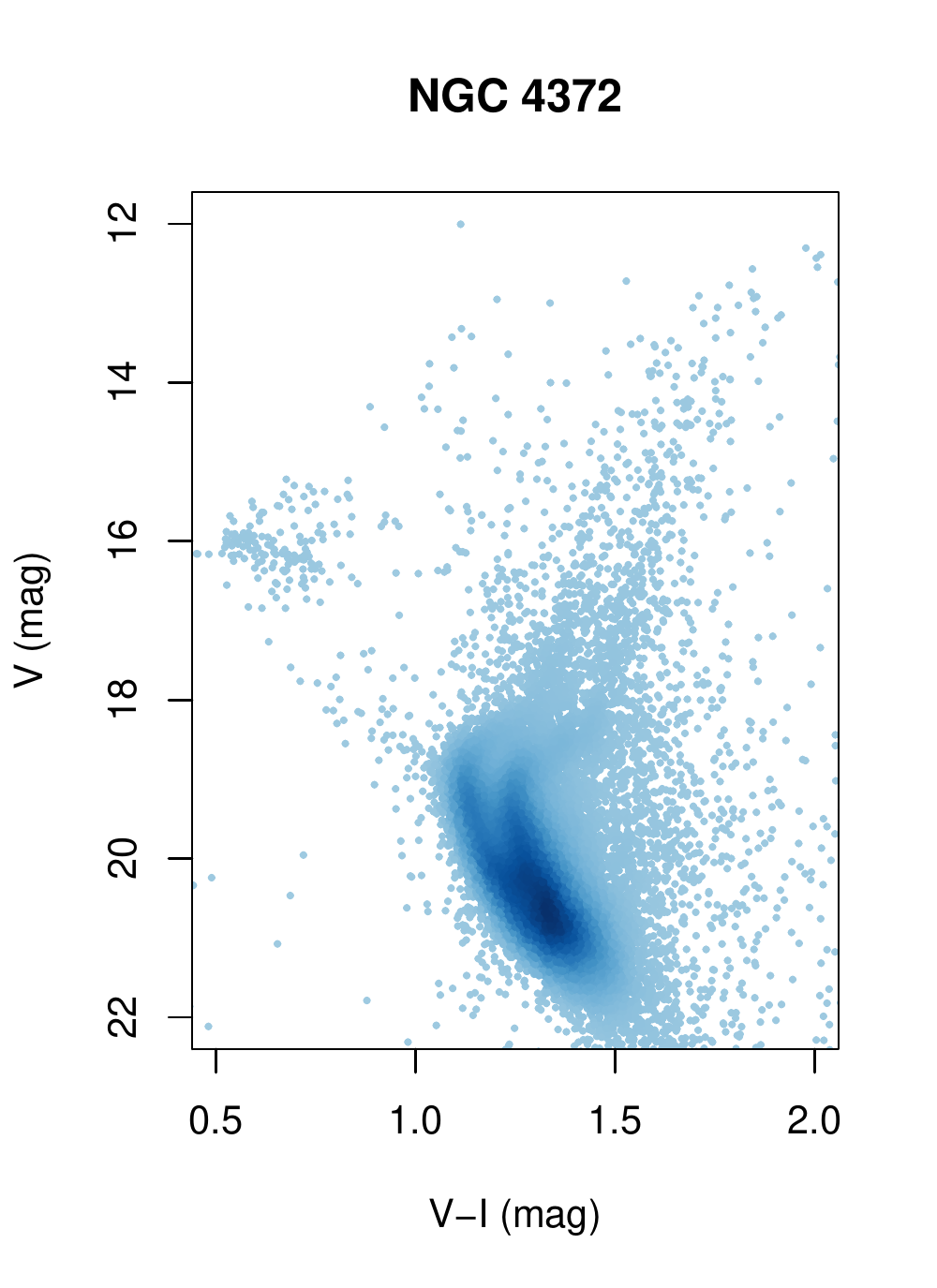}
      \caption{Examples of CMDs, cleaned with procedures such as the ones
      described in Section~\ref{sec:qc}. The adopted selections sacrifice
      photometric depth to enhance the CMD details for stars brighter than
      V$\simeq$20.5~mag (the approximate {\em Gaia} limit). {\em Left panel:}
      the well-defined CMD of NGC\,288, with its beautiful binary and blue
      straggler stars sequences. {\em Center panel:} the complexity and fine
      structure of the RGB of NGC\,1851 seen in $U-B$ colour. {\em Right panel:}
      the striking effect of differential reddening on the CMD of NGC\,4372,
      which appears bifurcated, with a redder turnoff dominated by stars north
      of the GC center and a bluer one dominated by stars south of it.}
   \label{fig:cmds}
   \end{figure*}

For each isopleth, we compute the ellipticity as $\varepsilon = 1 - b/a$. We
show the values of $\varepsilon$ as a function of the semi-major axis $a$ in the
center-right panel of Figure~\ref{fig:ellipses}, where we also compare our
results with those of \citet{lanzoni18} and \citet{white87}. We note how the
contours corresponding to smaller densities appear to be more flattened, in
agreement with what is found in the studies of NGC\,5904 already cited: a
similar behaviour was also observed in other globular clusters
\citep{geyer83,white87}. The improvement in the number and the quality of the
measurements in the ellipticity profile is evident, and the agreement with the
profile by \citet{white87} in the external parts of NGC\,5904 is remarkable.
Figure~\ref{fig:ellipses} (right panel) also illustrates the behaviour of
$\varphi$, the position angle of the fitted ellipses, measured as the
counterclockwise angle between the $x$-axis and the semimajor axis of the
ellipse. A small change in $\varphi$ is observed in the innermost part of the
profile, but the outermost appears to be consistent with a single orientation,
corresponding to $\varphi \lesssim 3 \pi/4$. The change in $\varphi$ in the
innermost region of the cluster needs to be taken with care because of the
completeness issues in the centre. A comparison with the global $\varphi$
estimates by \citet{white87}, \citet{lanzoni18}, and \citet{chen10} is also
shown, and reasonable agreement is found. The centers of the ellipses $(x_{\rm
e},y_{\rm e})$ obtained for the considered levels are always within some arcsec
from each other and from (X$_0$,Y$_0$): the largest distance from the GC
centroid is found for the innermost ellipse, and is of $\simeq 8$\Sec$7$,
possibly another manifestation of crowding issues in the GC central regions.

Often, the ellipticity of the contour containing half the mass of a star cluster
is assumed to be a good {\em global} indication of its morphology
\citep{kontizas89}. We therefore compute the fraction of the total number of
stars that is enclosed in every contour, $f_{\rm int}$
(Table~\ref{tab:ellipses}), and by interpolating it as a function of $f_{\rm
peak}$ we determine $\varepsilon = 0.11 \pm 0.01$ at $f_{\rm int} = 0.5$ for
NGC\,5904. Another possible estimate of the global ellipticity of the cluster can be
calculated as the mean of the ellipticities for all the considered contours,
which in the present case results to be $\varepsilon = 0.12 \pm 0.02$,
consistent with the value ($\varepsilon = 0.14$) calculated in a similar way by
\citet{white87}.

The results presented here show very good agreement with the estimates
available in the literature, but provide a much richer level of detail. The
ellipticity profile computed by \citet{white87} extends farther out than
the one calculated here, but it has significantly larger uncertainties. This
illustrates the impact that the photometry presented here can have in the study
of GC morphology and ellipticity profiles. Combining these catalogues with {\em
Gaia} data and with HST photometry \citep{anderson08,milone17}, it will be
possible to extend the profiles to cover both the most external regions and the
innermost crowded cores.


\subsection{Colour-magnitude diagrams}
\label{sec:clean}

The quality of the CMDs varies significantly depending on several factors,
including: the depth and number of the collected images; the atmospheric
conditions; the telescopes and detector quality; the sky coverage; galactic
coordinates and hence both field-star contamination and the amount of reddening,
especially differential reddening; and last but not least, the intrinsic
properties of each GC. Consequently, some of the GCs can only be studied after
filtering the data with techniques suited to the needs of the specific
scientific investigation. In Section~\ref{sec:qc}, we described the quality
parameters that are included in the catalogue and that can be used to pre-select
stars \citep[see also][for more details]{stetson88}. Additional external
constraints can be extremely useful: for example a differential reddening
correction can be computed following methods similar to that by
\citet{milone12}, or proper motions from {\em Gaia} or other sources can be used
to select probable members. 

   \begin{figure*}
   \centering
   \includegraphics[width=0.325\textwidth]{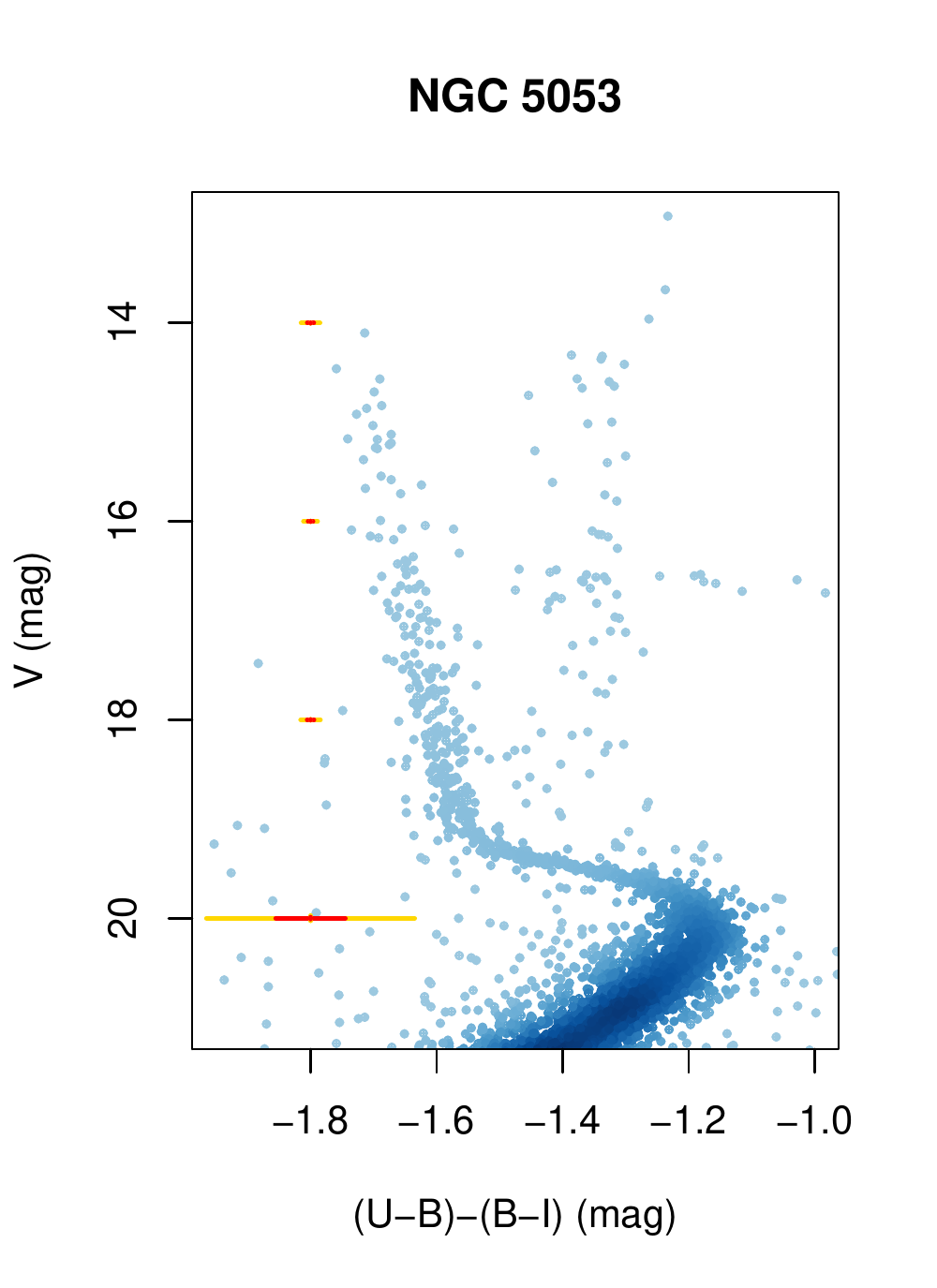}
   \includegraphics[width=0.325\textwidth]{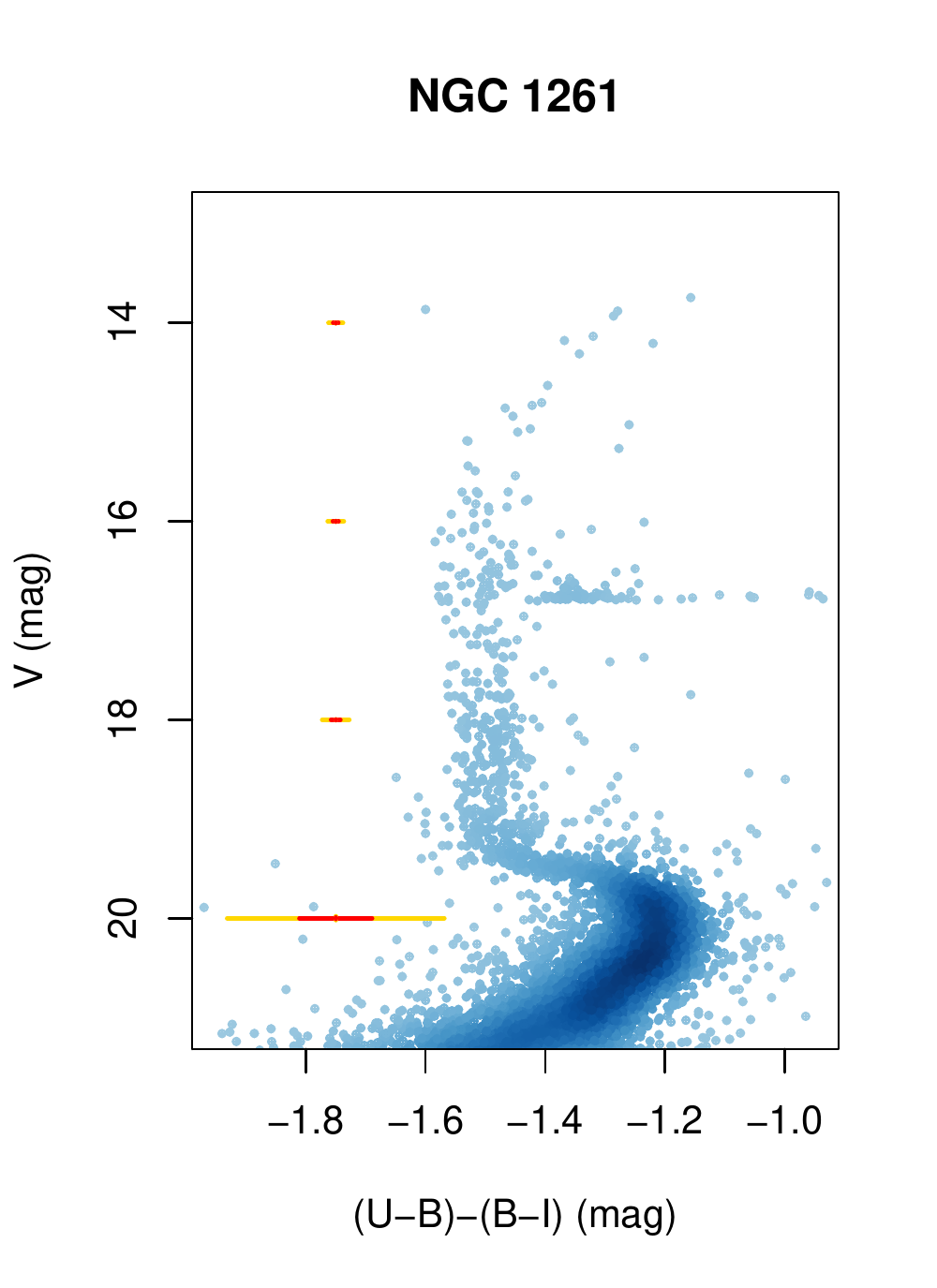}
   \includegraphics[width=0.325\textwidth]{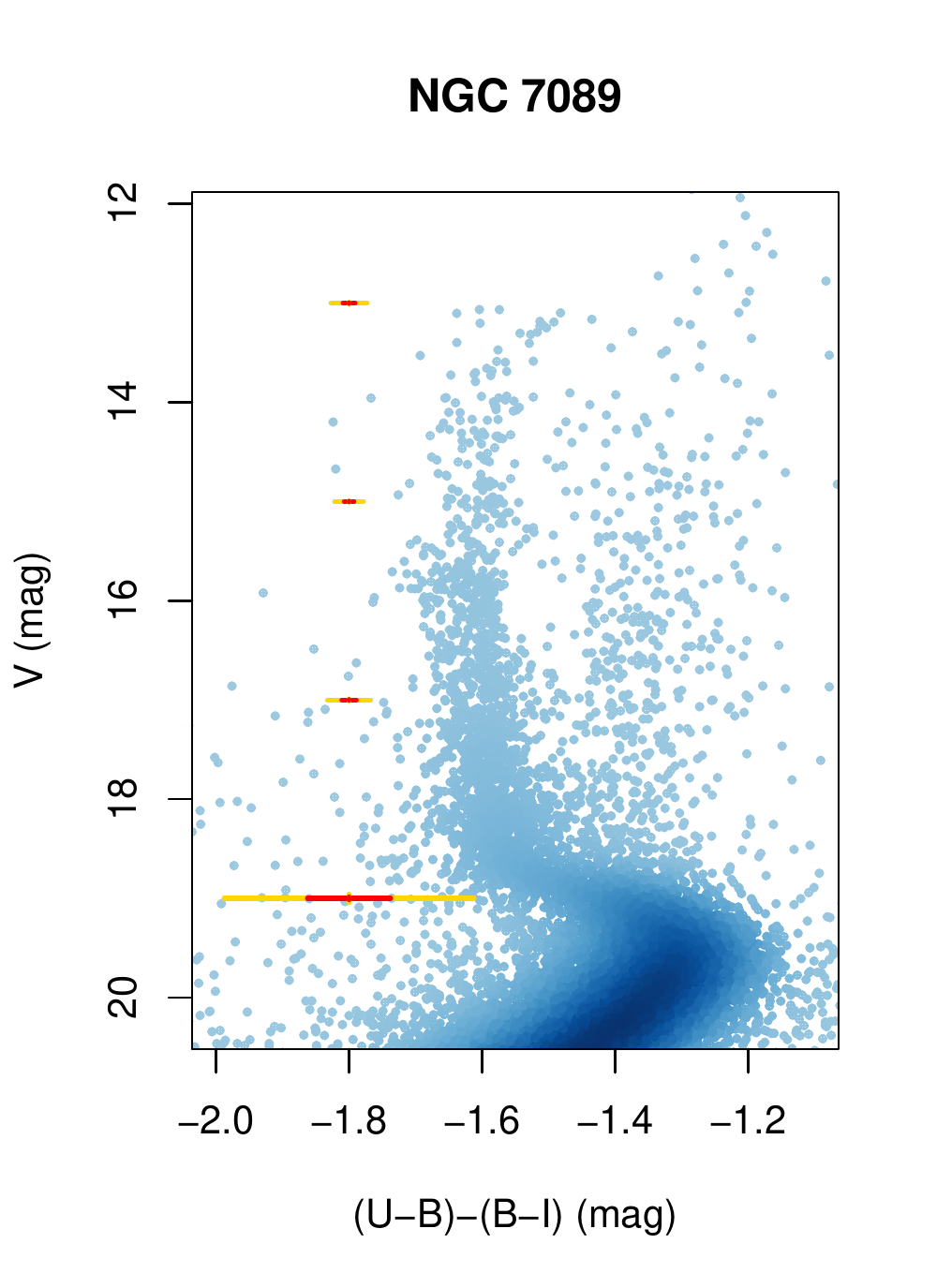}
      \caption{Examples of $V$, $C_{UBI}$ CMDs for three well-known GCs that were
      not included in the sample presented by \citet{monelli13}, sorted by RGB
      width and complexity. {\em Left panel:} NGC\,5053, one of the most
      metal-poor GCs displaying one of the thinnest RGBs in our sample, although
      still a few times wider that the typical photometric errors. {\em Center
      panel:} the beautiful substructure of the RGB of NGC\,1261, a low-reddening
      GC. {\em Right panel:} the complex structure of another anomalous GC,
      NGC\,7089 (M\,2), containing a small fraction of s-rich and C+N+O enhanced
      stars, which are however not clearly visible in $C_{UBI}$. In
      all panels, median errorbars are displayed in gold (3\,$\sigma$) and red
      (1\,$\sigma$) at different magnitude levels. }
   \label{fig:cubi}
   \end{figure*}

To illustrate the variety in our photometric catalogue, we present in
Figure~\ref{fig:cmds} the CMDs of three GCs in different passbands. Our $V$, $B-V$
CMD of NGC\,288 reveals in detail the beautiful binary sequence and blue-straggler
plume that were originally shown by \citet[][with ground-based
photometry]{bolte92} and \citet[][with HST photometry]{bellazzini02a}. NGC\,1851
is one of the so-called anomalous GCs that show an additional sub-population
compared to the typical GCs, in this case redder than the RGB and fainter than the
subgiant branch (SGB), with enhanced C+N+O and s-process element abundances
\citep{lardo12}. Some other well-known GCs in this category are, for example,
$\omega$~Cen \citep{lee99b,pancino00,ferraro04,marino12a}, M\,22
\citep{marino09,marino12b}, and M\,2 \citep{lardo13}. The $V$, $U-B$ CMD that we
present here shows the fine substructure of the RGB of NGC\,1851 with its
anomalous, additional red RGB that appears well separated from the remainder of
the RGB populations. The main RGB in turn is widened (and possibly bimodal), a
typical characteristic of all GCs, caused primarily by nitrogen variations
\citep{monelli13,sbordone11}. 

We also show in Figure~\ref{fig:cmds} the peculiar effect that differential
reddening has on the appearance of the $V$, $V-I$ CMD of NGC\,4372. A sharp
discontinuity in interstellar medium runs diagonally across the cluster's central
regions, splitting stars into two well-separated sets of sequences
\citep{gerashchenko04}. This is especially visible in the turnoff region of our
$V$, $V-I$ CMD, where a bluer turnoff is dominated by stars lying south of the GC
center, and a redder one by stars north of it.

\subsubsection{Multiple populations}
\label{sec:mps}

A full statistical analysis of the photometric properties of MPs in GCs is
clearly beyond the scope of the present paper. Nevertheless, we show the
potential of our catalogue for separating and classifying MPs using the
$C_{UBI}$ index defined by \citet{monelli13} --- based on the index by
\citet{milone13} --- as $(U-B)-(B-I)$. In normal stars of spectral classes G and
K (roughly speaking, $0.6 \lesssim B-V \lesssim 1.5$~mag, or $1.5 \lesssim B-I
\lesssim 3.7$~mag) both $B-I$ and $U-B$ are strongly dependent on temperature,
with a slope ${d(U-B)/ (B-I)} \sim 1$.\footnote{A plot of Landolt's mainly
Population I standard stars, including both giants and dwarfs, is well
represented by a line segment from $(B-I,U-B) = (1.5, 0.1)$ to $(3.7, 2.1)$. 
Metal-poor stars are shifted to smaller $U-B$ colors by their lower line
blanketing, but their sequence remains nearly parallel to the Population I
sequence in this temperature range.}  The $(U-B)-(B-I)$ color difference,
therefore, almost completely removes any temperature sensitivity, unmasking the
smaller effects due to variations in carbon- and nitrogen-sensitive
features---especially the strong CN band at 388~nm---and the effect of helium
abundance on the strength of the Balmer convergence and jump. It is noteworthy
that the reddening slope for stars of these colors is quite different: 
${E(U-B)/E(B-I)} \sim 0.4$, so significant differential reddening reduces the
effectiveness of the $C_{UBI}$ index. 

In the study by \citet{monelli13}, a variety of $C_{UBI}$ morphologies was
illustrated, and we find the same variety in the present sample. Some GCs
clearly show a split RGB in the $V$, $C_{UBI}$ CMD, with two well-separated
branches: the two most striking examples in our sample are NGC\,288 and
NGC\,6981. Some other GCs show (at least) three RGBs, like NGC\,6205, NGC\,1851,
NGC\,2808, $\omega$~Cen, or 47\,Tuc. However in several cases the RGB is not
clearly separated in distinct branches, and it rather shows a complex but not
fully discrete morphology. For a few GCs --- namely E\,3, NGC\,2298, NGC\,4833,
NGC\,5927, NGC\,6760, NGC\,6838, and NGC\,7006 --- differential reddening and/or
field contamination blur the RGB and thus a dedicated treatment of these effects
is needed before studying their MPs in details.

   \begin{figure*}
   \centering
   \includegraphics[width=0.325\textwidth]{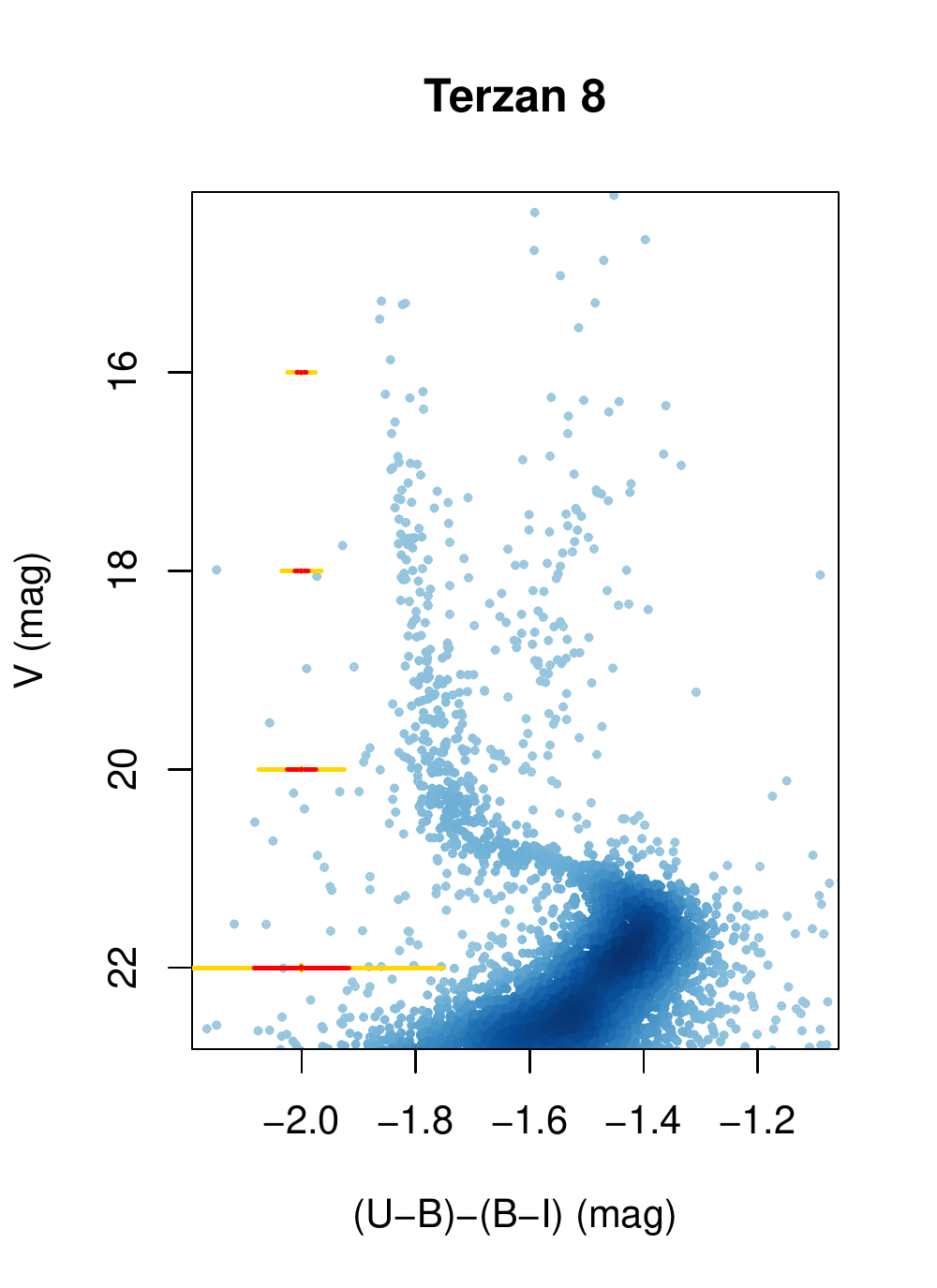}
   \includegraphics[width=0.325\textwidth]{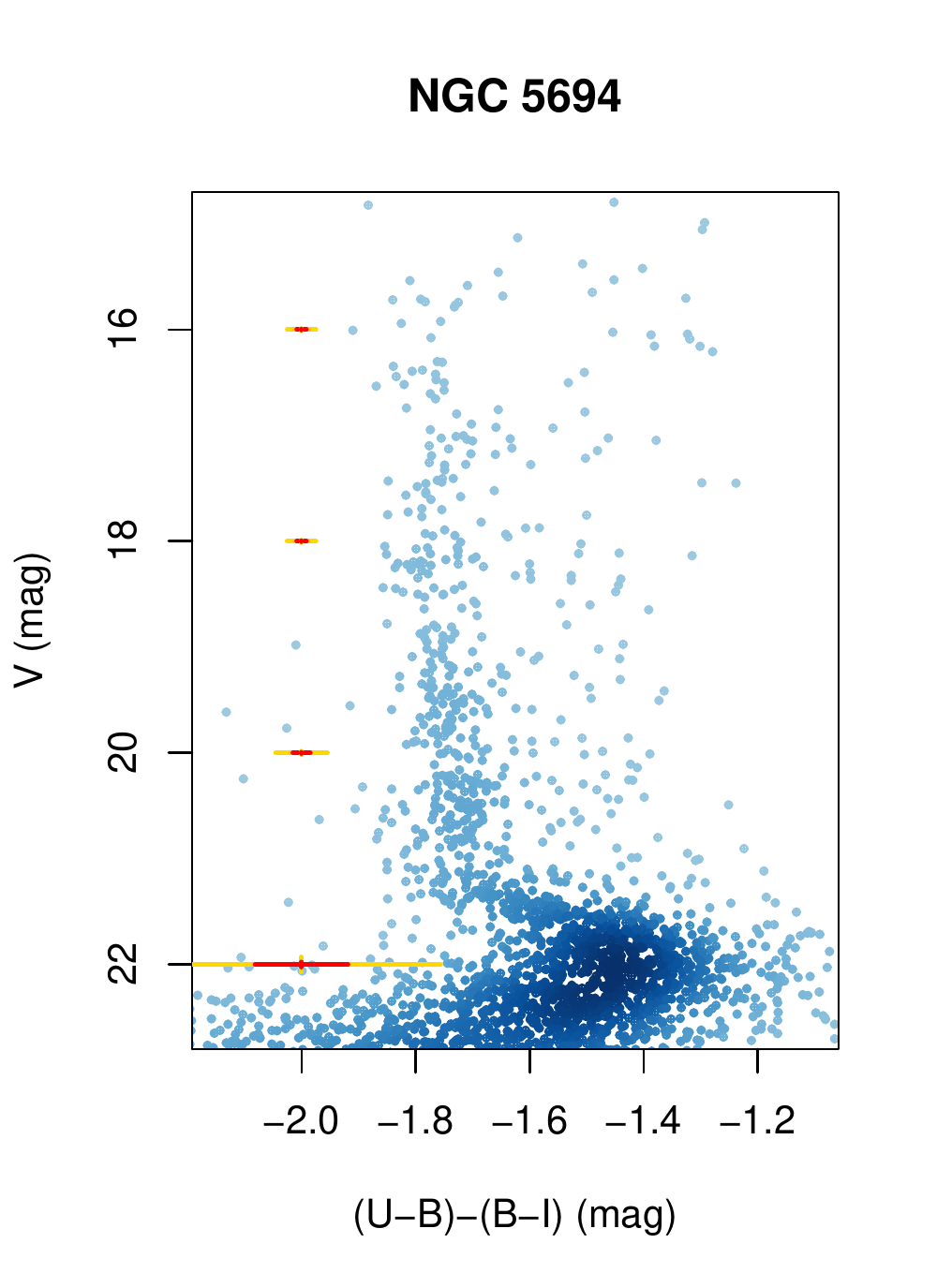}
   \includegraphics[width=0.325\textwidth]{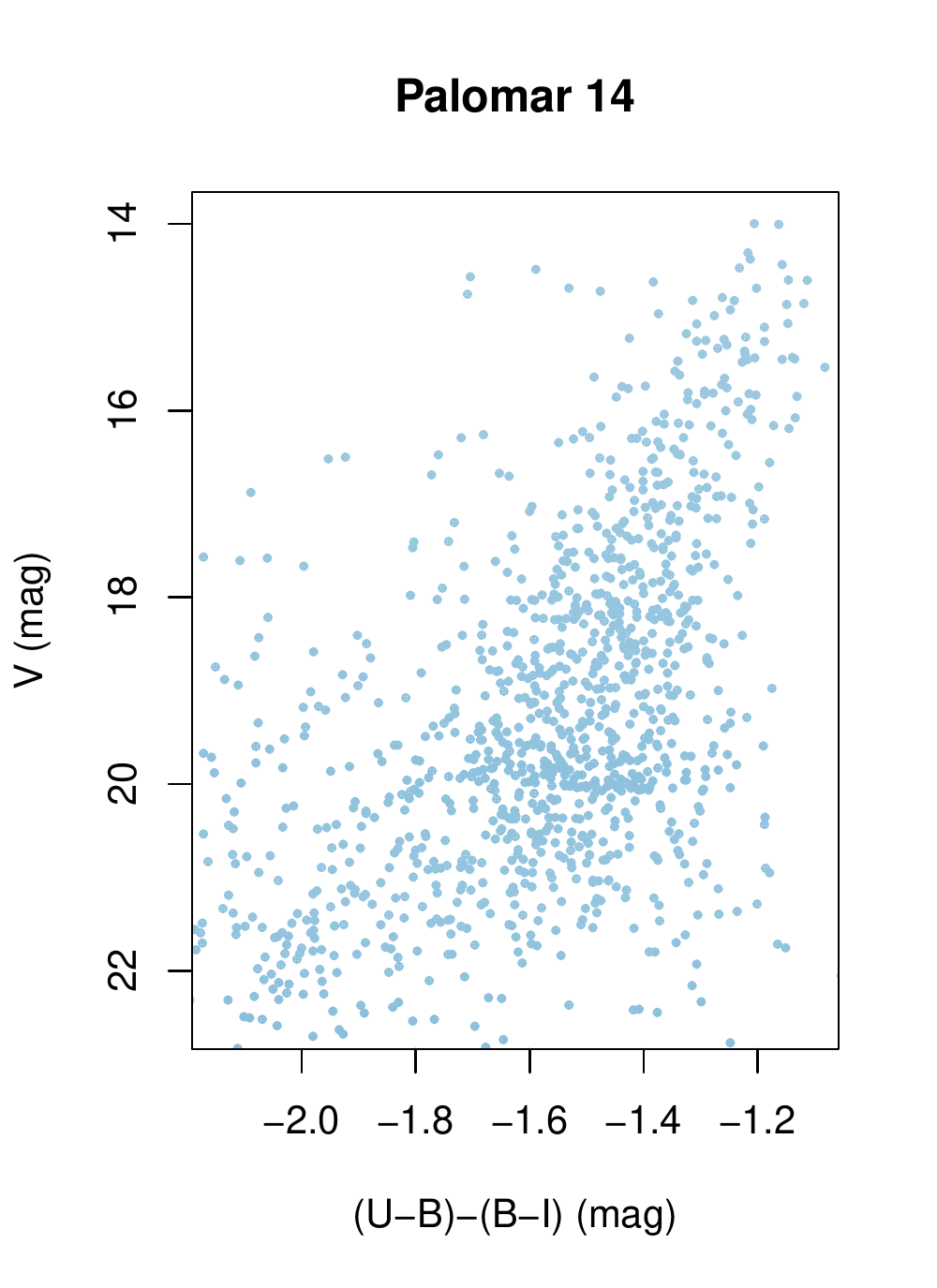}
      \caption{The $V$, $C_{UBI}$ CMDs of candidate single-population or mostly
      single-population GCs in our sample. As in Figure~\ref{fig:cubi}, median
      errorbars at different magnitude levels are shown in gold (3\,$\sigma$) and
      red (1\,$\sigma$). Unfortunately, the current $U$-band photometry of
      Palomar\,14 is not deep enough to study MPs.}
   \label{fig:single}
   \end{figure*}

We show in Figure~\ref{fig:cubi} three examples of $V$, $C_{UBI}$ CMDs of
well-known GCs that were not previously studied by \citep{monelli13}. The first is
NGC\,5053, a very metal-poor GC that may be associated with the disrupting Sgr
dwarf galaxy \citep[although the association is debated:][]{tang18}. It has one of
the narrowest RGBs (another such case is NGC\,4590), especially if compared with
similarly metal-poor GCs such as M\,15, but it still covers $\simeq$0.1~mag, which
is much wider than the typical errors of our photometry (Section~\ref{sec:phot},
Figure~\ref{fig:cubi}). NGC\,1261 is a metal-intermediate GC, with a bifurcated
sub-giant branch \citep{kravtsov10} and a diffuse stellar halo similar to that
observed around NGC\,1851 \citep{kuzma18}. It has low field contamination because
of its high Galactic latitude and therefore it shows very clean and well-defined
sequences. In the $V$, $C_{UBI}$ plane it clearly shows a wide and possibly
bimodal RGB, suggesting more a normal C-N anti-correlation rather than a
complexity comparable to that of NGC\,1851, but further analysis is required.
NGC\,7089 (M\,2) is an anomalous GC like NGC\,1851 but its $V$, $C_{UBI}$ CMD
appears only slightly more complex than that of NGC\,1261 and certainly not as
complex as that of NGC\,1851.

One thing that we noted, that can be also seen by comparing the CMDs presented in
Figures~\ref{fig:cmds} and \ref{fig:cubi}, is that the substructure of the RGB is
definitely clearer in the $V$, $U-B$ plane, because $C_{UBI}$ is the combination
of three magnitudes, with the accompanying error propagation. Moreover, the
combination of the $U-B$ colour, mostly affected by N variations, with $B-I$,
mostly affected by C+N+O, helium, and iron variations, does not represent the
chemistry of the populations as accurately as one would desire. Therefore, to
study multiple populations with ground-based photometry it might be better to
create a ground-based equivalent of the HST chromosome maps \citep{milone17},
i.e., a rectified colour-colour diagram of RGB stars, rather than combining the
two colours in a single index. 

\subsubsection{Single-population clusters}
\label{sec:single}

Finding even one GC that certainly does not host MPs would be very
important to define the boundaries of the MP phenomenon in the space of GC
global properties like mass, age, metallicity, orbit, structural parameters,
and so on. According to \citet{carretta10} or \citet{carrera13}, there seems to
be a minimum cluster mass that allows for the presence of MPs, of about
10$^4$~M$_{\odot}$. This could explain why open clusters generally do not show
MPs \citep[but see][]{pancino18} and why intermediate-age clusters in the
Magellanic Clouds do \citep{martocchia18}. Additionally, \citet{caloi11}
proposed a list of GCs that might contain (or be dominated by) only one
population, based on their HB morphology. 

However, most of the candidate single-population GCs proposed so far have been
shown to actually contain MPs when studied in sufficient detail. Some recent
examples are NGC\,5634, which was shown to contain the Na-O anti-correlation by
\citet{carretta17}; NGC\,6535, probably the lowest-mass GC to show
anti-correlations \citep{bragaglia17}; and NGC\,7099 \citep{omalley18}. 
NGC\,6101, proposed as candidate in the past, was shown to host MPs with HST
photometry \citep{milone17}. Very recently, another single-population candidate,
IC\,4499, was found to host MPs employing new, high-quality photometry and
spectroscopy data \citep{dalessandro18}. 

The GCs that still stand as single-population (or mostly single-population)
candidates are the ones for which the data are presently unsatisfactory. For
example, Palomar\,1 had only 4 stars spectroscopically analyzed
\citep{sakari11}; NGC\,5694 only 6 stars \citep{mucciarelli13}; Terzan\,7 only 5
stars \citep{sbordone05}; Terzan\,8 14 stars with only one having high Na and
low O \citep{carretta14}; Palomar\,12 was studied with high-resolution \citep [4
stars,][]{cohen04} and low-resolution spectroscopy \citep[23 stars with low S/N
ratio,][]{pancino10}, showing only hints of a bimodal C-N anti-correlation; and
Ruprecht\,106 had only 9 stars analyzed spectroscopically \citep{villanova13}
and shows evidence of a relatively small, but non-negligible, widening of the
RGB according to \citet{dotter18}.

Among the GCs in our sample, a few are still suspected of being single-population
or mostly single-population GCs: NGC\,5694, Terzan\,8, and Pal\,14
(Figure~\ref{fig:single}).
Unfortunately, the current $U$-band photometry of Palomar\,14 is not deep enough
to draw any firm conclusion, and we will try to gather more $U$-band data in the
future. While no clear multimodality can be discerned for Terzan\,8 and NGC\,5694
without a deeper analysis, the RGBs of these GCs are as wide as those observed in
other GCs in the sample, in particular they are as wide as that of IC\,4499: the
typical width ranges from 0.05 to 0.15~mag, while the typical propagated $C_{UBI}$
photometric uncertainties for red giants are of the order of 0.01--0.03~mag. 

To conclude, we note that even the GCs that show the thinnest RGBs in the $V$, $U-B$
or $V$, $C_{UBI}$ diagrams (such as NGC\,4590, NGC\,5053, IC\,4499) generally have an
RGB width that is at least a few times the typical photometric errors. Therefore it
is safe to say that none of the GCs studied here can be classified as a genuine
single-population GC, with the possible exception of a few GCs whose photometry is
difficult to interpret without further analysis because of differential reddening or
high field contamination.


\section{Summary and conclusions}
\label{sec:concl}

We have presented wide-field, Johnson-Cousins multiband photometry and astrometry
for 48 galactic GCs, based on a collection of 93\,272\footnote{They are 84\,106
images in $UBVRI$ and 9\,166 in other bands, that were used only to improve on
star's positions and deblending.} public and proprietary images, calibrated on the
basis of 61\,514 local secondary standard stars (Section~\ref{sec:data}). The full
data credits and observing information are provided in 
Appendix~\ref{sec:credits}. The internal and external calibration uncertainties
are estimated to be of the order of a few millimagnitudes, depending on the GC and
the photometric band as discussed in Section~\ref{sec:photcal} and
Appendix~\ref{sec:color}. Our error estimates are confirmed by extensive
literature comparisons and the residual ZP variations in the outer parts of the
photometry are of order 0.04~mag {\it at most\/} (Section~\ref{sec:phot}). For
several GCs in the sample, these are the first publicly available photometric
catalogues in the literature, at least in some of the photometric bands, typically
$U$ and $R$. In particular, among those in our sample, NGC\,6760 is the GC that
was least studied in the past.

We explore the photometry to illustrate its scientific potential
(Section~\ref{sec:cmd}) and find that:

\begin{itemize}
\item{the multiband photometry and the quality indicators provided allow for a
tailored selection of stars depending on the science goals of different
scientific studies; the majority of contaminating field stars can be removed not
only with radial selections from the GC centres, but also using colour-colour
plots; the catalogues provide quality indicators that allow for further cleaning
of the sample from stars with significant PSF-fitting residuals or with a high
chance of being variables;}
\item{the number density and surface brightness profiles that can be obtained
from our catalogues cover a large range of distances from the GC
centers and provide information as detailed as the one found in the available
literature studies; they will be computed for all the GCs in our database, i.e.,
eventually $\simeq$80\% of the \citet{harris10} catalogue;} 
\item{a study of the morphology of NGC\,5904 showed that we can derive
radial ellipticity profiles as accurate as those available in the literature,
but more detailed, thanks to the wide field and depth of our photometry; we
observe also the variation of the position angle of the elliptical contours with
more detail (i.e., more radial bins) than previous studies;}
\item{given the diversity of the collected photometric data in terms of
photometric depth, area coverage, measurement quality, and filter bandpasses
(especially $U$), we strongly recommend going through the image credits and
sources in Appendix~\ref{sec:credits} to identify the exact characteristics of
the data, and to allow for millimagnitude-scale residual uncertainties in our
calibration of the data to the Landolt system when only a few instrumental
setups are represented;}
\item{we show that none of the GCs in our sample can be safely assumed to be
free from MPs, except for a few GCs for which no clear conclusion can be made
because they have significant differential reddening, field contamination, or
insufficient $U$-band photometric depth; in particular, two GCs that were
considered good single-population or mostly single-population candidates,
Terzan\,8 and NGC\,5694, indeed have RGB widths that are significantly
wider than expected from photometric errors alone.}
\end{itemize}

Our photometry can bridge the gap between high-spatial-resolution studies carried
out with the HST and large photometric surveys like {\em Gaia}, SDSS, or LSST, for
which no PSF photometry was performed to study the crowded GC stellar fields.
Combining HST, large surveys, and our photometry with astrometry from {\em Gaia}
and spectroscopy from large surveys will allow GC research to enter the {\em Gaia}
era with a complete set of measurements to tackle the numerous open problems and
stimulate further theoretical advancements.


\section*{Acknowledgments}

We warmly thank the following colleagues for interesting discussions and
suggestions: P.~Bianchini, R.~Contreras, E.~Dalessandro, V.~H\'enault-Brunet,
C.~Lardo, C.~E.~Mart\'\i nez-V\'azquez, S.~Marinoni, P.~Marrese, L.~Monaco,
R.~M.~Murillo, A.~Sollima, A.~L.~Varri, and M.~Zoccali. AZ acknowledges support
through a ESA Research Fellowship. We thank the International Space Science
Institute (ISSI, Bern, CH) for welcoming the activities of the Team 407 {\em
``Globular Clusters in the Gaia Era"}. The full image sources and credits can be
found in Appendix~\ref{sec:credits}. This research has made ample use of the
Simbad astronomical database \citep{simbad} and the Vizier catalogue access tool
\citep{vizier}, both operated at the CDS in Strasbourg, and of Topcat
\citep{topcat}. Figures in this paper were produced either using the R
programming language \citep{R,data.table} and Rstudio
(https://www.rstudio.com/), or with Python (https://www.python.org/) and
specifically with the packages numpy, scipy, and matplotlib. This work presents
results from the European Space Agency (ESA) space mission {\em Gaia}. {\em
Gaia} data are being processed by the Gaia Data Processing and Analysis
Consortium (DPAC). Funding for the DPAC is provided by national institutions, in
particular the institutions participating in the Gaia Multi-Lateral Agreement
(MLA). This research has made use of the GaiaPortal catalogues access tool, ASI
- Space Science Data Center, Rome, Italy (http://gaiaportal.ssdc.asi.it).


\bibliographystyle{mn2e}

\bibliography{PeterGlobs}



\appendix


\section{Data logs and credits}
\label{sec:credits}

As mentioned, the catalogues presented are based on 
93\,272 individual CCD images (84\,106 in filters approximating Landolt's, and
another 9\,166 in other filters) collected over the course of more than 35 years
of observation at various telescopes (Table~\ref{tab:datasets}). 

The details of the data sources and credits are summarized in
Tables~\ref{tab:runs} and \ref{tab:gcimages}. The former contains a full list of
all the 390 observing runs used for the present study, including the program IDs
and, whenever known, the name of the observer. The latter lists, for each GC,
the runs from which the data are taken and the number of images used in each of
the five filters. The full tables can be found in the electronic version of the
journal and at CDS, here we report the list of columns and a brief description
of their content. 

\begin{table}
\caption{List of all the 384 observing runs included in the present study, with
some ancillary information, including the program ID or the observer, when
known. The table can be found in its entirety in the online version of the
journal, and at CDS.}            
\label{tab:runs}
\centering
\begin{tabular}{ll}
\hline\hline       
Column     & Description \\
\hline  
Run        & Unique run label (e.g., emmi8) \\
Date       & Observing date(s) covered (yyyy mmm dd-dd) \\
Site       & Observing site (as in Table~\ref{tab:datasets}) \\
Telescope  & Telescope (as in Table~\ref{tab:datasets}) \\
Instrument & Instrument (as in Table~\ref{tab:datasets}) \\
$n_U$      & Number of images in $U$ \\
$n_B$      & Number of images in $B$ \\
$n_V$      & Number of images in $V$ \\
$n_R$      & Number of images in $R$ \\
$n_I$      & Number of images in $I$ \\
$n_{\rm{other}}$ & Number of images in other filters \\
Multiplex  & For mosaic cameras: \# of CCDs in mosaic \\
Program    & Program ID (e.g., 083.D-0544(A), when known) \\
Observer   & Observer(s) name (when known) \\
Source     & Provider's name (when known) \\
\hline
\end{tabular}
\end{table}

\begin{table}
\caption{For each of the 48 GCs listed in Table~\ref{tab:clusters}, we report
here the data sources used (from Table~\ref{tab:runs}) and the number of images
analyzed in each of the five {\it UBVRI\/} passbands, where "other" refers to
non-standard passbands or other observations acquired in the same nights. The
table can be found in its entirety in the online version of the journal, and at
CDS.}            
\label{tab:gcimages}
\centering
\begin{tabular}{lll}
\hline\hline       
Column     & Description \\
\hline
Cluster    & GC name as in Table~\ref{tab:clusters} \\
Run        & Unique run label as in Table~\ref{tab:runs} \\
Date       & Observing date(s) covered (yyyy mmm dd-dd) \\
$n_U$      & Number of images in $U$ \\
$n_B$      & Number of images in $B$ \\
$n_V$      & Number of images in $V$ \\
$n_R$      & Number of images in $R$ \\
$n_I$      & Number of images in $I$ \\
$n_{\rm{other}}$ & Number of images in other filters \\
Multiplex  & For mosaic cameras: \# of CCDs \\
\hline
\end{tabular}
\end{table}


\section{Photometric calibration}
\label{sec:color}

Our ultimate goal is to provide final calibrated magnitudes on a photometric
system as similar to that of \citet{landolt92a} as possible. 

\subsection{Color and extinction corrections}

It is a
widely known fact that different combinations of glass filter and photosensitive
detector define bandpasses that differ from camera to camera.  Landolt has
published throughput curves for his filters and quantum efficiency curves for
his photomultipliers, but such information is not generally available for the
many different equipment setups represented in our database.  Other effects,
such as the instantaneous state of the reflective and transmissive elements in
the light path of the telescope and instrument, and the instantaneous
transparency of the terrestrial atmosphere, can also be important and are not
known to us in advance of the data analysis. 

Following standard practice that dates back {\it at least\/} to Johnson's  and
Hardie's definitive publications \citep[pages 157 and 178
in][respectively]{hiltner62} we compute a numerical mapping of the instrumental
photometric system of each night's observing to the fundamental photometric
system based upon the known broad-band colors of the pre-defined photometric
standards.  The approach exploits the fact that the family of stellar spectral
energy distributions for normal stars is close to a one-parameter family driven
primarily by effective temperature and reddening, and only slightly modulated by
other effects such as surface gravity, chemical abundance, and rotation.  A
typical example of one of our transformation equations is:

$$v = V + z_V + \alpha_V \cdot (B-V) + \beta_V \cdot (B-V)^2 + k_V\cdot X,$$

\noindent where $v$ is the instrumental magnitude obtained during the course of
our measurements, which scales as $-2.5~\log [
\rm{(counts)}/{\rm{(integration~time)}}]$; $V$ is the standard visual magnitude
as published by \citet{landolt92a}; $z_V$ is a photometric ZP which accounts for
the overall transmission of the telescope and instrument optics, filter
throughput, and detector quantum efficiency; \hbox{\it B--V\/} is the star's
measured color, again as published by \citet{landolt92a}; $X$ is the airmass at
which the observation is obtained, such that the airmass is defined to be unity
when observing at the zenith and goes very nearly as $\sec{z}$, where $z$ is the
angular zenith distance\footnote{In fact, a more accurate formula for $X$ that
takes into account the curvature of the atmosphere is employed, but the
distinction is not important for this discussion.}. In this equation, the
quantities $V$, \hbox{\it B--V\/}, and $X$ are presumed to be known quantities
with uncertainties negligible compared to the uncertainty of the observed
quantity $v$, while $\alpha_V$, $\beta_V$, and $k_V$ are treated as quantities
that are, initially, completely unknown and must be derived empirically from our
own observations of $v$ for many diverse stars by the method of (robust) least
squares.  Very similar equations are also employed for the $U$, $B$, $R$, and
$I$ photometric bandpasses, with some detailed modifications to be discussed
below.  

In this calibration, we take cognizance of the fact that the $v$ bandpass
defined by the total throughput of our combination of atmosphere, telescope, and
filter multiplied by detector sensitivity,
as a function of wavelength ($\lambda$) is not identical in detail
to Landolt's $V$ bandpass.  We ask the question: when Landolt observes a
star to have a particular $V$ and {\it B--V}, what $v$ magnitude do we observe? 
The calibration equation illustrated above represents a second-order
Taylor-expansion answer to that question, where {\it B--V} takes on the r\^ole
of $\rm d \log(\rm{flux})/ \rm d \lambda$ and $\alpha$ and $\beta$ are an
empirical second-order description of the $\Delta\lambda$ and $\Delta\lambda^2$
differences between Landolt's $V$ photometric bandpass and our $v$ bandpass.

In the case of the $B$ filter, we add a term in the calibration equal to $-0.016
\, (B-V) \, X$, to correct in gross fashion for the fact that hot blue
stars preferentially produce photons that pass through the shorter-wavelength
flank of the $B$ filter, while more of the light from red stars passes through
the longer-wavelength side of the filter.  The difference in the effective
wavelength of the $B$ filter for stars of extreme color is sufficient that the
hotter stars experience perceptibly  higher extinction in the terrestrial
atmosphere than cooler stars.  The value --0.016 is obtained by a numerical
analysis of observed stellar spectral-energy distributions multiplied by
Landolt's $B$ bandpass and a model terrestrial atmosphere for an observatory
roughly 2\,000~m above sea level. The effect is too subtle to be measured with
any precision in our actual photometric data.

The color terms in \hbox{\it B--V\/} are replaced by corresponding terms in
\hbox{\it V--R\/} in the equation for the $R$ band, and by terms in \hbox{\it
V--I\/} for the $I$ equation.  This ensures that the gross slope of the spectral
energy distribution is estimated at wavelengths close to the bandpass
being considered.  

The $U$ bandpass represents a particular problem. Landolt's $U$ bandpass is
defined on the short-wavelength side by the throughput of the terrestrial
atmosphere more than by either filter throughput or detector sensitivity. 
Furthermore, the $U$ filter as provided by most observatories is by far the most
poorly standardized.  Finally, the $U$ bandpass contains the Balmer convergence
and jump, as well as significant metal-line blanketing and molecular bands,
including most especially the 388\,nm band of CN.  Thus, unlike the case with the
other filters, the $U$ magnitude is not a monotonic function of temperature for
fixed apparent $V$ or other reference magnitude.  Our default calibration
equation for our instrumental $u$ observations is 

$$u = U + z_U + \alpha_U\cdot (U-B) + \beta_\cdot (B-V) + \gamma_U \cdot (B-V)^2 + k_U\cdot X.$$

\noindent To the extent that the classical ({\it U--B, B--V}) color-color
diagram can be approximated by three line segments---{\it U--B} increases with
{\it B--V}, {\it U--B} decreases as {\it B--V} increases, {\it U--B} increases
with {\it B--V}---with smooth transitions between, this formulation allows the
first-order Taylor expansion $u - U \sim \gamma(\hbox{\it U--B})$ to have a
different slope in each of the three regimes of \hbox{\it B--V}. In our
experience, the data are almost never good enough to determine reliable
coefficients for still higher-order terms. In particular, our limited attempts to
estimate a color-extinction term for $U$, similar to that employed for $B$, have not yet
led to perceptible improvements in the photometry.  The likelihood is that this
relatively subtle effect is being swamped by grosser systematics due to the
greater complexity of stellar spectral energy distributions at these wavelengths
and the diversity of the available $U$-band filters.

\subsection{Spatial corrections}

Wide-field cameras and other instruments with focal reducers often suffer from
vignetting and distortion effects \citep{andersen95} that introduce variations
in the measured stars' magnitudes as a function of position on the focal plane,
beyond what can be corrected by the flat-field corrections.  Scattered light that
enters the flat-field exposures but is not present during night-time science exposures
can also introduce a spurious positional gradient in the flux scale of flat-fielded
science images.  To the extent practical, vignetted regions were included in the
data masks derived for each chip for each observing run, and so should have been
largely omitted during our photometric reductions. To deal with residual photometric
variations that result from marginal vignetting and imperfect flat-field exposures, we routinely add
terms $+ \delta x + \epsilon y$ to the calibration equations presented in the
previous section, where $x$ and $y$ are the stars' coordinates in the natural
pixel grid of the CCD.  For mosaic cameras these coefficients are determined
completely independently for the individual chips.  As a result, calibration
effects that are primarily radial with respect to the center of the array can be
adequately approximated by effects that are planar over the individual
detectors.  These coefficients are also left to vary freely from night to night,
as are the ZPs, under the assumption that the different flat fields obtained for
different nights might not experience the same contamination.  

For very small detectors, a few arcminutes on a side and perhaps not containing
a large number of standard stars, these spatial terms are often omitted and any
uncorrected ZP gradient contributes to the observed scatter in the photometric
residuals.  Conversely, for large detectors where the photometric scatter is
large, we inspect residual plots and, where indicated, add calibration terms in
$x^2$, $xy$, and $y^2$. In the case of the ESO WFI, \citet{manfroid01} and
\citet{koch04} report 0.1--0.2~mag maximum variations that can be repaired with
a mosaic-wide quadratic form. In fact, for the WFI images employed here,
quadratic terms never led to a significant reduction in our fitting residuals,
presumably because our individual linear corrections for the different chips
were adequate. However, there were a few cases where quadratic positional
corrections improved the residuals for large, on-axis CCDs.

\subsection{Night-to-night differences}

Obviously, independent values of the various $\alpha$'s, $\beta$'s, and
$\gamma$'s, are obtained for each observing run; the $k$'s, $\delta$'s and
$\epsilon$'s are almost always determined independently for each individual
photometric night.  However, normally we will require the different CCDs of a
mosaic camera to have the same value of $k$ in a given filter on a given night
under the assumption that all are looking through effectively the same
atmosphere during an integration\footnote{And also assuming that the atmosphere
is the only source of airmass dependence in the observation:  instrumental
flexure, for instance, is assumed not to be a significant factor}. Usually we
will enforce the same value of $\beta$, and often the same value of $\alpha$ on
the different CCDs of a mosaic, under the assumption that different CCDs of
almost identical manufacture will have similar quantum efficiency curves, such
that the photometric bandpass is mostly defined by the atmosphere, telescope,
camera optics, filter, and the {\it average\/} CCD sensitivity dependence as a
function of $\lambda$; any residual differences in, say, the overall quantum
efficiency and gain of each detector, and gross positional variations in the
throughput of the filter are adequately absorbed into the different detectors'
ZPs, $z$---which are {\it always\/} left to be freely and independently
determined for each CCD on each night---and $\delta$ and $\epsilon$. 

Similarly, we usually require the coefficients $\alpha$ and $\beta$ to be the same
for a given detector on consecutive nights of an observing run where there is no
evidence that an instrument change has taken place, under the assumption that
these parameters are defined by physical properties of the equipment that should
not vary on timescales of days.  It also goes without saying that we examine the
residuals of our transformation fits and if we see evidence that our assumptions
fail to a degree that is significant compared to the random uncertainties of our
observed magnitudes, we leave the corresponding parameters free in the least-squares
reductions.  

On nights when the observed scatter in the standard-star residuals and/or
bizarre extinction coefficients imply that photometric conditions were not
obtained, we do not compute extinction coefficients $k$ or nightly ZPs $z$. 
Instead we compute an independent photometric ZP for each individual CCD image
based on the photometric standards contained within that image; if an image from
a non-photometric night contains no photometric standards, it is of no
scientific use to us. 

\subsection{Magnitude averages}

To obtain final Landolt-system magnitudes for our target stars, that is, for
stars for which these magnitudes are not known {\it a priori}, we use exactly
the same transformation equations, but the terms on the right sides of the
equations reverse their r\^oles.  The instrumental magnitudes ($v$ and $b$, for
example) continue to be observed quantities which alone contain uncertainty. 
Also as before, the airmass $X$ is presumed to be known with perfect geometric
accuracy for each observation.  Unlike before, however, the $z$'s, $\alpha$'s,
$\beta$'s, $\gamma$'s $\delta$'s, $\epsilon$'s and $k$'s are now presumed to be
known quantities, with uncertainties that are negligible compared to the
observational uncertainties in $v$ and $b$.  The standard-system magnitudes ($V$
and $B$ in this example) and the implied value of {\it B--V} are now the unknown
quantities to be determined via least squares.  

All of the available values of $v$ from all of the images on all of the
nights from all of the observing runs where that star has been observed in
each observatory's approximation of the $V$ filter, and likewise all of the available
values of $b$ are employed in a numerical solution
to derive that singular value of $V$ and that singular value of $B$ that best
explain all the observations in a robust least-square sense.  Since {\it B--V}
is not known ahead of time, a typical value must be assumed initially, and the
pair of transformation equations solved iteratively.  As long as
$\left|\alpha\right|,\left|\beta\right|,\left|\gamma\right| \ll 1$ convergence
to an unambiguous solution of the system of equations is extremely rapid.  Even
if $\left|\alpha,\beta,\gamma\right| \lesssim 1$, convergence occurs eventually
unless few observations are available and some of them are highly defective.  This
approach presumes, of course, that the star is not variable on time scales
relevant to the available data, but the root-mean-square residuals of the
solutions to these equations is always computed and is available as an index of
intrinsic variability or photometric blunder.  Stars with repeatability poorer
than expected from the estimated uncertainties of the individual values of $v$
and $b$ are flagged and identified as either candidate variable stars or stars with
seriously defective observations. 

Overall, our network of calibration equations is both empirical and approximate,
necessarily, given the nature of the data and the problem.  For average stars
observed under average conditions, the calibration approaches exactitude; for
extreme stars observed under extreme conditions, it is approximate.  Unmodeled
complexity in the color-dependent extinction beyond the corrections described
here, for instance, might mean that very red stars observed at high airmass will
have a residual calibration error that is of one sign, while stars much bluer
than average would have a calibration error of the opposite sign.  For those
same stars observed at very low airmass the signs of the residual errors will
reverse, and the average results, interpolated to average airmass, will be
nearly correct.  The behavior of the same two stars observed during a different
observing run might be similar, or opposite, or indeed the residual calibration
errors might be effectively zero.  In the long run, over many observing runs
with many equipment setups, the net results of neglected complexity in the model
transformations should decay away, in the average.  

These residual calibration errors will still contribute, however, to the
observation-to-observation repeatability of the magnitudes for individual stars,
and will therefore be reflected in the standard errors of the mean magnitudes
that we report with each measured magnitude.  In general, therefore, we expect
that stars of ``typical'' colors, say $0.5 < B-V < 1.1$~mag, will have very
reliably determined magnitudes.  Results for stars of extreme color, $B-V
\lesssim 0$ or $B-V\ \gtrsim 2.0$, for example, will be more uncertain and will
have larger reported uncertainties.  Stars near the celestial poles may have
larger systematic errors than stars that pass overhead, because they can never
be observed at low airmass.  But all these problems should (a) be reflected in
our reported standard errors of the mean magnitudes for each star, and (b)
decrease as the number of independent times and nights and observing runs during
which the star was observed increases.  

In general, our magnitudes will be least reliable in $U$, where the total number
of observations is small and the diversity of available filters is large.  In
particular, there is the possibility of a drift in the ZP of our $U$
magnitude scale for very blue and very red stars near the two celestial poles, where our neglect of a
color-extinction correction in the $U$ transformation equation becomes a
systematic, rather than a random error.  The magnitudes will be most reliable in
$V$, where the number of available images is the largest, and the homogeneity of
the photometry is high due to the relative simplicity of stellar spectral energy
distributions at these wavelengths and the general similarity of the various
observatories' $V$ filters.  $B$ and $I$ should be nearly as good as $V$, while
$R$ suffers mainly from a paucity of available data.  These statements are
further quantified in Section~\ref{sec:photcal}.


\label{lastpage}

\end{document}